\documentclass[longauth,traditabstract]{aa} % for the abstract without structuration  % (traditional abstract)

\usepackage{natbib}
\bibpunct{(}{)}{;}{a}{}{,} %
\usepackage{graphicx}
\usepackage{txfonts}
\usepackage[breaklinks, colorlinks, citecolor=blue]{hyperref} %TURN THIS OFF TO PRODUCE REFEREE VERSION ! (ONE COLUMN)

\usepackage{overpic}
\usepackage{fixltx2e}
\usepackage{url}
\usepackage{rotating}

\def\setsymbol#1#2{\expandafter\def\csname #1\endcsname{#2}}
\def\getsymbol#1{\csname #1\endcsname}

\def\Planck{\textit{Planck}}

\def\HeJT{$^4$He-JT}

\def\allearlypapers{\nocite{planck2011-1.1, planck2011-1.3, planck2011-1.4, planck2011-1.5, planck2011-1.6, planck2011-1.7, planck2011-1.10, planck2011-1.10sup, planck2011-5.1a, planck2011-5.1b, planck2011-5.2a, planck2011-5.2b, planck2011-5.2c, planck2011-6.1, planck2011-6.2, planck2011-6.3a, planck2011-6.4a, planck2011-6.4b, planck2011-6.6, planck2011-7.0, planck2011-7.2, planck2011-7.3, planck2011-7.7a, planck2011-7.7b, planck2011-7.12, planck2011-7.13}}

\def\all2013resultspapers{\nocite{planck2013-p01, planck2013-p02, planck2013-p02a, planck2013-p02d, planck2013-p02b, planck2013-p03, planck2013-p03c, planck2013-p03f, planck2013-p03d, planck2013-p03e, planck2013-p01a, planck2013-p06, planck2013-p03a, planck2013-pip88, planck2013-p08, planck2013-p11, planck2013-p12, planck2013-p13, planck2013-p14, planck2013-p15, planck2013-p05b, planck2013-p17, planck2013-p09, planck2013-p09a, planck2013-p20, planck2013-p19, planck2013-pipaberration, planck2013-p05, planck2013-p05a, planck2013-pip56, planck2013-p06b}}

\newbox\tablebox    \newdimen\tablewidth
\def\leaderfil{\leaders\hbox to 5pt{\hss.\hss}\hfil}
\def\endPlancktable{\tablewidth=\columnwidth 
    $$\hss\copy\tablebox\hss$$
    \vskip-\lastskip\vskip -2pt}
\def\endPlancktablewide{\tablewidth=\textwidth 
    $$\hss\copy\tablebox\hss$$
    \vskip-\lastskip\vskip -2pt}
\def\tablenote#1 #2\par{\begingroup \parindent=0.8em
    \abovedisplayshortskip=0pt\belowdisplayshortskip=0pt
    \noindent
    $$\hss\vbox{\hsize\tablewidth \hangindent=\parindent \hangafter=1 \noindent
    \hbox to \parindent{$^#1$\hss}\strut#2\strut\par}\hss$$
    \endgroup}
\def\doubleline{\vskip 3pt\hrule \vskip 1.5pt \hrule \vskip 5pt}

\def\L2{\ifmmode L_2\else $L_2$\fi}

\def\DeltaT{\ifmmode \Delta T\else $\Delta T$\fi}
\def\deltat{\ifmmode \Delta t\else $\Delta t$\fi}
\def\fknee{\ifmmode f_{\rm knee}\else $f_{\rm knee}$\fi}
\def\Fmax{\ifmmode F_{\rm max}\else $F_{\rm max}$\fi}
\def\solar{\ifmmode{\rm M}_{\mathord\odot}\else${\rm M}_{\mathord\odot}$\fi}
\def\Msolar{\ifmmode{\rm M}_{\mathord\odot}\else${\rm M}_{\mathord\odot}$\fi}
\def\Lsolar{\ifmmode{\rm L}_{\mathord\odot}\else${\rm L}_{\mathord\odot}$\fi}

\def\inv{\ifmmode^{-1}\else$^{-1}$\fi}
\def\mo{\ifmmode^{-1}\else$^{-1}$\fi}
\def\sup#1{\ifmmode ^{\rm #1}\else $^{\rm #1}$\fi}
\def\expo#1{\ifmmode \times 10^{#1}\else $\times 10^{#1}$\fi}
\def\,{\thinspace}
\def\lsim{\mathrel{\raise .4ex\hbox{\rlap{$<$}\lower 1.2ex\hbox{$\sim$}}}}
\def\gsim{\mathrel{\raise .4ex\hbox{\rlap{$>$}\lower 1.2ex\hbox{$\sim$}}}}

\def\simprop{\mathrel{\raise .4ex\hbox{\rlap{$\propto$}\lower 1.2ex\hbox{$\sim$}}}}
\def\deg{\ifmmode^\circ\else$^\circ$\fi}
\def\pdeg{\ifmmode $\setbox0=\hbox{$^{\circ}$}\rlap{\hskip.11\wd0 .}$^{\circ}
          \else \setbox0=\hbox{$^{\circ}$}\rlap{\hskip.11\wd0 .}$^{\circ}$\fi}
\def\arcs{\ifmmode {^{\scriptstyle\prime\prime}}
          \else $^{\scriptstyle\prime\prime}$\fi}
\def\arcm{\ifmmode {^{\scriptstyle\prime}}
          \else $^{\scriptstyle\prime}$\fi}
\newdimen\sa  \newdimen\sb
\def\parcs{\sa=.07em \sb=.03em
     \ifmmode \hbox{\rlap{.}}^{\scriptstyle\prime\kern -\sb\prime}\hbox{\kern -\sa}
     \else \rlap{.}$^{\scriptstyle\prime\kern -\sb\prime}$\kern -\sa\fi}
\def\parcm{\sa=.08em \sb=.03em
     \ifmmode \hbox{\rlap{.}\kern\sa}^{\scriptstyle\prime}\hbox{\kern-\sb}
     \else \rlap{.}\kern\sa$^{\scriptstyle\prime}$\kern-\sb\fi}
\def\ra[#1 #2 #3.#4]{#1\sup{h}#2\sup{m}#3\sup{s}\llap.#4}
\def\dec[#1 #2 #3.#4]{#1\deg#2\arcm#3\arcs\llap.#4}
\def\deco[#1 #2 #3]{#1\deg#2\arcm#3\arcs}
\def\rra[#1 #2]{#1\sup{h}#2\sup{m}}

\def\dots{\relax\ifmmode \ldots\else $\ldots$\fi}
\def\WHzsr{\ifmmode $W\,Hz\mo\,sr\mo$\else W\,Hz\mo\,sr\mo\fi}
\def\mHz{\ifmmode $\,mHz$\else \,mHz\fi}
\def\GHz{\ifmmode $\,GHz$\else \,GHz\fi}
\def\mKs{\ifmmode $\,mK\,s$^{1/2}\else \,mK\,s$^{1/2}$\fi}
\def\muKs{\ifmmode \,\mu$K\,s$^{1/2}\else \,$\mu$K\,s$^{1/2}$\fi}
\def\muKRJs{\ifmmode \,\mu$K$_{\rm RJ}$\,s$^{1/2}\else \,$\mu$K$_{\rm RJ}$\,s$^{1/2}$\fi}
\def\muKHz{\ifmmode \,\mu$K\,Hz$^{-1/2}\else \,$\mu$K\,Hz$^{-1/2}$\fi}
\def\MJysr{\ifmmode \,$MJy\,sr\mo$\else \,MJy\,sr\mo\fi}
\def\MJysrmK{\ifmmode \,$MJy\,sr\mo$\,mK$_{\rm CMB}\mo\else \,MJy\,sr\mo\,mK$_{\rm CMB}\mo$\fi}
\def\microns{\ifmmode \,\mu$m$\else \,$\mu$m\fi}
\def\micron{\microns}
\def\muK{\ifmmode \,\mu$K$\else \,$\mu$\hbox{K}\fi}
\def\microK{\ifmmode \,\mu$K$\else \,$\mu$\hbox{K}\fi}
\def\muW{\ifmmode \,\mu$W$\else \,$\mu$\hbox{W}\fi}
\def\kms{\ifmmode $\,km\,s$^{-1}\else \,km\,s$^{-1}$\fi}
\def\kmsMpc{\ifmmode $\,\kms\,Mpc\mo$\else \,\kms\,Mpc\mo\fi}
\setsymbol{LFI:center:frequency:70GHz:units}{70.3\,GHz}
\setsymbol{LFI:center:frequency:44GHz:units}{44.1\,GHz}
\setsymbol{LFI:center:frequency:30GHz:units}{28.5\,GHz}

\setsymbol{LFI:center:frequency:70GHz}{70.3}
\setsymbol{LFI:center:frequency:44GHz}{44.1}
\setsymbol{LFI:center:frequency:30GHz}{28.5}

\setsymbol{LFI:center:frequency:LFI18:Rad:M:units}{71.7\GHz}
\setsymbol{LFI:center:frequency:LFI19:Rad:M:units}{67.5\GHz}
\setsymbol{LFI:center:frequency:LFI20:Rad:M:units}{69.2\GHz}
\setsymbol{LFI:center:frequency:LFI21:Rad:M:units}{70.4\GHz}
\setsymbol{LFI:center:frequency:LFI22:Rad:M:units}{71.5\GHz}
\setsymbol{LFI:center:frequency:LFI23:Rad:M:units}{70.8\GHz}
\setsymbol{LFI:center:frequency:LFI24:Rad:M:units}{44.4\GHz}
\setsymbol{LFI:center:frequency:LFI25:Rad:M:units}{44.0\GHz}
\setsymbol{LFI:center:frequency:LFI26:Rad:M:units}{43.9\GHz}
\setsymbol{LFI:center:frequency:LFI27:Rad:M:units}{28.3\GHz}
\setsymbol{LFI:center:frequency:LFI28:Rad:M:units}{28.8\GHz}
\setsymbol{LFI:center:frequency:LFI18:Rad:S:units}{70.1\GHz}
\setsymbol{LFI:center:frequency:LFI19:Rad:S:units}{69.6\GHz}
\setsymbol{LFI:center:frequency:LFI20:Rad:S:units}{69.5\GHz}
\setsymbol{LFI:center:frequency:LFI21:Rad:S:units}{69.5\GHz}
\setsymbol{LFI:center:frequency:LFI22:Rad:S:units}{72.8\GHz}
\setsymbol{LFI:center:frequency:LFI23:Rad:S:units}{71.3\GHz}
\setsymbol{LFI:center:frequency:LFI24:Rad:S:units}{44.1\GHz}
\setsymbol{LFI:center:frequency:LFI25:Rad:S:units}{44.1\GHz}
\setsymbol{LFI:center:frequency:LFI26:Rad:S:units}{44.1\GHz}
\setsymbol{LFI:center:frequency:LFI27:Rad:S:units}{28.5\GHz}
\setsymbol{LFI:center:frequency:LFI28:Rad:S:units}{28.2\GHz}

\setsymbol{LFI:center:frequency:LFI18:Rad:M}{71.7}
\setsymbol{LFI:center:frequency:LFI19:Rad:M}{67.5}
\setsymbol{LFI:center:frequency:LFI20:Rad:M}{69.2}
\setsymbol{LFI:center:frequency:LFI21:Rad:M}{70.4}
\setsymbol{LFI:center:frequency:LFI22:Rad:M}{71.5}
\setsymbol{LFI:center:frequency:LFI23:Rad:M}{70.8}
\setsymbol{LFI:center:frequency:LFI24:Rad:M}{44.4}
\setsymbol{LFI:center:frequency:LFI25:Rad:M}{44.0}
\setsymbol{LFI:center:frequency:LFI26:Rad:M}{43.9}
\setsymbol{LFI:center:frequency:LFI27:Rad:M}{28.3}
\setsymbol{LFI:center:frequency:LFI28:Rad:M}{28.8}
\setsymbol{LFI:center:frequency:LFI18:Rad:S}{70.1}
\setsymbol{LFI:center:frequency:LFI19:Rad:S}{69.6}
\setsymbol{LFI:center:frequency:LFI20:Rad:S}{69.5}
\setsymbol{LFI:center:frequency:LFI21:Rad:S}{69.5}
\setsymbol{LFI:center:frequency:LFI22:Rad:S}{72.8}
\setsymbol{LFI:center:frequency:LFI23:Rad:S}{71.3}
\setsymbol{LFI:center:frequency:LFI24:Rad:S}{44.1}
\setsymbol{LFI:center:frequency:LFI25:Rad:S}{44.1}
\setsymbol{LFI:center:frequency:LFI26:Rad:S}{44.1}
\setsymbol{LFI:center:frequency:LFI27:Rad:S}{28.5}
\setsymbol{LFI:center:frequency:LFI28:Rad:S}{28.2}

\setsymbol{LFI:white:noise:sensitivity:70GHz:units}{134.7\muKs}
\setsymbol{LFI:white:noise:sensitivity:44GHz:units}{164.7\muKs}
\setsymbol{LFI:white:noise:sensitivity:30GHz:units}{143.4\muKs}

\setsymbol{LFI:white:noise:sensitivity:70GHz}{134.7}
\setsymbol{LFI:white:noise:sensitivity:44GHz}{164.7}
\setsymbol{LFI:white:noise:sensitivity:30GHz}{143.4}

\setsymbol{LFI:white:noise:sensitivity:LFI18:Rad:M:units}{512.0\muKs}
\setsymbol{LFI:white:noise:sensitivity:LFI19:Rad:M:units}{581.4\muKs}
\setsymbol{LFI:white:noise:sensitivity:LFI20:Rad:M:units}{590.8\muKs}
\setsymbol{LFI:white:noise:sensitivity:LFI21:Rad:M:units}{455.2\muKs}
\setsymbol{LFI:white:noise:sensitivity:LFI22:Rad:M:units}{492.0\muKs}
\setsymbol{LFI:white:noise:sensitivity:LFI23:Rad:M:units}{507.7\muKs}
\setsymbol{LFI:white:noise:sensitivity:LFI24:Rad:M:units}{462.2\muKs}
\setsymbol{LFI:white:noise:sensitivity:LFI25:Rad:M:units}{413.6\muKs}
\setsymbol{LFI:white:noise:sensitivity:LFI26:Rad:M:units}{478.6\muKs}
\setsymbol{LFI:white:noise:sensitivity:LFI27:Rad:M:units}{277.7\muKs}
\setsymbol{LFI:white:noise:sensitivity:LFI28:Rad:M:units}{312.3\muKs}
\setsymbol{LFI:white:noise:sensitivity:LFI18:Rad:S:units}{465.7\muKs}
\setsymbol{LFI:white:noise:sensitivity:LFI19:Rad:S:units}{555.6\muKs}
\setsymbol{LFI:white:noise:sensitivity:LFI20:Rad:S:units}{623.2\muKs}
\setsymbol{LFI:white:noise:sensitivity:LFI21:Rad:S:units}{564.1\muKs}
\setsymbol{LFI:white:noise:sensitivity:LFI22:Rad:S:units}{534.4\muKs}
\setsymbol{LFI:white:noise:sensitivity:LFI23:Rad:S:units}{542.4\muKs}
\setsymbol{LFI:white:noise:sensitivity:LFI24:Rad:S:units}{399.2\muKs}
\setsymbol{LFI:white:noise:sensitivity:LFI25:Rad:S:units}{392.6\muKs}
\setsymbol{LFI:white:noise:sensitivity:LFI26:Rad:S:units}{418.6\muKs}
\setsymbol{LFI:white:noise:sensitivity:LFI27:Rad:S:units}{302.9\muKs}
\setsymbol{LFI:white:noise:sensitivity:LFI28:Rad:S:units}{285.3\muKs}

\setsymbol{LFI:white:noise:sensitivity:LFI18:Rad:M}{512.0}
\setsymbol{LFI:white:noise:sensitivity:LFI19:Rad:M}{581.4}
\setsymbol{LFI:white:noise:sensitivity:LFI20:Rad:M}{590.8}
\setsymbol{LFI:white:noise:sensitivity:LFI21:Rad:M}{455.2}
\setsymbol{LFI:white:noise:sensitivity:LFI22:Rad:M}{492.0}
\setsymbol{LFI:white:noise:sensitivity:LFI23:Rad:M}{507.7}
\setsymbol{LFI:white:noise:sensitivity:LFI24:Rad:M}{462.2}
\setsymbol{LFI:white:noise:sensitivity:LFI25:Rad:M}{413.6}
\setsymbol{LFI:white:noise:sensitivity:LFI26:Rad:M}{478.6}
\setsymbol{LFI:white:noise:sensitivity:LFI27:Rad:M}{277.7}
\setsymbol{LFI:white:noise:sensitivity:LFI28:Rad:M}{312.3}
\setsymbol{LFI:white:noise:sensitivity:LFI18:Rad:S}{465.7}
\setsymbol{LFI:white:noise:sensitivity:LFI19:Rad:S}{555.6}
\setsymbol{LFI:white:noise:sensitivity:LFI20:Rad:S}{623.2}
\setsymbol{LFI:white:noise:sensitivity:LFI21:Rad:S}{564.1}
\setsymbol{LFI:white:noise:sensitivity:LFI22:Rad:S}{534.4}
\setsymbol{LFI:white:noise:sensitivity:LFI23:Rad:S}{542.4}
\setsymbol{LFI:white:noise:sensitivity:LFI24:Rad:S}{399.2}
\setsymbol{LFI:white:noise:sensitivity:LFI25:Rad:S}{392.6}
\setsymbol{LFI:white:noise:sensitivity:LFI26:Rad:S}{418.6}
\setsymbol{LFI:white:noise:sensitivity:LFI27:Rad:S}{302.9}
\setsymbol{LFI:white:noise:sensitivity:LFI28:Rad:S}{285.3}

\setsymbol{LFI:knee:frequency:70GHz:units}{29.5\mHz}
\setsymbol{LFI:knee:frequency:44GHz:units}{56.2\mHz}
\setsymbol{LFI:knee:frequency:30GHz:units}{113.7\mHz}

\setsymbol{LFI:knee:frequency:70GHz}{29.5}
\setsymbol{LFI:knee:frequency:44GHz}{56.2}
\setsymbol{LFI:knee:frequency:30GHz}{113.7}

\setsymbol{LFI:knee:frequency:LFI18:Rad:M:units}{16.3\mHz}
\setsymbol{LFI:knee:frequency:LFI19:Rad:M:units}{15.1\mHz}
\setsymbol{LFI:knee:frequency:LFI20:Rad:M:units}{18.7\mHz}
\setsymbol{LFI:knee:frequency:LFI21:Rad:M:units}{37.2\mHz}
\setsymbol{LFI:knee:frequency:LFI22:Rad:M:units}{12.7\mHz}
\setsymbol{LFI:knee:frequency:LFI23:Rad:M:units}{34.6\mHz}
\setsymbol{LFI:knee:frequency:LFI24:Rad:M:units}{46.2\mHz}
\setsymbol{LFI:knee:frequency:LFI25:Rad:M:units}{24.9\mHz}
\setsymbol{LFI:knee:frequency:LFI26:Rad:M:units}{67.6\mHz}
\setsymbol{LFI:knee:frequency:LFI27:Rad:M:units}{187.4\mHz}
\setsymbol{LFI:knee:frequency:LFI28:Rad:M:units}{122.2\mHz}
\setsymbol{LFI:knee:frequency:LFI18:Rad:S:units}{17.7\mHz}
\setsymbol{LFI:knee:frequency:LFI19:Rad:S:units}{22.0\mHz}
\setsymbol{LFI:knee:frequency:LFI20:Rad:S:units}{8.7\mHz}
\setsymbol{LFI:knee:frequency:LFI21:Rad:S:units}{25.9\mHz}
\setsymbol{LFI:knee:frequency:LFI22:Rad:S:units}{15.8\mHz}
\setsymbol{LFI:knee:frequency:LFI23:Rad:S:units}{129.8\mHz}
\setsymbol{LFI:knee:frequency:LFI24:Rad:S:units}{100.9\mHz}
\setsymbol{LFI:knee:frequency:LFI25:Rad:S:units}{38.9\mHz}
\setsymbol{LFI:knee:frequency:LFI26:Rad:S:units}{58.9\mHz}
\setsymbol{LFI:knee:frequency:LFI27:Rad:S:units}{104.4\mHz}
\setsymbol{LFI:knee:frequency:LFI28:Rad:S:units}{40.7\mHz}

\setsymbol{LFI:knee:frequency:LFI18:Rad:M}{16.3}
\setsymbol{LFI:knee:frequency:LFI19:Rad:M}{15.1}
\setsymbol{LFI:knee:frequency:LFI20:Rad:M}{18.7}
\setsymbol{LFI:knee:frequency:LFI21:Rad:M}{37.2}
\setsymbol{LFI:knee:frequency:LFI22:Rad:M}{12.7}
\setsymbol{LFI:knee:frequency:LFI23:Rad:M}{34.6}
\setsymbol{LFI:knee:frequency:LFI24:Rad:M}{46.2}
\setsymbol{LFI:knee:frequency:LFI25:Rad:M}{24.9}
\setsymbol{LFI:knee:frequency:LFI26:Rad:M}{67.6}
\setsymbol{LFI:knee:frequency:LFI27:Rad:M}{187.4}
\setsymbol{LFI:knee:frequency:LFI28:Rad:M}{122.2}
\setsymbol{LFI:knee:frequency:LFI18:Rad:S}{17.7}
\setsymbol{LFI:knee:frequency:LFI19:Rad:S}{22.0}
\setsymbol{LFI:knee:frequency:LFI20:Rad:S}{8.7}
\setsymbol{LFI:knee:frequency:LFI21:Rad:S}{25.9}
\setsymbol{LFI:knee:frequency:LFI22:Rad:S}{15.8}
\setsymbol{LFI:knee:frequency:LFI23:Rad:S}{129.8}
\setsymbol{LFI:knee:frequency:LFI24:Rad:S}{100.9}
\setsymbol{LFI:knee:frequency:LFI25:Rad:S}{38.9}
\setsymbol{LFI:knee:frequency:LFI26:Rad:S}{58.9}
\setsymbol{LFI:knee:frequency:LFI27:Rad:S}{104.4}
\setsymbol{LFI:knee:frequency:LFI28:Rad:S}{40.7}

\setsymbol{LFI:slope:70GHz:units}{$-1.03$\mHz}
\setsymbol{LFI:slope:44GHz:units}{$-0.89$\mHz}
\setsymbol{LFI:slope:30GHz:units}{$-0.87$\mHz}

\setsymbol{LFI:slope:70GHz}{$-1.03$}
\setsymbol{LFI:slope:44GHz}{$-0.89$}
\setsymbol{LFI:slope:30GHz}{$-0.87$}

\setsymbol{LFI:slope:LFI18:Rad:M:units}{$-1.04$\mHz}
\setsymbol{LFI:slope:LFI19:Rad:M:units}{$-1.09$\mHz}
\setsymbol{LFI:slope:LFI20:Rad:M:units}{$-0.69$\mHz}
\setsymbol{LFI:slope:LFI21:Rad:M:units}{$-1.56$\mHz}
\setsymbol{LFI:slope:LFI22:Rad:M:units}{$-1.01$\mHz}
\setsymbol{LFI:slope:LFI23:Rad:M:units}{$-0.96$\mHz}
\setsymbol{LFI:slope:LFI24:Rad:M:units}{$-0.83$\mHz}
\setsymbol{LFI:slope:LFI25:Rad:M:units}{$-0.91$\mHz}
\setsymbol{LFI:slope:LFI26:Rad:M:units}{$-0.95$\mHz}
\setsymbol{LFI:slope:LFI27:Rad:M:units}{$-0.87$\mHz}
\setsymbol{LFI:slope:LFI28:Rad:M:units}{$-0.88$\mHz}
\setsymbol{LFI:slope:LFI18:Rad:S:units}{$-1.15$\mHz}
\setsymbol{LFI:slope:LFI19:Rad:S:units}{$-1.00$\mHz}
\setsymbol{LFI:slope:LFI20:Rad:S:units}{$-0.95$\mHz}
\setsymbol{LFI:slope:LFI21:Rad:S:units}{$-0.92$\mHz}
\setsymbol{LFI:slope:LFI22:Rad:S:units}{$-1.01$\mHz}
\setsymbol{LFI:slope:LFI23:Rad:S:units}{$-0.95$\mHz}
\setsymbol{LFI:slope:LFI24:Rad:S:units}{$-0.73$\mHz}
\setsymbol{LFI:slope:LFI25:Rad:S:units}{$-1.16$\mHz}
\setsymbol{LFI:slope:LFI26:Rad:S:units}{$-0.79$\mHz}
\setsymbol{LFI:slope:LFI27:Rad:S:units}{$-0.82$\mHz}
\setsymbol{LFI:slope:LFI28:Rad:S:units}{$-0.91$\mHz}

\setsymbol{LFI:slope:LFI18:Rad:M}{$-1.04$}
\setsymbol{LFI:slope:LFI19:Rad:M}{$-1.09$}
\setsymbol{LFI:slope:LFI20:Rad:M}{$-0.69$}
\setsymbol{LFI:slope:LFI21:Rad:M}{$-1.56$}
\setsymbol{LFI:slope:LFI22:Rad:M}{$-1.01$}
\setsymbol{LFI:slope:LFI23:Rad:M}{$-0.96$}
\setsymbol{LFI:slope:LFI24:Rad:M}{$-0.83$}
\setsymbol{LFI:slope:LFI25:Rad:M}{$-0.91$}
\setsymbol{LFI:slope:LFI26:Rad:M}{$-0.95$}
\setsymbol{LFI:slope:LFI27:Rad:M}{$-0.87$}
\setsymbol{LFI:slope:LFI28:Rad:M}{$-0.88$}
\setsymbol{LFI:slope:LFI18:Rad:S}{$-1.15$}
\setsymbol{LFI:slope:LFI19:Rad:S}{$-1.00$}
\setsymbol{LFI:slope:LFI20:Rad:S}{$-0.95$}
\setsymbol{LFI:slope:LFI21:Rad:S}{$-0.92$}
\setsymbol{LFI:slope:LFI22:Rad:S}{$-1.01$}
\setsymbol{LFI:slope:LFI23:Rad:S}{$-0.95$}
\setsymbol{LFI:slope:LFI24:Rad:S}{$-0.73$}
\setsymbol{LFI:slope:LFI25:Rad:S}{$-1.16$}
\setsymbol{LFI:slope:LFI26:Rad:S}{$-0.79$}
\setsymbol{LFI:slope:LFI27:Rad:S}{$-0.82$}
\setsymbol{LFI:slope:LFI28:Rad:S}{$-0.91$}

\setsymbol{LFI:FWHM:70GHz:units}{13\parcm01}
\setsymbol{LFI:FWHM:44GHz:units}{27\parcm92}
\setsymbol{LFI:FWHM:30GHz:units}{32\parcm65}

\setsymbol{LFI:FWHM:70GHz}{13.01}
\setsymbol{LFI:FWHM:44GHz}{27.92}
\setsymbol{LFI:FWHM:30GHz}{32.65}

\setsymbol{LFI:FWHM:LFI18:units}{13\parcm39}
\setsymbol{LFI:FWHM:LFI19:units}{13\parcm01}
\setsymbol{LFI:FWHM:LFI20:units}{12\parcm75}
\setsymbol{LFI:FWHM:LFI21:units}{12\parcm74}
\setsymbol{LFI:FWHM:LFI22:units}{12\parcm87}
\setsymbol{LFI:FWHM:LFI23:units}{13\parcm27}
\setsymbol{LFI:FWHM:LFI24:units}{22\parcm98}
\setsymbol{LFI:FWHM:LFI25:units}{30\parcm46}
\setsymbol{LFI:FWHM:LFI26:units}{30\parcm31}
\setsymbol{LFI:FWHM:LFI27:units}{32\parcm65}
\setsymbol{LFI:FWHM:LFI28:units}{32\parcm66}

\setsymbol{LFI:FWHM:LFI18}{13.39}
\setsymbol{LFI:FWHM:LFI19}{13.01}
\setsymbol{LFI:FWHM:LFI20}{12.75}
\setsymbol{LFI:FWHM:LFI21}{12.74}
\setsymbol{LFI:FWHM:LFI22}{12.87}
\setsymbol{LFI:FWHM:LFI23}{13.27}
\setsymbol{LFI:FWHM:LFI24}{22.98}
\setsymbol{LFI:FWHM:LFI25}{30.46}
\setsymbol{LFI:FWHM:LFI26}{30.31}
\setsymbol{LFI:FWHM:LFI27}{32.65}
\setsymbol{LFI:FWHM:LFI28}{32.66}

\setsymbol{LFI:FWHM:uncertainty:LFI18:units}{0.170\arcm}
\setsymbol{LFI:FWHM:uncertainty:LFI19:units}{0.174\arcm}
\setsymbol{LFI:FWHM:uncertainty:LFI20:units}{0.170\arcm}
\setsymbol{LFI:FWHM:uncertainty:LFI21:units}{0.156\arcm}
\setsymbol{LFI:FWHM:uncertainty:LFI22:units}{0.164\arcm}
\setsymbol{LFI:FWHM:uncertainty:LFI23:units}{0.171\arcm}
\setsymbol{LFI:FWHM:uncertainty:LFI24:units}{0.652\arcm}
\setsymbol{LFI:FWHM:uncertainty:LFI25:units}{1.075\arcm}
\setsymbol{LFI:FWHM:uncertainty:LFI26:units}{1.131\arcm}
\setsymbol{LFI:FWHM:uncertainty:LFI27:units}{1.266\arcm}
\setsymbol{LFI:FWHM:uncertainty:LFI28:units}{1.287\arcm}

\setsymbol{LFI:FWHM:uncertainty:LFI18}{0.170}
\setsymbol{LFI:FWHM:uncertainty:LFI19}{0.174}
\setsymbol{LFI:FWHM:uncertainty:LFI20}{0.170}
\setsymbol{LFI:FWHM:uncertainty:LFI21}{0.156}
\setsymbol{LFI:FWHM:uncertainty:LFI22}{0.164}
\setsymbol{LFI:FWHM:uncertainty:LFI23}{0.171}
\setsymbol{LFI:FWHM:uncertainty:LFI24}{0.652}
\setsymbol{LFI:FWHM:uncertainty:LFI25}{1.075}
\setsymbol{LFI:FWHM:uncertainty:LFI26}{1.131}
\setsymbol{LFI:FWHM:uncertainty:LFI27}{1.266}
\setsymbol{LFI:FWHM:uncertainty:LFI28}{1.287}

\setsymbol{HFI:center:frequency:100GHz:units}{100\,GHz}
\setsymbol{HFI:center:frequency:143GHz:units}{143\,GHz}
\setsymbol{HFI:center:frequency:217GHz:units}{217\,GHz}
\setsymbol{HFI:center:frequency:353GHz:units}{353\,GHz}
\setsymbol{HFI:center:frequency:545GHz:units}{545\,GHz}
\setsymbol{HFI:center:frequency:857GHz:units}{857\,GHz}

\setsymbol{HFI:center:frequency:100GHz}{100}
\setsymbol{HFI:center:frequency:143GHz}{143}
\setsymbol{HFI:center:frequency:217GHz}{217}
\setsymbol{HFI:center:frequency:353GHz}{353}
\setsymbol{HFI:center:frequency:545GHz}{545}
\setsymbol{HFI:center:frequency:857GHz}{857}

\setsymbol{HFI:Ndetectors:100GHz}{8}
\setsymbol{HFI:Ndetectors:143GHz}{11}
\setsymbol{HFI:Ndetectors:217GHz}{12}
\setsymbol{HFI:Ndetectors:353GHz}{12}
\setsymbol{HFI:Ndetectors:545GHz}{3}
\setsymbol{HFI:Ndetectors:857GHz}{4}

\setsymbol{HFI:FWHM:Maps:100GHz:units}{9\parcm88}
\setsymbol{HFI:FWHM:Maps:143GHz:units}{7\parcm18}
\setsymbol{HFI:FWHM:Maps:217GHz:units}{4\parcm87}
\setsymbol{HFI:FWHM:Maps:353GHz:units}{4\parcm65}
\setsymbol{HFI:FWHM:Maps:545GHz:units}{4\parcm72}
\setsymbol{HFI:FWHM:Maps:857GHz:units}{4\parcm39}
\setsymbol{HFI:FWHM:Maps:100GHz}{9.88}
\setsymbol{HFI:FWHM:Maps:143GHz}{7.18}
\setsymbol{HFI:FWHM:Maps:217GHz}{4.87}
\setsymbol{HFI:FWHM:Maps:353GHz}{4.65}
\setsymbol{HFI:FWHM:Maps:545GHz}{4.72}
\setsymbol{HFI:FWHM:Maps:857GHz}{4.39}

\setsymbol{HFI:beam:ellipticity:Maps:100GHz}{1.15}
\setsymbol{HFI:beam:ellipticity:Maps:143GHz}{1.01}
\setsymbol{HFI:beam:ellipticity:Maps:217GHz}{1.06}
\setsymbol{HFI:beam:ellipticity:Maps:353GHz}{1.05}
\setsymbol{HFI:beam:ellipticity:Maps:545GHz}{1.14}
\setsymbol{HFI:beam:ellipticity:Maps:857GHz}{1.19}

\setsymbol{HFI:FWHM:Mars:100GHz:units}{9\parcm37}
\setsymbol{HFI:FWHM:Mars:143GHz:units}{7\parcm04}
\setsymbol{HFI:FWHM:Mars:217GHz:units}{4\parcm68}
\setsymbol{HFI:FWHM:Mars:353GHz:units}{4\parcm43}
\setsymbol{HFI:FWHM:Mars:545GHz:units}{3\parcm80}
\setsymbol{HFI:FWHM:Mars:857GHz:units}{3\parcm67}

\setsymbol{HFI:FWHM:Mars:100GHz}{9.37}
\setsymbol{HFI:FWHM:Mars:143GHz}{7.04}
\setsymbol{HFI:FWHM:Mars:217GHz}{4.68}
\setsymbol{HFI:FWHM:Mars:353GHz}{4.43}
\setsymbol{HFI:FWHM:Mars:545GHz}{3.80}
\setsymbol{HFI:FWHM:Mars:857GHz}{3.67}

\setsymbol{HFI:beam:ellipticity:Mars:100GHz}{1.18}
\setsymbol{HFI:beam:ellipticity:Mars:143GHz}{1.03}
\setsymbol{HFI:beam:ellipticity:Mars:217GHz}{1.14}
\setsymbol{HFI:beam:ellipticity:Mars:353GHz}{1.09}
\setsymbol{HFI:beam:ellipticity:Mars:545GHz}{1.25}
\setsymbol{HFI:beam:ellipticity:Mars:857GHz}{1.03}

\setsymbol{HFI:CMB:relative:calibration:100GHz}{$\lsim 1\%$}
\setsymbol{HFI:CMB:relative:calibration:143GHz}{$\lsim 1\%$}
\setsymbol{HFI:CMB:relative:calibration:217GHz}{$\lsim 1\%$}
\setsymbol{HFI:CMB:relative:calibration:353GHz}{$\lsim 1\%$}
\setsymbol{HFI:CMB:relative:calibration:545GHz}{}
\setsymbol{HFI:CMB:relative:calibration:857GHz}{}

\setsymbol{HFI:CMB:absolute:calibration:100GHz}{$\lsim 2\%$}
\setsymbol{HFI:CMB:absolute:calibration:143GHz}{$\lsim 2\%$}
\setsymbol{HFI:CMB:absolute:calibration:217GHz}{$\lsim 2\%$}
\setsymbol{HFI:CMB:absolute:calibration:353GHz}{$\lsim 2\%$}
\setsymbol{HFI:CMB:absolute:calibration:545GHz}{}
\setsymbol{HFI:CMB:absolute:calibration:857GHz}{}

\setsymbol{HFI:FIRAS:gain:calibration:accuracy:statistical:100GHz}{}
\setsymbol{HFI:FIRAS:gain:calibration:accuracy:statistical:143GHz}{}
\setsymbol{HFI:FIRAS:gain:calibration:accuracy:statistical:217GHz}{}
\setsymbol{HFI:FIRAS:gain:calibration:accuracy:statistical:353GHz}{2.5\%}
\setsymbol{HFI:FIRAS:gain:calibration:accuracy:statistical:545GHz}{1\%}
\setsymbol{HFI:FIRAS:gain:calibration:accuracy:statistical:857GHz}{0.5\%}

\setsymbol{HFI:FIRAS:gain:calibration:accuracy:systematic:100GHz}{}
\setsymbol{HFI:FIRAS:gain:calibration:accuracy:systematic:143GHz}{}
\setsymbol{HFI:FIRAS:gain:calibration:accuracy:systematic:217GHz}{}
\setsymbol{HFI:FIRAS:gain:calibration:accuracy:systematic:353GHz}{}
\setsymbol{HFI:FIRAS:gain:calibration:accuracy:systematic:545GHz}{7\%}
\setsymbol{HFI:FIRAS:gain:calibration:accuracy:systematic:857GHz}{7\%}

\setsymbol{HFI:FIRAS:zero:point:accuracy:100GHz:units}{0.8\MJysr}
\setsymbol{HFI:FIRAS:zero:point:accuracy:143GHz:units}{}
\setsymbol{HFI:FIRAS:zero:point:accuracy:217GHz:units}{}
\setsymbol{HFI:FIRAS:zero:point:accuracy:353GHz:units}{1.4\MJysr}
\setsymbol{HFI:FIRAS:zero:point:accuracy:545GHz:units}{2.2\MJysr}
\setsymbol{HFI:FIRAS:zero:point:accuracy:857GHz:units}{1.7\MJysr}

\setsymbol{HFI:FIRAS:zero:point:accuracy:100GHz}{0.8}
\setsymbol{HFI:FIRAS:zero:point:accuracy:143GHz}{}
\setsymbol{HFI:FIRAS:zero:point:accuracy:217GHz}{}
\setsymbol{HFI:FIRAS:zero:point:accuracy:353GHz}{1.4}
\setsymbol{HFI:FIRAS:zero:point:accuracy:545GHz}{2.2}
\setsymbol{HFI:FIRAS:zero:point:accuracy:857GHz}{1.7}

\setsymbol{HFI:unit:conversion:100GHz:units}{0.2415\MJysrmK}
\setsymbol{HFI:unit:conversion:143GHz:units}{0.3694\MJysrmK}
\setsymbol{HFI:unit:conversion:217GHz:units}{0.4811\MJysrmK}
\setsymbol{HFI:unit:conversion:353GHz:units}{0.2883\MJysrmK}
\setsymbol{HFI:unit:conversion:545GHz:units}{0.05826\MJysrmK}
\setsymbol{HFI:unit:conversion:857GHz:units}{0.002238\MJysrmK}

\setsymbol{HFI:unit:conversion:100GHz}{0.2415}
\setsymbol{HFI:unit:conversion:143GHz}{0.3694}
\setsymbol{HFI:unit:conversion:217GHz}{0.4811}
\setsymbol{HFI:unit:conversion:353GHz}{0.2883}
\setsymbol{HFI:unit:conversion:545GHz}{0.05826}
\setsymbol{HFI:unit:conversion:857GHz}{0.002238}

\setsymbol{HFI:colour:correction:alpha=-2:V1.01:100GHz}{0.9893}
\setsymbol{HFI:colour:correction:alpha=-2:V1.01:143GHz}{0.9759}
\setsymbol{HFI:colour:correction:alpha=-2:V1.01:217GHz}{1.0007}
\setsymbol{HFI:colour:correction:alpha=-2:V1.01:353GHz}{1.0028}
\setsymbol{HFI:colour:correction:alpha=-2:V1.01:545GHz}{1.0019}
\setsymbol{HFI:colour:correction:alpha=-2:V1.01:857GHz}{0.9889}

\setsymbol{HFI:colour:correction:alpha=0:V1.01:100GHz}{1.0008}
\setsymbol{HFI:colour:correction:alpha=0:V1.01:143GHz}{1.0148}
\setsymbol{HFI:colour:correction:alpha=0:V1.01:217GHz}{0.9909}
\setsymbol{HFI:colour:correction:alpha=0:V1.01:353GHz}{0.9888}
\setsymbol{HFI:colour:correction:alpha=0:V1.01:545GHz}{0.9878}
\setsymbol{HFI:colour:correction:alpha=0:V1.01:857GHz}{1.0014}

\providecommand{\sorthelp}[1]{}

\def\result#1{\vspace{5pt}\noindent\textbf{#1}---}

\def\LCDM{$\Lambda$CDM}
\newcommand{\planck}{\Planck}
\newcommand{\WMAP}{\textit{WMAP}}
\newcommand{\WP}{WP}
\newcommand{\highL}{highL}

\newcommand{\As}{A_{\rm s}}

\newcommand{\ns}{n_{\rm s}}
\newcommand{\lcdm}{$\Lambda$CDM}

\newcommand{\Aphiphi}{A_{\rm L}^{\phi\phi}}

\newcommand{\rstar}{r_{\ast}}
\newcommand{\rs}{r_{\rm s}}

\begin{document}

\title{\vglue -17mm\Planck\ 2013 results. I. Overview of products and scientific results}		 

%This author list corresponds to \title{Author list for Proj. Ref. 1.1: The Planck mission}

%This author list corresponds to \title{Author list for SVN P01\_Mission, Proj. Ref. 1\_1: Overview of Planck Products and Scientific Results}
%Prepared by R. Leonardi (rleonardi@sciops.esa.int), ESAC/ESA
%This version is from Thu Dec 19 11:47:26 2013 CET
%\subtitle{There are 400 co-authors in this list}
\author{\small
Planck Collaboration:
P.~A.~R.~Ade\inst{116}
\and
N.~Aghanim\inst{79}
\and
M.~I.~R.~Alves\inst{79}
\and
C.~Armitage-Caplan\inst{122}
\and
M.~Arnaud\inst{96}
\and
M.~Ashdown\inst{93, 8}
\and
F.~Atrio-Barandela\inst{23}
\and
J.~Aumont\inst{79}
\and
H.~Aussel\inst{96}
\and
C.~Baccigalupi\inst{114}
% ...
}
\institute{\small
APC, AstroParticule et Cosmologie, Universit\'{e} Paris Diderot, CNRS/IN2P3, CEA/lrfu, Observatoire de Paris, Sorbonne Paris Cit\'{e}, 10, rue Alice Domon et L\'{e}onie Duquet, 75205 Paris Cedex 13, France\\
\and
Aalto University Mets\"{a}hovi Radio Observatory and Dept of Radio Science and Engineering, P.O. Box 13000, FI-00076 AALTO, Finland\\
\and
Academy of Sciences of Tatarstan, Bauman Str., 20, Kazan, 420111, Republic of Tatarstan, Russia\\
\and
African Institute for Mathematical Sciences, 6-8 Melrose Road, Muizenberg, Cape Town, South Africa\\
\and
Agenzia Spaziale Italiana Science Data Center, Via del Politecnico snc, 00133, Roma, Italy\\
\and
Agenzia Spaziale Italiana, Viale Liegi 26, Roma, Italy\\
\and
Argelander-Institut f\"{u}r Astronomie, Universit\"{a}t Bonn, Auf dem H\"{u}gel 71, D-53121 Bonn, Germany\\
\and
Astrophysics Group, Cavendish Laboratory, University of Cambridge, J J Thomson Avenue, Cambridge CB3 0HE, U.K.\\
\and
Astrophysics \& Cosmology Research Unit, School of Mathematics, Statistics \& Computer Science, University of KwaZulu-Natal, Westville Campus, Private Bag X54001, Durban 4000, South Africa\\
\and
Atacama Large Millimeter/submillimeter Array, ALMA Santiago Central Offices, Alonso de Cordova 3107, Vitacura, Casilla 763 0355, Santiago, Chile\\
\and
CITA, University of Toronto, 60 St. George St., Toronto, ON M5S 3H8, Canada\\
\and
CNR - ISTI, Area della Ricerca, via G. Moruzzi 1, Pisa, Italy\\
\and
CNRS, IRAP, 9 Av. colonel Roche, BP 44346, F-31028 Toulouse cedex 4, France\\
\and
California Institute of Technology, Pasadena, California, U.S.A.\\
\and
Centre for Theoretical Cosmology, DAMTP, University of Cambridge, Wilberforce Road, Cambridge CB3 0WA, U.K.\\
\and
Centro de Astrof\'{\i}sica, Universidade do Porto, Rua das Estrelas, 4150-762 Porto, Portugal\\
\and
Centro de Estudios de F\'{i}sica del Cosmos de Arag\'{o}n (CEFCA), Plaza San Juan, 1, planta 2, E-44001, Teruel, Spain\\
\and
Computational Cosmology Center, Lawrence Berkeley National Laboratory, Berkeley, California, U.S.A.\\
\and
Consejo Superior de Investigaciones Cient\'{\i}ficas (CSIC), Madrid, Spain\\
\and
DSM/Irfu/SPP, CEA-Saclay, F-91191 Gif-sur-Yvette Cedex, France\\
\and
DTU Space, National Space Institute, Technical University of Denmark, Elektrovej 327, DK-2800 Kgs. Lyngby, Denmark\\
\and
D\'{e}partement de Physique Th\'{e}orique, Universit\'{e} de Gen\`{e}ve, 24, Quai E. Ansermet,1211 Gen\`{e}ve 4, Switzerland\\
\and
Departamento de F\'{\i}sica Fundamental, Facultad de Ciencias, Universidad de Salamanca, 37008 Salamanca, Spain\\
\and
Departamento de F\'{\i}sica, Universidad de Oviedo, Avda. Calvo Sotelo s/n, Oviedo, Spain\\
\and
Departamento de Matem\'{a}ticas, Estad\'{\i}stica y Computaci\'{o}n, Universidad de Cantabria, Avda. de los Castros s/n, Santander, Spain\\
\and
Department of Astronomy and Astrophysics, University of Toronto, 50 Saint George Street, Toronto, Ontario, Canada\\
\and
Department of Astronomy and Geodesy, Kazan Federal University,  Kremlevskaya Str., 18, Kazan, 420008, Russia\\
\and
Department of Astrophysics/IMAPP, Radboud University Nijmegen, P.O. Box 9010, 6500 GL Nijmegen, The Netherlands\\
\and
Department of Electrical Engineering and Computer Sciences, University of California, Berkeley, California, U.S.A.\\
\and
Department of Physics \& Astronomy, University of British Columbia, 6224 Agricultural Road, Vancouver, British Columbia, Canada\\
\and
Department of Physics and Astronomy, Dana and David Dornsife College of Letter, Arts and Sciences, University of Southern California, Los Angeles, CA 90089, U.S.A.\\
\and
Department of Physics and Astronomy, University College London, London WC1E 6BT, U.K.\\
\and
Department of Physics and Astronomy, University of Sussex, Brighton BN1 9QH, U.K.\\
\and
Department of Physics, Carnegie Mellon University, 5000 Forbes Ave, Pittsburgh, PA 15213, U.S.A.\\
\and
Department of Physics, Florida State University, Keen Physics Building, 77 Chieftan Way, Tallahassee, Florida, U.S.A.\\
\and
Department of Physics, Gustaf H\"{a}llstr\"{o}min katu 2a, University of Helsinki, Helsinki, Finland\\
\and
Department of Physics, Princeton University, Princeton, New Jersey, U.S.A.\\
\and
Department of Physics, University of Alberta, 11322-89 Avenue, Edmonton, Alberta, T6G 2G7, Canada\\
\and
Department of Physics, University of California, Berkeley, California, U.S.A.\\
\and
Department of Physics, University of California, One Shields Avenue, Davis, California, U.S.A.\\
\and
Department of Physics, University of California, Santa Barbara, California, U.S.A.\\
\and
Department of Physics, University of Illinois at Urbana-Champaign, 1110 West Green Street, Urbana, Illinois, U.S.A.\\
\and
Dipartimento di Fisica e Astronomia G. Galilei, Universit\`{a} degli Studi di Padova, via Marzolo 8, 35131 Padova, Italy\\
\and
Dipartimento di Fisica e Astronomia, Universit\`{a} degli Studi di Bologna, viale Berti Pichat 6/2, I-40127, Bologna, Italy\\
\and
Dipartimento di Fisica e Scienze della Terra, Universit\`{a} di Ferrara, Via Saragat 1, 44122 Ferrara, Italy\\
\and
Dipartimento di Fisica, Universit\`{a} La Sapienza, P. le A. Moro 2, Roma, Italy\\
\and
Dipartimento di Fisica, Universit\`{a} degli Studi di Milano, Via Celoria, 16, Milano, Italy\\
\and
Dipartimento di Fisica, Universit\`{a} degli Studi di Trieste, via A. Valerio 2, Trieste, Italy\\
\and
Dipartimento di Fisica, Universit\`{a} di Roma Tor Vergata, Via della Ricerca Scientifica, 1, Roma, Italy\\
\and
Dipartimento di Matematica, Universit\`{a} di Roma Tor Vergata, Via della Ricerca Scientifica, 1, Roma, Italy\\
\and
Discovery Center, Niels Bohr Institute, Blegdamsvej 17, Copenhagen, Denmark\\
\and
Dpto. Astrof\'{i}sica, Universidad de La Laguna (ULL), E-38206 La Laguna, Tenerife, Spain\\
\and
European Southern Observatory, ESO Vitacura, Alonso de Cordova 3107, Vitacura, Casilla 19001, Santiago, Chile\\
\and
European Space Agency, ESAC, Camino bajo del Castillo, s/n, Urbanizaci\'{o}n Villafranca del Castillo, Villanueva de la Ca\~{n}ada, Madrid, Spain\\
\and
European Space Agency, ESAC, Planck Science Office, Camino bajo del Castillo, s/n, Urbanizaci\'{o}n Villafranca del Castillo, Villanueva de la Ca\~{n}ada, Madrid, Spain\\
\and
European Space Agency, ESOC, Robert-Bosch-Str. 5, Darmstadt, Germany\\
\and
European Space Agency, ESTEC, Keplerlaan 1, 2201 AZ Noordwijk, The Netherlands\\
\and
Finnish Centre for Astronomy with ESO (FINCA), University of Turku, V\"{a}is\"{a}l\"{a}ntie 20, FIN-21500, Piikki\"{o}, Finland\\
\and
GEPI, Observatoire de Paris, Section de Meudon, 5 Place J. Janssen, 92195 Meudon Cedex, France\\
\and
Haverford College Astronomy Department, 370 Lancaster Avenue, Haverford, Pennsylvania, U.S.A.\\
\and
Helsinki Institute of Physics, Gustaf H\"{a}llstr\"{o}min katu 2, University of Helsinki, Helsinki, Finland\\
\and
INAF - Osservatorio Astrofisico di Catania, Via S. Sofia 78, Catania, Italy\\
\and
INAF - Osservatorio Astronomico di Padova, Vicolo dell'Osservatorio 5, Padova, Italy\\
\and
INAF - Osservatorio Astronomico di Roma, via di Frascati 33, Monte Porzio Catone, Italy\\
\and
INAF - Osservatorio Astronomico di Trieste, Via G.B. Tiepolo 11, Trieste, Italy\\
\and
INAF Istituto di Radioastronomia, Via P. Gobetti 101, 40129 Bologna, Italy\\
\and
INAF/IASF Bologna, Via Gobetti 101, Bologna, Italy\\
\and
INAF/IASF Milano, Via E. Bassini 15, Milano, Italy\\
\and
INFN, Sezione di Bologna, Via Irnerio 46, I-40126, Bologna, Italy\\
\and
INFN, Sezione di Roma 1, Universit\`{a} di Roma Sapienza, Piazzale Aldo Moro 2, 00185, Roma, Italy\\
\and
INFN/National Institute for Nuclear Physics, Via Valerio 2, I-34127 Trieste, Italy\\
\and
IPAG: Institut de Plan\'{e}tologie et d'Astrophysique de Grenoble, Universit\'{e} Joseph Fourier, Grenoble 1 / CNRS-INSU, UMR 5274, Grenoble, F-38041, France\\
\and
ISDC Data Centre for Astrophysics, University of Geneva, ch. d'Ecogia 16, Versoix, Switzerland\\
\and
IUCAA, Post Bag 4, Ganeshkhind, Pune University Campus, Pune 411 007, India\\
\and
Imperial College London, Astrophysics group, Blackett Laboratory, Prince Consort Road, London, SW7 2AZ, U.K.\\
\and
Infrared Processing and Analysis Center, California Institute of Technology, Pasadena, CA 91125, U.S.A.\\
\and
Institut N\'{e}el, CNRS, Universit\'{e} Joseph Fourier Grenoble I, 25 rue des Martyrs, Grenoble, France\\
\and
Institut Universitaire de France, 103, bd Saint-Michel, 75005, Paris, France\\
\and
Institut d'Astrophysique Spatiale, CNRS (UMR8617) Universit\'{e} Paris-Sud 11, B\^{a}timent 121, Orsay, France\\
\and
Institut d'Astrophysique de Paris, CNRS (UMR7095), 98 bis Boulevard Arago, F-75014, Paris, France\\
\and
Institut de Ci\`{e}ncies de l'Espai, CSIC/IEEC, Facultat de Ci\`{e}ncies, Campus UAB, Torre C5 par-2, Bellaterra 08193, Spain\\
\and
Institute for Space Sciences, Bucharest-Magurale, Romania\\
\and
Institute of Astro and Particle Physics, Technikerstrasse 25/8, University of Innsbruck, A-6020, Innsbruck, Austria\\
\and
Institute of Astronomy and Astrophysics, Academia Sinica, Taipei, Taiwan\\
\and
Institute of Astronomy, University of Cambridge, Madingley Road, Cambridge CB3 0HA, U.K.\\
\and
Institute of Mathematics and Physics, Centre for Cosmology, Particle Physics and Phenomenology, Louvain University, Louvain-la-Neuve, Belgium\\
\and
Institute of Theoretical Astrophysics, University of Oslo, Blindern, Oslo, Norway\\
\and
Instituto de Astrof\'{\i}sica de Canarias, C/V\'{\i}a L\'{a}ctea s/n, La Laguna, Tenerife, Spain\\
\and
Instituto de F\'{\i}sica de Cantabria (CSIC-Universidad de Cantabria), Avda. de los Castros s/n, Santander, Spain\\
\and
Istituto di Fisica del Plasma, CNR-ENEA-EURATOM Association, Via R. Cozzi 53, Milano, Italy\\
\and
Jet Propulsion Laboratory, California Institute of Technology, 4800 Oak Grove Drive, Pasadena, California, U.S.A.\\
\and
Jodrell Bank Centre for Astrophysics, Alan Turing Building, School of Physics and Astronomy, The University of Manchester, Oxford Road, Manchester, M13 9PL, U.K.\\
\and
Kavli Institute for Cosmology Cambridge, Madingley Road, Cambridge, CB3 0HA, U.K.\\
\and
LAL, Universit\'{e} Paris-Sud, CNRS/IN2P3, Orsay, France\\
\and
LERMA, CNRS, Observatoire de Paris, 61 Avenue de l'Observatoire, Paris, France\\
\and
Laboratoire AIM, IRFU/Service d'Astrophysique - CEA/DSM - CNRS - Universit\'{e} Paris Diderot, B\^{a}t. 709, CEA-Saclay, F-91191 Gif-sur-Yvette Cedex, France\\
\and
Laboratoire Traitement et Communication de l'Information, CNRS (UMR 5141) and T\'{e}l\'{e}com ParisTech, 46 rue Barrault F-75634 Paris Cedex 13, France\\
\and
Laboratoire de Physique Subatomique et de Cosmologie, Universit\'{e} Joseph Fourier Grenoble I, CNRS/IN2P3, Institut National Polytechnique de Grenoble, 53 rue des Martyrs, 38026 Grenoble cedex, France\\
\and
Laboratoire de Physique Th\'{e}orique, Universit\'{e} Paris-Sud 11 \& CNRS, B\^{a}timent 210, 91405 Orsay, France\\
\and
Lawrence Berkeley National Laboratory, Berkeley, California, U.S.A.\\
\and
Leung Center for Cosmology and Particle Astrophysics, National Taiwan University, Taipei 10617, Taiwan\\
\and
MPA Partner Group, Key Laboratory for Research in Galaxies and Cosmology, Shanghai Astronomical Observatory, Chinese Academy of Sciences, Nandan Road 80, Shanghai 200030, China\\
\and
Max-Planck-Institut f\"{u}r Astrophysik, Karl-Schwarzschild-Str. 1, 85741 Garching, Germany\\
\and
Max-Planck-Institut f\"{u}r Extraterrestrische Physik, Giessenbachstra{\ss}e, 85748 Garching, Germany\\
\and
McGill Physics, Ernest Rutherford Physics Building, McGill University, 3600 rue University, Montr\'{e}al, QC, H3A 2T8, Canada\\
\and
MilliLab, VTT Technical Research Centre of Finland, Tietotie 3, Espoo, Finland\\
\and
Moscow Institute of Physics and Technology, Dolgoprudny, Institutsky per., 9, 141700, Russia\\
\and
Mullard Space Science Laboratory, University College London, Surrey RH5 6NT, U.K.\\
\and
National University of Ireland, Department of Experimental Physics, Maynooth, Co. Kildare, Ireland\\
\and
Niels Bohr Institute, Blegdamsvej 17, Copenhagen, Denmark\\
\and
Observational Cosmology, Mail Stop 367-17, California Institute of Technology, Pasadena, CA, 91125, U.S.A.\\
\and
Optical Science Laboratory, University College London, Gower Street, London, U.K.\\
\and
SB-ITP-LPPC, EPFL, CH-1015, Lausanne, Switzerland\\
\and
SISSA, Astrophysics Sector, via Bonomea 265, 34136, Trieste, Italy\\
\and
SUPA, Institute for Astronomy, University of Edinburgh, Royal Observatory, Blackford Hill, Edinburgh EH9 3HJ, U.K.\\
\and
School of Physics and Astronomy, Cardiff University, Queens Buildings, The Parade, Cardiff, CF24 3AA, U.K.\\
\and
School of Physics and Astronomy, University of Nottingham, Nottingham NG7 2RD, U.K.\\
\and
Space Research Institute (IKI), Russian Academy of Sciences, Profsoyuznaya Str, 84/32, Moscow, 117997, Russia\\
\and
Space Sciences Laboratory, University of California, Berkeley, California, U.S.A.\\
\and
Special Astrophysical Observatory, Russian Academy of Sciences, Nizhnij Arkhyz, Zelenchukskiy region, Karachai-Cherkessian Republic, 369167, Russia\\
\and
Stanford University, Dept of Physics, Varian Physics Bldg, 382 Via Pueblo Mall, Stanford, California, U.S.A.\\
\and
Sub-Department of Astrophysics, University of Oxford, Keble Road, Oxford OX1 3RH, U.K.\\
% ...
}

%\author{\small
%Planck Collaboration:
%J.~A.~Tauber\inst{1}~\thanks{Corresponding author: J. A. Tauber, jtauber@rssd.esa.int}
%}
%\institute{\small
%European Space Agency, ESTEC, Keplerlaan 1, 2201 AZ Noordwijk, The Netherlands\\
%}
%

\date{\vglue -1.5mm Received XX, 2012; accepted XX, 2013\vglue -5mm}
%\vglue -4mm
\abstract{\vglue -3mm 
The European Space Agency's \Planck\ satellite, dedicated to studying the early Universe and its subsequent evolution,  was launched 14~May 2009 and has been scanning the microwave and submillimetre sky continuously since 12~August 2009.  In March~2013, ESA and the \Planck\ Collaboration released the initial cosmology products based on the the first 15.5\,months of \Planck\ data, along with a set of scientific and technical papers and a web-based explanatory supplement.  This paper gives an overview of the mission and its performance, the processing, analysis, and characteristics of the data, the scientific results, and the science data products and papers in the release.  The science products include maps of the cosmic microwave background (CMB) and diffuse extragalactic foregrounds, a catalogue of compact Galactic and extragalactic sources, and a list of sources detected through the Sunyaev-Zeldovich effect.  The likelihood code used to assess cosmological models against the \Planck\ data and a lensing likelihood are described.   Scientific results include robust support for the standard six-parameter \LCDM\ model of cosmology and improved measurements of its parameters, including a highly significant deviation from scale invariance of the primordial power spectrum.  The \Planck\ values for these parameters and others derived from them are significantly different from those previously determined.  Several large-scale anomalies in the temperature distribution of the CMB, first detected by \WMAP, are confirmed with higher confidence. \Planck\ sets new limits on the number and mass of neutrinos, and has measured gravitational lensing of CMB anisotropies at greater than $25\sigma$.  \Planck\ finds no evidence for non-Gaussianity in the CMB.  \Planck's results agree well with results from the measurements of baryon acoustic oscillations.   \Planck\ finds a lower Hubble constant than found in some more local measures. Some tension is also present between the amplitude of matter fluctuations ($\sigma_8$) derived from CMB data and that derived from Sunyaev-Zeldovich data. The\Planck\ and \WMAP\  power spectra are offset from each other by an average level of about 2\% around the first acoustic peak.
Analysis of \Planck\ polarization data is not yet mature, therefore polarization results are not released, although the robust detection of $E$-mode polarization around CMB hot and cold spots is shown graphically.
}

\authorrunning{Planck Collaboration}
\titlerunning{The \Planck\ mission}
   \maketitle
%
%============================================================

%============================================================

\all2013resultspapers

\section{Introduction}

\allearlypapers

The \Planck\ satellite\footnote{\Planck\
(http://www.esa.int/\Planck) is a project of the European Space Agency (ESA) with instruments provided by two scientific consortia funded by ESA member states (in particular the lead countries, France and Italy) with contributions from NASA (USA), and telescope reflectors provided in a collaboration between ESA and a scientific consortium led and funded by Denmark.} 
\citep{tauber2010a, planck2011-1.1} was launched on 14 May 2009 and observed the sky stably and continuously from 12~August 2009 to 23~October 2013.  \Planck's scientific payload comprised an array of 74~detectors sensitive to frequencies between 25 and 1000\,GHz, which scanned the sky with  angular resolution between 33\arcm\ and 5\arcm.  The detectors of the Low Frequency Instrument \citep[LFI;][]{Bersanelli2010, planck2011-1.4} are pseudo-correlation radiometers, covering bands centred at 30, 44, and 70\,GHz.  The detectors of the High Frequency Instrument \citep[HFI;][]{Lamarre2010, planck2011-1.5} are bolometers, covering bands centred at 100, 143, 217, 353, 545, and $857\,$GHz.  \Planck\ images the whole sky twice in one year, with a combination of sensitivity, angular resolution, and frequency coverage never before achieved.  \Planck, its payload, and its performance as predicted at the time of launch are described in 13~papers included in a special issue of Astronomy \& Astrophysics (Volume 520).

\begin{figure*}
   \centering
   \includegraphics[width=0.95\textwidth]{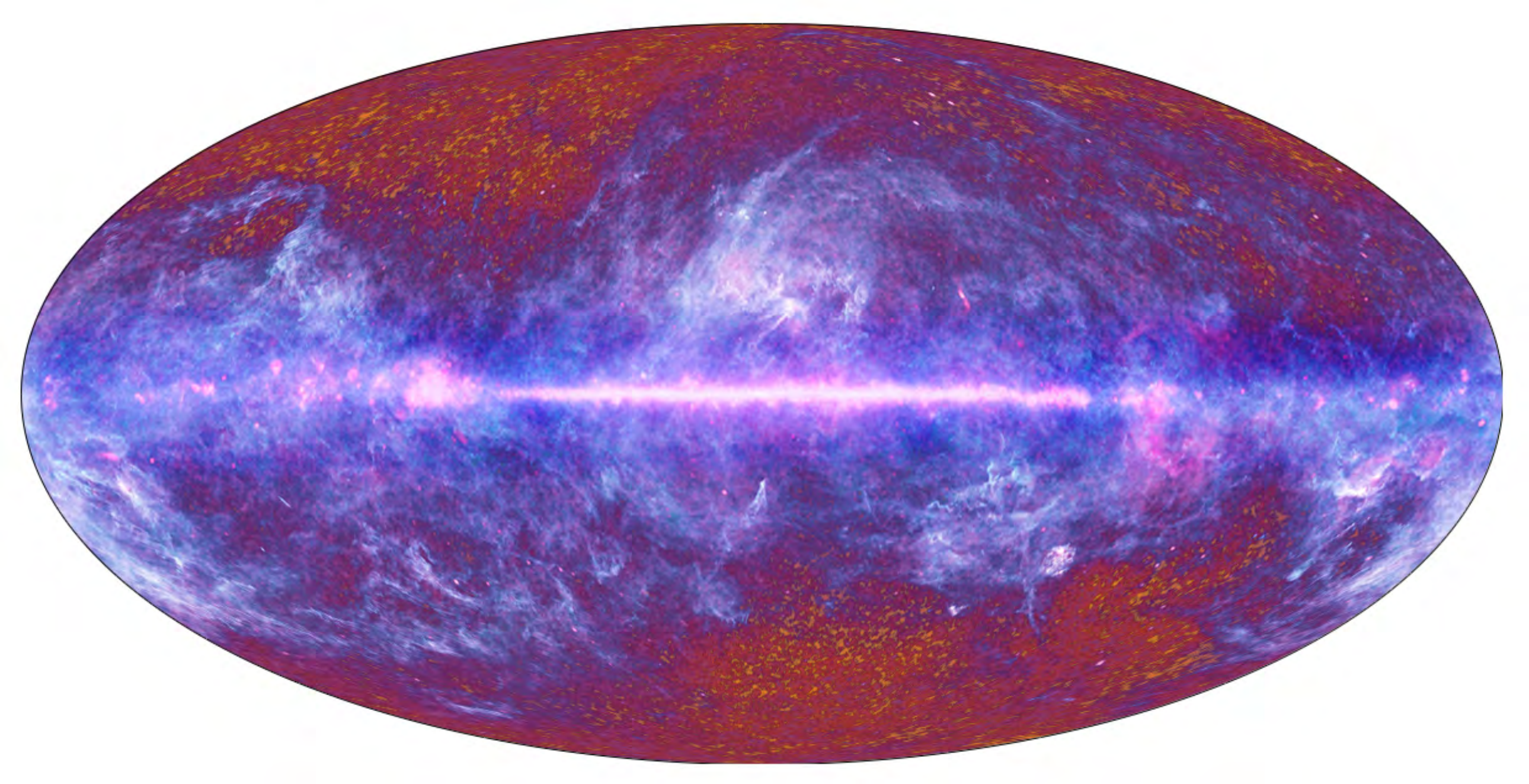}
   \caption{Composite, multi-frequency, full-sky image released by \Planck\ in 2010.  Made from the first nine months of the data, it illustrates artistically the multitude of Galactic, extragalactic, and cosmological components of the radiation detected by its payload.  Unless otherwise specified, all full-sky images in this paper are Mollweide projections in Galactic coordinates, pixelised according to the {\tt HEALPix} \citep{gorski2005} scheme.}
   \label{FigAllSky}
   \end{figure*}

The main objective of \Planck, defined in 1995, is to measure the spatial anisotropies in the temperature of the cosmic microwave background (CMB), with an accuracy set by fundamental astrophysical limits, thereby extracting essentially all the cosmological information embedded in the temperature anisotropies of the CMB. \Planck\ was also designed to measure to high accuracy the CMB polarization anisotropies, which encode not only a wealth of cosmological information, but also provide a unique probe of the early history of the Universe during the time when the first stars and galaxies formed.  Finally, \Planck\ produces a wealth of information on the properties of extragalactic sources and on the dust and gas in the Milky Way (see Fig.~\ref{FigAllSky}).  The scientific objectives of \Planck\ are described in detail in \cite{planck2005-bluebook}.  With the results presented here and in a series of accompanying papers (see Fig.~\ref{FigPapers}), \Planck\ has already achieved many of its planned science goals.

This paper presents an overview of the \Planck\ mission, and the main data products and scientific results of \Planck's second release%
\footnote{In January of 2011, ESA and the \Planck\ Collaboration released to the public a first set of scientific data, the Early Release Compact Source Catalogue (ERCSC), a list of unresolved and compact sources extracted from the first complete all-sky survey carried out by \Planck\ \citep{planck2011-1.10}.  At the same time, initial scientific results related to astrophysical foregrounds were published in a special issue of Astronomy and Astrophysics (Vol 520, 2011). Since then, 12 ``Intermediate" papers have been submitted for publication to A\&A containing further astrophysical investigations by the Collaboration.}, 
based on data acquired in the period 12~August 2009 to 28~November 2010.

\subsection{Overview of 2013 science results}

\result{Cosmology}A major goal of \Planck\ is to measure the key cosmological parameters describing our Universe.  \Planck's combination of sensitivity, angular resolution, and frequency coverage enables it to measure anisotropies on intermediate and small angular scales over the whole sky much more accurately than previous experiments.  This leads to improved constraints on individual parameters, the breaking of degeneracies between combinations of other parameters, and less reliance on supplementary astrophysical data than previous CMB experiments.  Cosmological parameters are presented and discussed in Sect.~\ref{sec:CMBcosmology} and in \citet{planck2013-p11}.  

The Universe observed by \Planck\ is well-fit by a six-parameter, vacuum-dominated, cold dark matter ($\Lambda$CDM) model, and we provide strong constraints on deviations from this model.  The values of key parameters in this model are summarized in Table~\ref{tab:LCDMparams}.  In some cases we find significant changes compared to previous measurements, as discussed in detail in \citet{planck2013-p11}.

With the \Planck\ data, we: 
(a) firmly establish deviation from scale invariance of the primordial matter perturbations, a key indicator of cosmic inflation; 
(b) detect with high significance lensing of the CMB by intervening matter, providing evidence for dark energy from the CMB alone;
(c) find no evidence for significant  deviations from Gaussianity in the statistics of CMB anisotropies; 
(d) find a deficit of power on large angular scales with respect to our best-fit model; 
(e) confirm the anomalies at large angular scales first detected by \WMAP; and 
(f) establish the number of neutrino species to be consistent with three.

The \planck\ data are in remarkable accord with a flat $\Lambda$CDM model; however, there are tantalizing hints of tensions both internal to the \planck\ data and with other data sets.  From the CMB, \Planck\ determines a lower value of the Hubble constant than some more local measures, and a higher value for the amplitude of matter fluctuations ($\sigma_8$) than that derived from Sunyaev-Zeldovich data.
While such tensions are model-dependent, none of the extensions of the six-parameter $\Lambda$CDM cosmology that we explored resolves them.  More data and further analysis may shed light on such tensions.  Along these lines, we expect significant improvement in data quality and the level of systematic error control, plus the addition of polarization data, from \planck\ in 2014.

A more extensive summary of cosmology results is given in Sect.~\ref{sec:CMBcosmology}.

\result{Foregrounds}The astrophysical foregrounds measured by \Planck\ to be separated from the CMB are interesting in their own right.  Compact and point-like sources consist mainly of extragalactic infrared and radio sources, and are released in the \Planck\ Catalogue of Compact Sources (PCCS; \citealt{planck2013-p05}).  An all-sky catalogue of sources detected via the Sunyaev-Zeldovich (SZ) effect, which will become a reference for studies of SZ-detected galaxy clusters, is given in \cite{planck2013-p05a}. 

Seven types of unresolved foregrounds must be removed or controlled for CMB analysis:  thermal dust emission; anomalous microwave emission (likely due to tiny spinning dust grains); CO rotational emission lines (significant in at least three HFI bands); free-free emission; synchrotron emission; the clustered cosmic infrared background (CIB); and SZ secondary CMB distortions.   For cosmological purposes, we achieve robust separation of the CMB from foregrounds using only \Planck\ data with multiple independent methods.  We release maps of: thermal dust + fluctuations of the cosmic infrared background; integrated emission of carbon monoxide; and synchrotron + free-free + spinning dust emission. These maps  provide a rich source for studies of the interstellar medium.  Other maps are released that use ancillary data in addition to the \Planck\ data to achieve more physically meaningful analysis.

These foreground products are described in Sect.~\ref{sec:AstroProds}.

\subsection{Features of the \Planck\ mission}

\Planck\ has an unprecedented combination of sensitivity, angular resolution, and frequency coverage.  For example, the \Planck\ detector array at 143\,GHz has instantaneous sensitivity  and angular resolution 25 and three times better, respectively, than the \WMAP\ V~band \citep{bennett2003a, hinshaw2012}.  Considering the final mission durations (nine years for \WMAP, 29~months for \Planck\ HFI, and 50~months for \Planck\ LFI), the white noise at map level, for example, is 12~times lower at 143\,GHz for the same resolution.  In harmonic space, the noise level in the \Planck\ power spectra is two orders of magnitude lower than in those of \WMAP\ at angular scales where beams are unimportant ($\ell<700$ for \WMAP\ and 2500 for \Planck). \Planck\ measures 2.6 times as many independent multipoles as \WMAP, corresponding to 6.8 times as many independent modes ($\ell,m$) when comparing the same leading CMB channels for the two missions.  This increase in angular resolution and sensitivity results in a large gain for analysis of CMB non-Gaussianity and cosmological parameters.  In addition, \Planck\ has a large overlap in $\ell$ with the high resolution ground-based experiments ACT \citep{2013arXiv1301.0824S} and SPT \citep{2011ApJ...743...28K}.  The noise spectra of SPT and \Planck\ cross at $\ell\sim 2000$, allowing an excellent check of the relative calibrations and transfer functions. 

Increased sensitivity places \Planck\ in a new situation.  Earlier satellite experiments ({\it COBE}/DMR, \citealt{1992ApJ...396L...1S}; \WMAP, \citealt{bennett2012}) were limited by detector noise more than systematic effects and foregrounds.  Ground-based and balloon-borne experiments ongoing or under development (e.g., ACT, \citealt{2003NewAR..47..939K}; SPT, \citealt{2004SPIE.5498...11R}; SPIDER, \citealt{2011arXiv1106.3087F}; and EBEX, \citealt{2010SPIE.7741E..37R}), have far larger numbers of detectors and higher angular resolution than \Planck, but can survey only a fraction of the sky over a limited frequency range.  They are therefore sensitive to foregrounds and limited to analysing only the cleanest regions of the sky.  Considering the impact of cosmic variance, Galactic foregrounds are not a serious limitation for CMB temperature-based cosmology at the largest spatial scales over a limited part ($<0.5$) of the sky.  Diffuse Galactic emission components  have steep frequency and angular spectra, and are very bright at frequencies below 70 and above 100\,GHz at low spatial frequencies.  At intermediate and small angular scales, extragalactic foregrounds, such as unresolved compact sources, the SZ effect from unresolved galaxy clusters and diffuse hot gas, and the correlated CIB, become important and cannot be ignored when carrying out CMB cosmology studies. \Planck's all-sky, wide-frequency coverage is key, allowing it to measure these foregrounds and remove them to below intrinsic detector noise levels, helped by  higher resolution experiments in characterizing the statistics of discrete foregrounds.

When detector noise is very low, systematic effects that arise from the instrument, telescope, scanning strategy, or calibration approach may dominate over noise in specific spatial or frequency ranges.  The analysis of redundancy is the main tool used by \Planck\ to understand and quantify the effect of systematics.  Redundancy on short timescales comes from the scanning strategy (Sect.~\ref{sec:ScanStrat}), which has particular advantages in this respect, especially for the largest scales.  When first designed, this strategy was considered ambitious because it required low $1/f$ noise near 0.0167\,Hz (the spin frequency) and very stable instruments over the whole mission.  Redundancy on long timescales comes in two versions: 1)~\Planck\ scans approximately the same circle on the sky every six months, alternating in the direction of the scan; and 2)~\Planck\ scans exactly (within arcminutes) the same circle on the sky every one year.  The ability to compare maps made in individual all-sky ``Surveys'' (covering approximately six month intervals, see Sect.~\ref{sec:ScanStrat} and Table~\ref{TabSurveys}) and year-by-year is invaluable in identifying specific systematic effects and calibration errors.  Although \Planck\ was designed to cover the whole sky twice over, its superb in-flight performance has enabled it to complete nearly five full-sky maps with the HFI instrument, and more than eight with the LFI instrument. The redundancy provided by such a large number of Surveys is a major asset for \Planck, allowing tests of the overall stability of the instruments over the mission and sensitive measurements of systematic residuals on the sky.

Redundancy of a different sort is provided by multiple detectors within frequency bands.  HFI includes four independent pairs of polarization-sensitive detectors in each channel from 100 to 353\,GHz, in addition to the four total intensity (spider web) detectors at all frequencies except 100\,GHz.  LFI includes six independent pairs of polarization-sensitive detectors at 70\,GHz, with three at 44\,GHz and two at 30\,GHz.   The different technologies used in the two instruments provide an additional powerful tool to identify and remove systematic effects.

Overall, the combination of scanning strategy and instrumental redundancy has allowed identification and removal of most systematic effects affecting CMB temperature measurements.  This can be seen in the fact that additional Surveys have led to significant improvements, at a rate greater than the square root of the integration time, in the signal-to-noise ratio (SNR) achieved in the combined maps.   Given that the two instruments have achieved their expected intrinsic sensitivity, and that most systematics have been brought below the noise (detector or cosmic variance) for intensity, it is a fact that cosmological results derived from the \Planck\ temperature data are already being limited by the foregrounds, fulfilling one of the main objectives of the mission.

\subsection{Status of \Planck\ polarization measurements}
\label{subsec:polstatus}

The situation for CMB polarization, whose amplitude is typically 4\,\% of intensity, is less mature.  At present, \Planck's sensitivity to the CMB polarization power spectrum at low multipoles ($\ell < 20$) is significantly limited by residual systematics.  These are of a different nature than those of temperature because polarization measurement with \Planck\ requires differencing between detector pairs.  Furthermore, the component separation problem is different, on the one hand simpler because only three polarized foregrounds have been identified so far (diffuse synchrotron and thermal dust emission, and radio sources), on the other hand more complicated because the diffuse foregrounds are more highly polarized than the CMB, and therefore more dominant over a larger fraction of the sky.  Moreover, no external templates exist for the polarized foregrounds.  These factors are currently restricting \Planck's ability to meet its most ambitious goals, e.g., to measure or set stringent upper limits on cosmological $B$-mode amplitudes.  Although this situation is being improved at the present time, the possibility remains that these effects will be the final limitation for cosmology using the polarized \Planck\ data. The situation is much better at high multipoles, where the polarization data are already close to being limited by  intrinsic detector noise.

These considerations have led to the strategy adopted by the \Planck\ Collaboration for the 2013 release of using only \Planck\ temperature data for scientific results.  To reduce the uncertainty on the reionization optical depth, $\tau$, we sometimes supplement the \Planck\ temperature data with the \WMAP\ low-$\ell$ polarization likelihood (the data designation in such cases includes ``WP'').  And we give two examples of polarization data at higher multipoles to demonstrate the quality already achieved.  The first example shows that the measured high-$\ell$ $EE$ spectrum agrees extremely well with that expected from the best-fit model derived from temperature data alone \citep{planck2013-p11}.  The second uses stacking techniques on the peaks and troughs of the CMB intensity (Sect.~\ref{sec:PolCMB}), giving a direct and spectacular visualization of the $E$-mode polarization induced by matter oscillating in the potential well of dark matter at recombination.  

Cosmological analysis using the full 29- and 50-month data sets, including polarization, will be published with the second major release of data in 2014.  Scientific investigations of diffuse Galactic polarized emissions at frequencies and angular scales where the polarized emission is strong compared to residual systematics will be released in the coming months (see Sect.~\ref{sec:DustPol} for a description).  The sensitivity and accuracy of \Planck's polarized maps is already well beyond that of any previous survey in this frequency range.

\section{Data products in the 2013 release}

The 2013 distribution of released products (hereafter the ``2013 products''), which can be freely accessed via the \Planck\ Legacy Archive interface\footnote{\url{http://archives.esac.esa.int/pla2}}, 
 is based on data acquired by \Planck\ during the ``nominal mission", defined as 12~August 2009 to 28~November 2010, and comprises:

\begin{itemize}

\item Maps of the sky at nine frequencies (Sect.~\ref{sec:FreqMaps}).  

\item Additional products that serve to quantify the characteristics of the maps to a level adequate for the science results being presented, such as noise maps, masks, and instrument characteristics. 

\item Four high-resolution maps of the CMB sky and accompanying characterization products (Sect.~\ref{subsec:CMBmapNG}).  Non-Gaussianity results are based on one of the maps; the others demonstrate the robustness of the results and their insensitivity to different methods of analysis.

\item A low-resolution CMB map (Sect.~\ref{subsec:CMBmapNG}) used in the low $\ell$ likelihood code, with an associated set of foreground maps produced in the process of separating the low-resolution CMB from foregrounds, with accompanying characterization products.

\item Maps of foreground components at high resolution, including: thermal dust + residual CIB; CO; synchrotron + free-free + spinning dust emission; and maps of dust temperature and opacity (Sect.~\ref{sec:AstroProds}).

\item A likelihood code and data package used for testing cosmological models against the \Planck\ data, including both the CMB (Sect.~\ref{sec:CMBLike}) and CMB lensing (Sect.~\ref{sec:LensLike}) . The CMB part is  based at $\ell < 50$ on the low-resolution CMB map just described and on the \WMAP-9 polarized likelihood (to reduce the uncertainty in $\tau$), and at $\ell\ge50$ on cross-power spectra of individual detector sets.  The lensing part is based on the 143 and 217\,GHz maps. 

\item The \Planck\ Catalogue of Compact Sources (PCCS, Sect.~\ref{sec:PCCS}), comprising lists of compact sources over the entire sky at the nine \Planck\ frequencies. The PCCS supersedes the previous Early Release Compact Source Catalogue \citep{planck2011-6.2}.

\item The \Planck\ Catalogue of Sunyaev-Zeldovich Sources (PSZ, Sect.~\ref{sec:PSZ}), comprising a list of sources detected by their SZ distortion of the CMB spectrum.  The PSZ supersedes the previous Early Sunyaev-Zeldovich Catalogue \citep{planck2013-p05a}.

\end{itemize}

\section{Papers accompanying the 2013 release }

\begin{figure*}
   \centering
   \includegraphics[width=180mm]{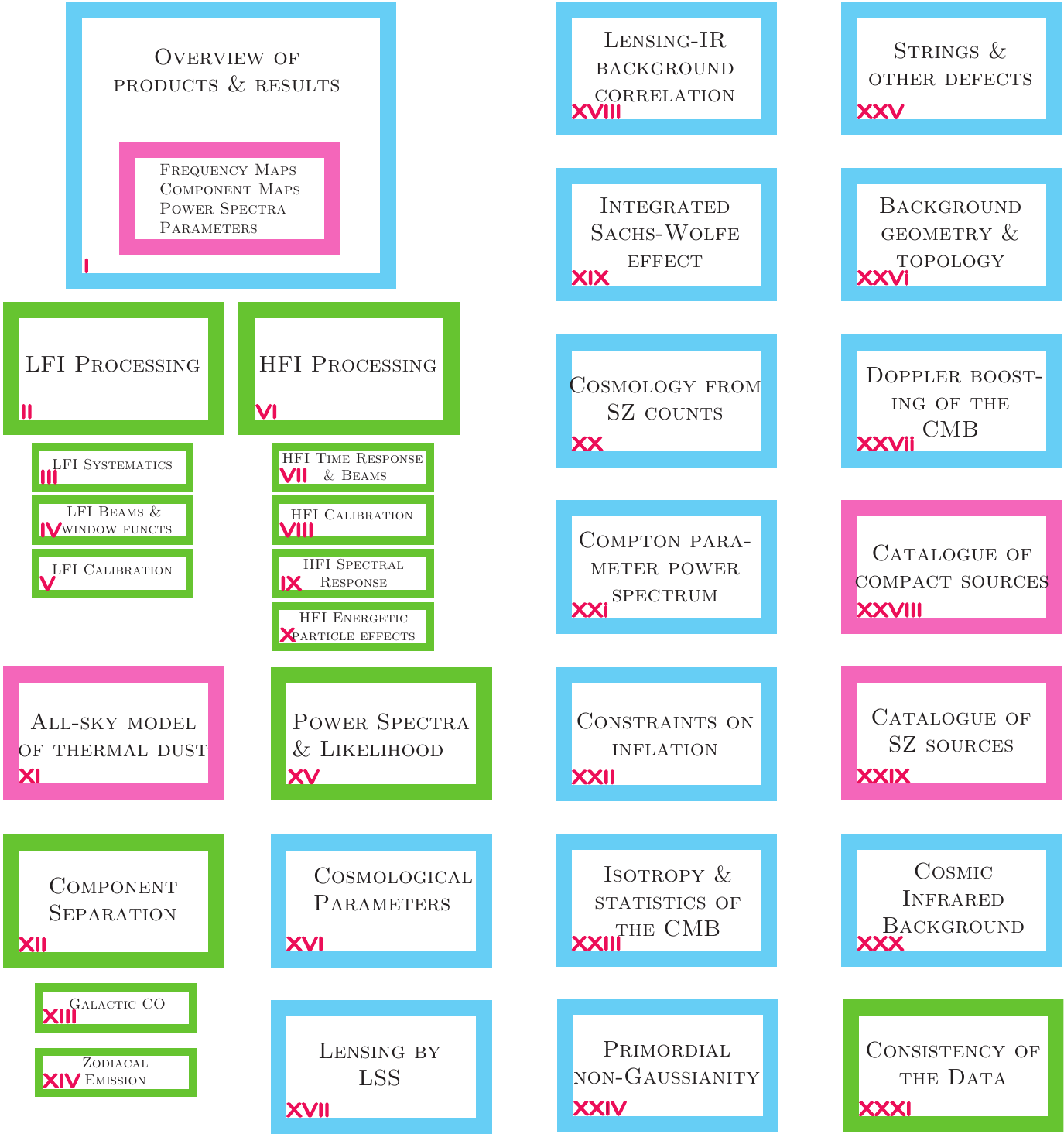}
   \caption{\Planck\ papers published simultaneously with the release of the 2013 products. The title of each paper is abbreviated. The roman numerals correspond to the sequence number assigned to each of the papers in the series; references include this number. Green boxes refer to papers describing aspects of data processing and the 2013 \Planck\ products. Blue boxes refer to papers mainly dedicated to scientific analysis of the products. Pink boxes describe specific 2013 \Planck\ products.
}
              \label{FigPapers}
    \end{figure*}

The characteristics, processing, and analysis of the \Planck\ data as well as a number of scientific results are described in a series of papers released simultaneously with the data.  The titles of the papers begin with ``\Planck\ 2013 results.'', followed by the  specific titles below.  Figure~\ref{FigPapers} gives a graphical view of the papers, divided into product, processing, and scientific result categories.

\def\CPPtitle#1{\vskip 2pt\noindent\vbox{\raggedright\hsize=\columnwidth\hangafter=1\hangindent=3em\noindent\strut#1\strut\par}}
\bigskip

\CPPtitle{I. Overview of products and results (\textit{this paper})}
\CPPtitle{II. Low Frequency Instrument data processing}
\CPPtitle{III. LFI systematic uncertainties}
\CPPtitle{IV.  LFI beams and window functions}
\CPPtitle{V. LFI calibration}
\CPPtitle{VI. High Frequency Instrument data processing}
\CPPtitle{VII. HFI time response and beams}
\CPPtitle{VIII. HFI photometric calibration and mapmaking}
\CPPtitle{IX. HFI spectral response}
\CPPtitle{X. HFI energetic particle effects: characterization, removal, and simulation}
\CPPtitle{XI. All-sky model of dust emission based on Planck data}
\CPPtitle{XII. Diffuse component separation}
\CPPtitle{XIII. Galactic CO emission}
\CPPtitle{XIV. Zodiacal emission}
\CPPtitle{XV. CMB power spectra and likelihood}
\CPPtitle{XVI. Cosmological parameters}
\CPPtitle{XVII. Gravitational lensing by large-scale structure}
\CPPtitle{XVIII. The gravitational lensing-infrared background correlation}
\CPPtitle{XIX. The integrated Sachs-Wolfe effect}
\CPPtitle{XX. Cosmology from Sunyaev-Zeldovich cluster counts}
\CPPtitle{XXI. Cosmology with the all-sky Compton-parameter power spectrum}
\CPPtitle{XXII. Constraints on inflation}
\CPPtitle{XXIII. Isotropy and statistics of the CMB}
\CPPtitle{XXIV. Constraints on primordial non-Gaussianity}
\CPPtitle{XXV. Searches for cosmic strings and other topological defects}
\CPPtitle{XXVI. Background geometry and topology of the Universe}
\CPPtitle{XXVII. Doppler boosting of the CMB: Eppur si muove}
\CPPtitle{XXVIII. The \Planck\ Catalogue of Compact Sources}
\CPPtitle{XXIX. The \Planck\ catalogue of Sunyaev-Zeldovich sources}
\CPPtitle{XXX. Cosmic infrared background measurements and implications for star formation}
\CPPtitle{XXXI. Consistency of the \Planck\ data}

\bigskip

In the next few months additional papers will be released concentrating on Galactic foregrounds in both temperature and polarization.

This paper contains an overview of the main aspects of the \Planck\ project that have contributed to the 2013 release, and points to the papers (Fig. \ref{FigPapers}) that contain full descriptions.  It proceeds as follows:

\begin{itemize}

\item Section~\ref{sec:Mission} summarizes the operations of \Planck\ and the performance of the spacecraft and  instruments.

\item Sections~\ref{sec:DataProcessing} and \ref{sec:FreqMaps} describe the processing steps carried out in the generation of the nine \Planck\ frequency maps and their characteristics.

\item Section~\ref{sec:SciProds} describes the \Planck\ 2013 products related to the Cosmic Microwave Background, namely the CMB maps, the lensing products, and the likelihood code.

\item Section~\ref{sec:AstroProds} describes the \Planck\ 2013 astrophysical products, namely catalogues of compact sources and maps of diffuse foreground emission.

\item Section~\ref{sec:CMBcosmology} describes the main cosmological science results based on the 2013 CMB products.

\item Section~\ref{sec:summary} concludes with a summary and a look towards the next generation of \Planck\ products.

\end{itemize}

%============================================================

%________________________________________________________________

\section{The \Planck\ mission}
\label{sec:Mission}

\begin{figure*}
   \centering
   \includegraphics[width=0.95\textwidth]{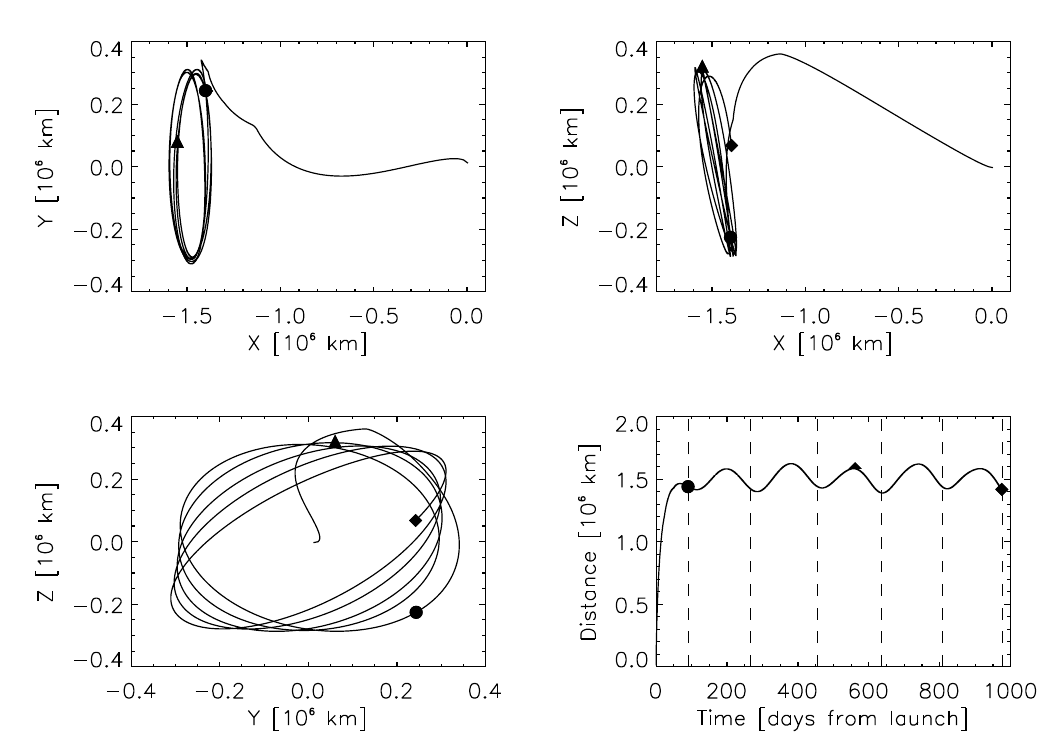}
   \caption{The trajectory of \Planck\ from launch until 13~January 2012, in Earth-centred rotating coordinates ($X$ is in the Sun-Earth direction; $Z$ points to the north ecliptic pole).  Symbols indicate the start of routine operations (circle), the end of the nominal mission (triangle), and the end of HFI data acquisition (diamond). The orbital periodicity is 6\,months. The distance from the Earth-Moon barycentre is shown in the bottom right panel, together with Survey boundaries.}
    \label{FigOrbit}
\end{figure*}

\Planck\ was launched from Kourou, French Guiana, on 14~May 2009 on an Ariane~5 ECA rocket, together with the {\it Herschel} Space Observatory.   After separation from the rocket and from {\it Herschel}, \Planck\ followed a trajectory to the $L_2$ point of the Sun-Earth system.  It was injected into a 6-month Lissajous orbit around $L_2$ in early July 2009 (Fig.~\ref{FigOrbit}).  Small manoeuvres are required at approximately monthly intervals (totalling around 1\,m\,s\mo\ per year) to keep \Planck\ from drifting away from $L_2$.  

The first three months of operations focused on commissioning (during which \Planck\ cooled down to the operating temperatures of the coolers and the instruments), calibration, and performance verification.  Routine operations and science observations began 12~August 2009.  Detailed information about the first phases of operations may be found in \cite{planck2011-1.1} and \citet{planck2013-p28}.

\subsection{Scanning strategy}
\label{sec:ScanStrat}

\begin{figure*}
   \centering
   \includegraphics[width=180mm]{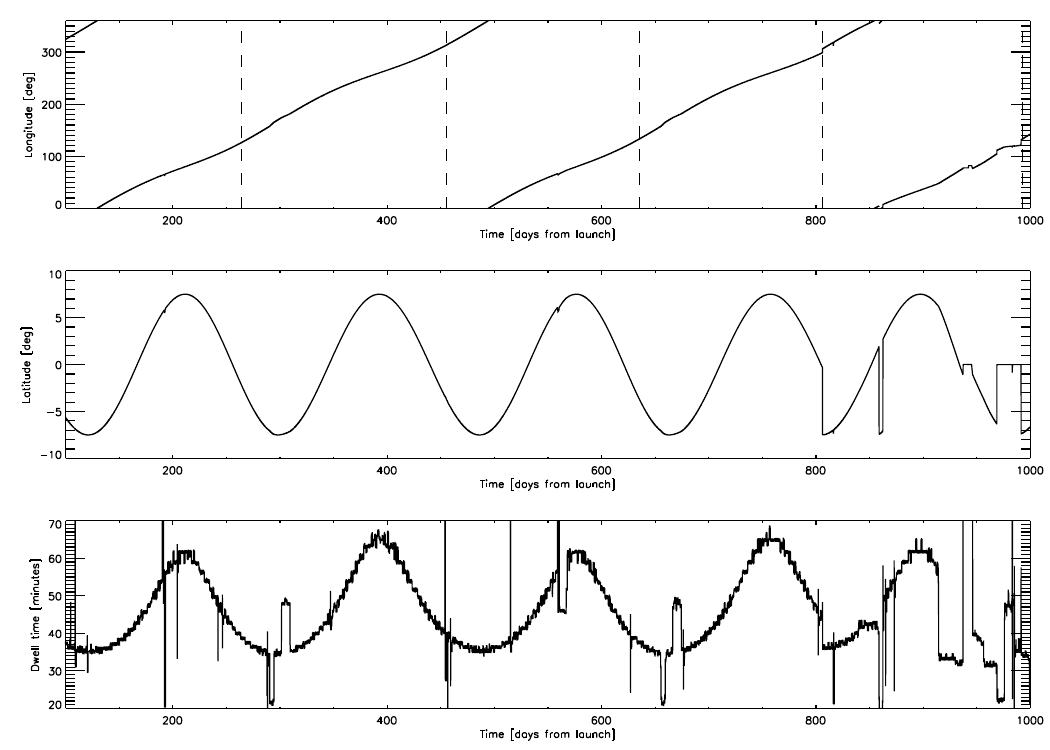}   
\caption{{\it Top two panels\/}: the path of the spin axis of \Planck\ (in ecliptic longitude and latitude) over the period 12 August 2009 (91 days after launch) to 13 January 2012, the ``0.1-K mission'' period (Table~\ref{TabSurveys}).  {\it Bottom panel\/}: the evolution of the dwell time during the same period.  Intervals of  acceleration/deceleration (e.g., around observations of the Crab) are clearly visible as symmetric temporary increases and reductions of dwell time. Survey boundaries are indicated by vertical dashed lines in the upper plot.  The change in cycloid phase is clearly visible at operational day (OD) 807. The disturbances around OD\,950 are due to the ``spin-up campaign".}
    \label{FigScanning}
\end{figure*}

\Planck\ spins at 1\,rpm about the symmetry axis of the spacecraft.  The spin axis follows a cycloidal path across the sky in step-wise displacements of 2\arcm\ (Fig.~\ref{FigScanning}).   To maintain a steady advance of the projected position of the spin axis along the ecliptic plane, the time interval between two manoeuvres varies between 2360\,s and 3904\,s.  Details of the scanning strategy are given in \cite{tauber2010a} and \cite{planck2011-1.1}.  

The fraction of time used by the manoeuvres themselves (typical duration of five minutes) varies between 6\,\% and 12\,\%, depending on the phase of the cycloid.  At present, the reconstructed position of the spin axis during manoeuvres has not been determined accurately enough for scientific work (but see Sect.~\ref{sec:Point}), and the data taken during manoeuvres are not used in the analysis.  Over the nominal mission, the total reduction of scientific data due to manoeuvres was 9.2\,\%.

The boresight of the telescope is 85\deg\ from the spin axis.  As \Planck\ spins, the instrument beams cover nearly great circles in the sky.  The spin axis remains fixed (except for a small drift due to Solar radiation pressure) for between 39 and 65 spins (corresponding to the dwell times given above), depending on which part of the cycloid \Planck\ is in.  To high accuracy, any one beam covers precisely the same sky between 39 and 65 times.  The set of observations made during a period of fixed spin axis pointing is often referred to as a ``ring.''  This redundancy plays a key role in the analysis of the data, as will be seen below, and is an important feature of the scan strategy.

As the Earth and \Planck\ orbit the Sun, the nearly-great circles that are observed rotate about the ecliptic poles.  The amplitude of the spin-axis cycloid is chosen so that all beams of both instruments cover the entire sky in one year.  In effect, \Planck\ tilts to cover first one Ecliptic pole, then tilts the other way to cover the other pole six months later.  If the spin axis stayed exactly on the ecliptic plane, the telescope boresight were perpendicular to the spin axis, the Earth were in a precisely circular orbit, and \Planck\ had only one detector with a beam aligned precisely with the telescope boresight, that beam would cover the full sky in six months.  In the next six months, it would cover the same sky, but with the opposite sense of rotation on a given great circle.  However, since the spin axis is steered in a cycloid, the telescope is 85\deg\ to the spin axis, the focal plane is several degrees wide, and the Earth's orbit is slightly elliptical, the symmetry of the scanning is (slightly) broken.   Thus the \Planck\ beams scan the entire sky exactly twice in one year, but scan only 93\,\% of the sky in six months.  For convenience, we call an approximately six month period one ``survey'', and use that term as an inexact shorthand for one coverage of the sky.  Nine numbered ``Surveys'' are defined precisely in Table~\ref{TabSurveys}.  It is important to remember that as long as the phase of the cycloid remains constant, one year corresponds to exactly two coverages of the sky, while one Survey has an exact meaning only as defined in Table~\ref{TabSurveys}.  Null tests between 1-year periods with the same cycloid phase are extremely powerful.  Null tests between Surveys are also useful for many types of tests, particularly in revealing differences due to beam orientation.

\subsection{Routine operations}

Routine operations started on 12~August 2009.  The beginning  and end dates of each Survey are listed in Table~\ref{TabSurveys}, which also shows the fraction of the sky covered by all frequencies.  The fourth Survey was shortened somewhat so that the slightly different scanning strategy adopted for Surveys~5--8 (see below) could be started before the Crab Nebula, an important polarization calibration source, was observed. The coverage of the fifth Survey is smaller than the others because several weeks of integration time were dedicated to ``deep rings" (defined below) covering sources of special importance.

\begin{table*}
\caption{\Planck\ Surveys (defined in Sect.~\ref{sec:ScanStrat}).  Times are UT.   } 
\label{TabSurveys} 
\vskip -18pt
\setbox\tablebox=\vbox{
\newdimen\digitwidth
\setbox0=\hbox{\rm 0}
\digitwidth=\wd0
\catcode`*=\active
\def*{\kern\digitwidth}
\newdimen\signwidth
\setbox0=\hbox{+}
\signwidth=\wd0
\catcode`!=\active
\def!{\kern\signwidth}
\newdimen\pointwidth
\setbox0=\hbox{.}
\pointwidth=\wd0
\catcode`@=\active
\def@{\kern\pointwidth}
\newdimen\monthwidth
\setbox0=\hbox{Aug }
\monthwidth=\wd0
\def\Jan{\hbox to \monthwidth{Jan\hfil}}
\def\Feb{\hbox to \monthwidth{Feb\hfil}}
\def\Mar{\hbox to \monthwidth{Mar\hfil}}
\def\Apr{\hbox to \monthwidth{Apr\hfil}}
\def\May{\hbox to \monthwidth{May\hfil}}
\def\Jun{\hbox to \monthwidth{Jun\hfil}}
\def\Jul{\hbox to \monthwidth{Jul\hfil}}
\def\Aug{\hbox to \monthwidth{Aug\hfil}}
\def\Sep{\hbox to \monthwidth{Sep\hfil}}
\def\Oct{\hbox to \monthwidth{Oct\hfil}}
\def\Nov{\hbox to \monthwidth{Nov\hfil}}
\def\Dec{\hbox to \monthwidth{Dec\hfil}}
\halign{\hbox to 3.5cm{#\leaderfil}\tabskip 2.5em&
\hfil#\hfil&
\hfil#\hfil&
\hfil#\hfil&
\hfil#\hfil\tabskip 0pt\cr
\noalign{\vskip 3pt\hrule\vskip 1.5pt\hrule\vskip 5pt}
\omit\hfil Survey\hfil&  Instrument& Beginning& End& Coverage$^{\rm a}$\cr
\noalign{\vskip 4pt\hrule\vskip 6pt}
1&LFI \& HFI&12 \Aug 2009 (14:16:51)&02 \Feb 2010 (20:51:04)&93.1\,\%\cr
2&LFI \& HFI&02 \Feb 2010 (20:54:43)&12 \Aug 2010 (19:27:20)&93.1\,\%\cr
3&LFI \& HFI&12 \Aug 2010 (19:30:44)&08 \Feb 2011 (20:55:55)&93.1\,\%\cr
4&LFI \& HFI&08 \Feb 2011 (20:59:10)&29 \Jul 2011 (17:13:32)&86.6\,\%\cr
5&LFI \& HFI&29 \Jul 2011 (18:04:49)&01 \Feb 2012 (05:25:59)&80.1\,\%\cr
6&       LFI&01 \Feb 2012 (05:26:29)&03 \Aug 2012 (16:48:51)&79.2\,\%\cr
7&       LFI&03 \Aug 2012 (16:48:53)&31 \Jan 2013 (10:32:08)&73.7\,\%\cr
8&       LFI&31 \Jan 2013 (10:32:10)&03 \Aug 2013 (21:53:37)&70.6\,\%\cr
9&       LFI&03 \Aug 2013 (21:53:39)&03 \Oct 2013 (21:13:38)&21.2\,\%\cr
\noalign{\vskip 5pt}
``Nominal mission''&LFI \& HFI&12 \Aug 2009 (14:16:51)&28 \Nov 2010 (12:00:53)&\dots\cr
``0.1-K mission''&  LFI \& HFI&12 \Aug 2009 (14:16:51)&13 \Jan 2012 (14:54:07)&\dots\cr
\noalign{\vskip 3pt\hrule\vskip 4pt}
}}
\endPlancktablewide
\tablenote {{\rm a}} {Fraction of sky covered by all frequencies}  \par
\end{table*}

During routine scanning, the \Planck\ instruments naturally observe objects of special interest for calibration.  These include Mars, Jupiter, Saturn, Uranus, Neptune, and the Crab Nebula.  Different types of observations of these objects were performed:

\begin{itemize}

\item Normal scans on Solar System objects and the Crab Nebula.  The complete list of observing dates for these objects can be found in \citet{planck2013-p28}.

\item ``Deep rings."  These special scans are performed on observations of Jupiter and the Crab Nebula from January 2012 onward.  They comprise deeply and finely sampled (step size 0\parcm5) observations  with the spin axis along the Ecliptic plane, lasting typically two to three weeks.  Since the Crab is crucial for calibration of both instruments,  the average longitudinal speed of the pointing steps was increased before scanning the Crab, to improve operational margins and ease recovery in case of problems.

\item ``Drift scans."  These special observations are performed on Mars, making use of its proper motion.  They allow finely-sampled measurements of the beams, particularly for HFI.  The rarity of Mars observations during the mission gives them high priority.

\end{itemize}

The cycloid phase was shifted by 90\deg\ for Surveys~5--8 to optimize the range of polarization angles on key sources in the combination of Surveys 1--8, thereby helping in the treatment of systematic effects and improving polarization calibration.

As stated in Sect.~2, the 2013 products are based on the 15.5-month nominal mission, and include data acquired during Surveys 1, 2, and part of 3. 

The scientific lifetime of the HFI bolometers ended on 13~January 2012 when the supply of $^3$He needed to cool them to 0.1\,K ran out.  LFI continued to operate and acquire scientific data through 3~October 2013.  \Planck\ operations ended 23~October 2013.  Data from the remaining part of Survey 3, Surveys 4 and 5 (both LFI and HFI), and Surveys 6--9 (LFI only) will be released in 2014.

Routine operations were significantly modified twice more:

\begin{itemize}

\item The sorption cooler switchover from the nominal to the redundant unit took place on~11 August 2010, leading to an interruption of acquisition of useful scientific data for about two days (one for the operation itself, and one for re-tuning of the cooling chain).

\item The satellite's rotation speed was increased to 1.4\,rpm between 8 and 16~December 2011 for observations of Mars,  to measure possible systematic effects on the scientific data linked to the spin rate. 

\end{itemize}

 Data were acquired in the normal way during the above two periods, but were not used in the 2013 products.

\begin{figure*}
\centering
\includegraphics[width=180mm]{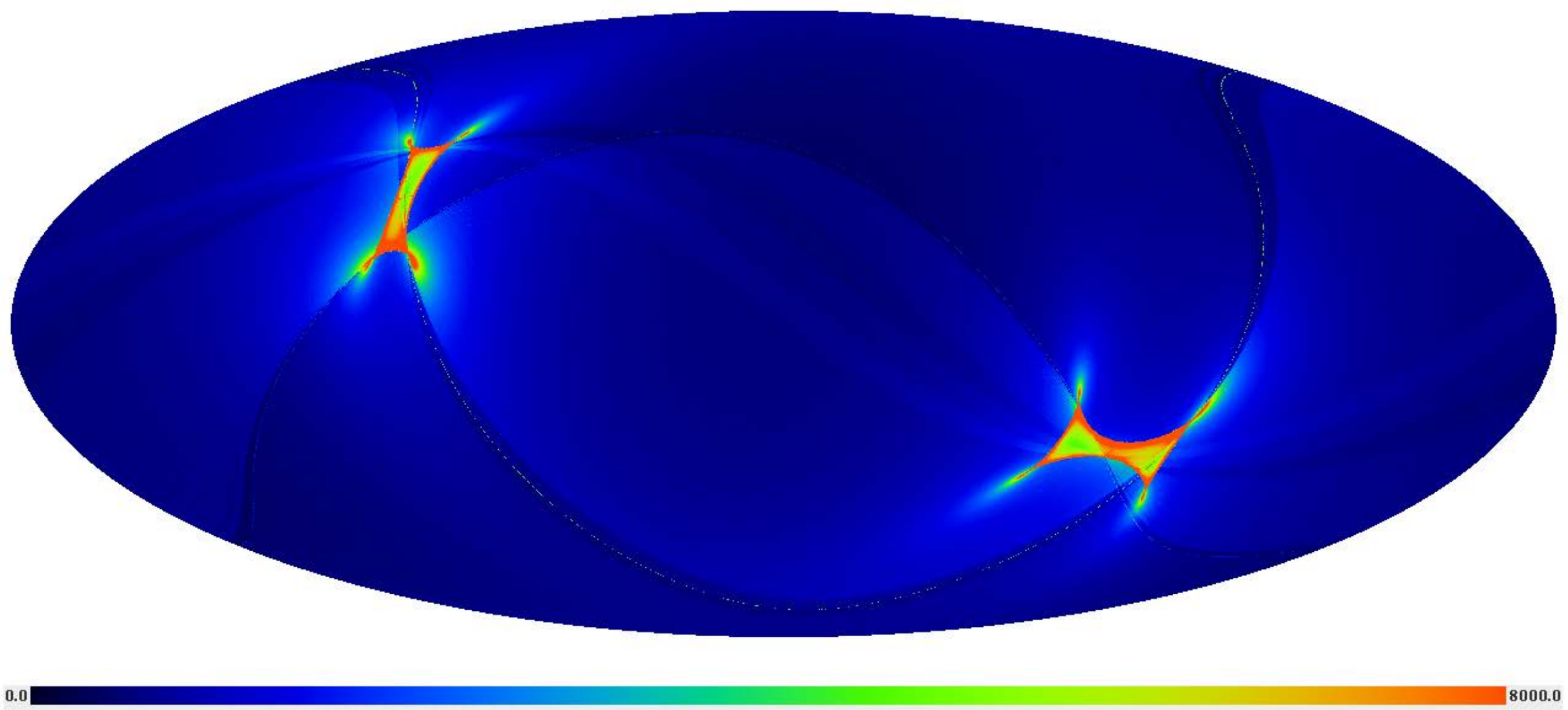}
\includegraphics[width=180mm]{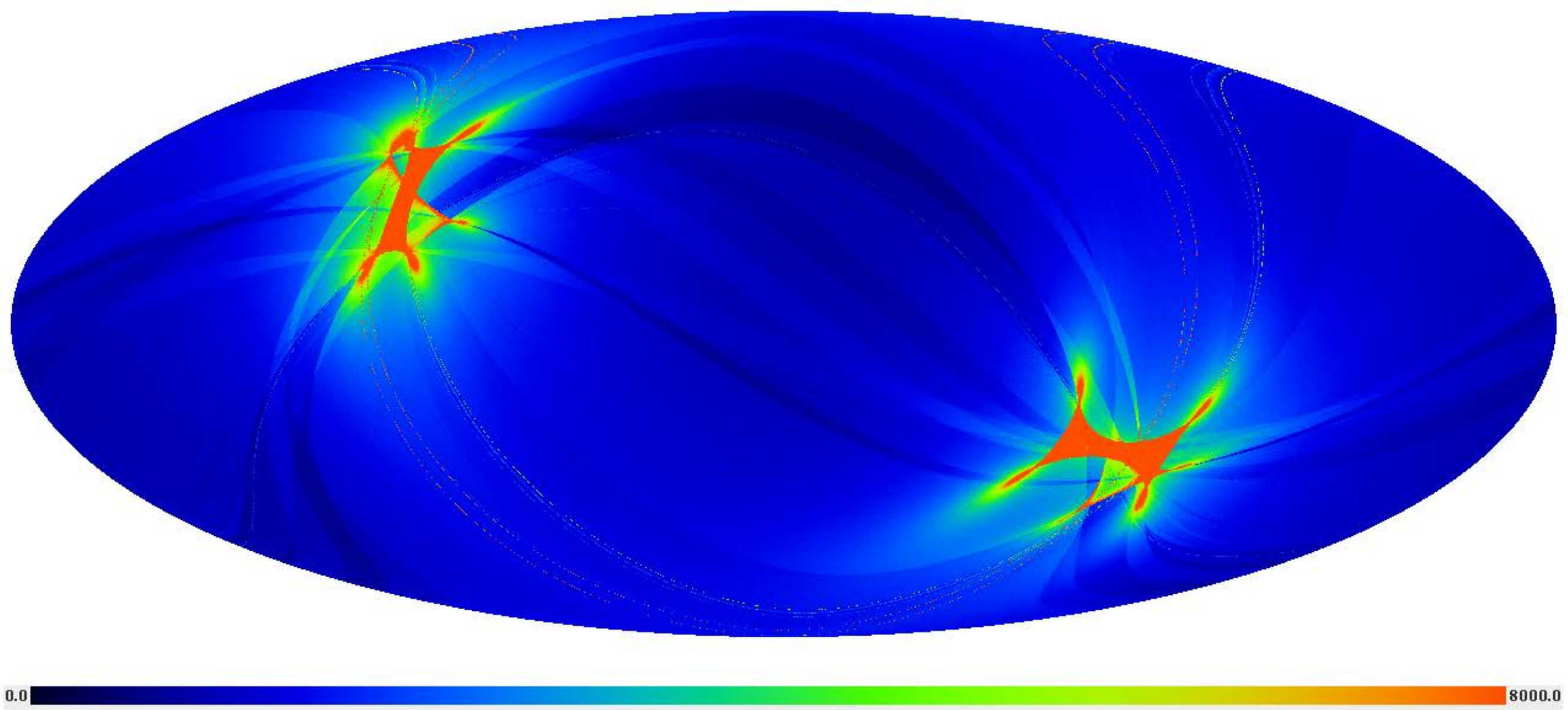}
\caption{ Survey coverage for the nominal (\textit{top})  and 0.1-K (\textit{bottom}) missions (see Table~\ref{TabSurveys}). The colour scale represents total integration time (varying between 50 and 8000\,s\,deg$^{-2}$) for the 353\,GHz channel.  The maps are at $N_{\rm side}$ = 1024.}
\label{FigCoverage}
\end{figure*}

The distribution of integration time over the sky for the nominal and ``0.1-K" (i.e., until the $^3$He ran out, see Table~\ref{TabSurveys}) missions is illustrated in Fig.~\ref{FigCoverage} for a representative frequency channel.  More details can be found in the Explanatory Supplement \citep{planck2013-p28}.

Operations have been extremely smooth throughout the mission.  The total observation time lost due to a few anomalies is about 5\,days, spread over the 15.5\,months of the nominal mission.

\subsection{Satellite environment}

\begin{figure*}
   \centering
   \includegraphics[width=180mm]{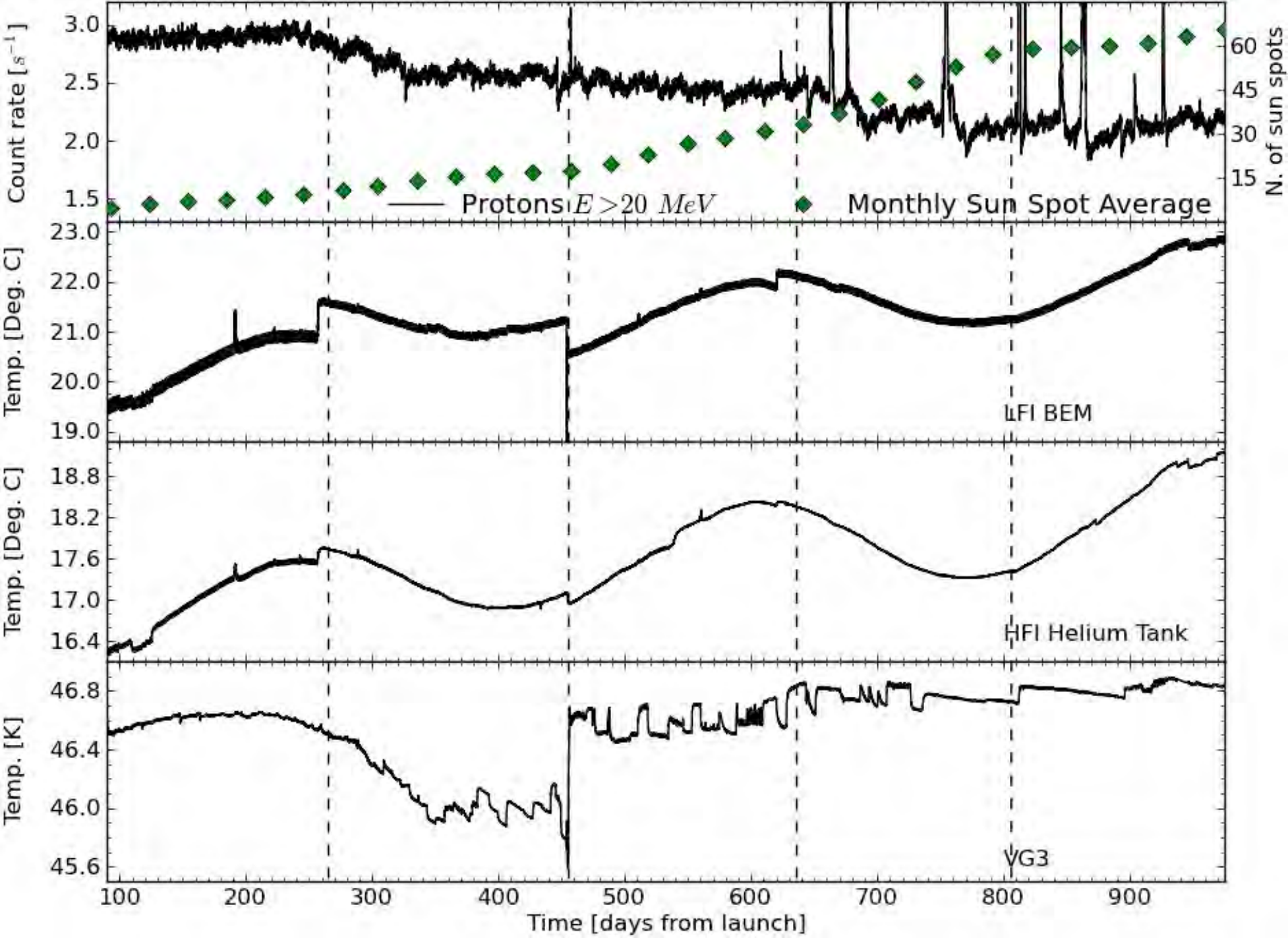} 
   \caption{The thermal and radiation environment of \Planck.  Vertical lines indicate boundaries between Surveys.   The top panel shows the cosmic ray flux as measured by the onboard SREM; its decrease over time is due to the corresponding increase in Solar activity, indicated by the sunspot number, from {\tt http://www.sidc.be/sunspot-data/}. Solar flares show up as spikes in the proton flux.   The second and third panels show the temperature variation at two representative locations in the room-temperature SVM, i.e., on one of the (HFI) helium tanks and on one of the LFI back-end modules (BEMs). The sine-wave modulation tracks the variation of distance from the Sun.  The bottom panel shows the temperature evolution of VG3, the coldest of three V-grooves, to which the sorption cooler is heat-sunk. The disturbances on the curve are due to adjustments of the operational parameters of this cooler.}
    \label{FigThermalEnvironment}
    \end{figure*}

The thermal and radiation environment of the satellite during the routine phase is illustrated in Fig.~\ref{FigThermalEnvironment}. The dominant long-timescale thermal modulation is driven by variations in Solar power absorbed by the satellite in its elliptical orbit around Sun.  The thermal environment is sensitive to various satellite operations. For example, before day 257, the communications transmitter was turned on only during the daily data transmission period, causing a daily temperature variation clearly visible at all locations in the Service Module (Fig.~\ref{FigThermalEnvironment}).   Some operational events%
\footnote{Most notably: a)~the ``catbed'' event between 110 and 126\,days after launch; b)~the ``day \Planck\ stood still'' 191 days after launch; c)~the sorption cooler switchover (OD\,460); d) the change in the thermal control loop (OD\,540) of the LFI radiometer electronics assembly box; and e)~the spin-up campaign around OD\,950.}  
had a significant thermal impact as shown in Fig.~\ref{FigThermalEnvironment} and detailed in \citet{planck2013-p28}.

The sorption cooler dissipates a large amount of power and drives temperature variations at multiple levels in the satellite.  The bottom panel of Fig.~\ref{FigThermalEnvironment} shows the temperature evolution of the coldest of the three stacked conical structures or V-grooves that thermally isolate the warm service module (SVM) from the cold payload module.  Most variations of this structure are due to quasi-weekly power input adjustments of the sorption cooler, whose tube-in-tube heat-exchanger supplying high pressure gas to the 20-K Joule-Thomson valve and returning low pressure gas to the compressor assembly is heat-sunk to it.  Many adjustments are seen in the roughly three months leading up to switchover.  After switchover to the redundant cooler (Sect.~\ref{sec:LFIOps}), thermal instabilities were present in the newly operating sorption cooler, which required frequent adjustment, until they reduced significantly around day 750.

Figure~\ref{FigThermalEnvironment} also shows the radiation environment history.  As \Planck\ started operations, Solar activity was extremely low, and Galactic cosmic rays (which produce sharp ``glitches'' in the HFI bolometer signals, see Sect.~\ref{sec:HFIOps}) were more easily able to enter the heliosphere and hit the satellite.  As Solar activity increased the cosmic ray flux measured by the onboard standard radiation environment monitor (SREM; \citealt{planck2013-p28}) decreased correspondingly, but Solar flares increased.

\subsection{Instrument environment, operations, and performance}

\subsubsection{LFI}
\label{sec:LFIOps}

The front-end of the LFI array is cooled to $20\,$K by a sorption cooler system, which included a nominal and a redundant unit \citep{planck2011-1.3}.   In early August of 2010, the gas-gap heat switch of one compressor element on the active cooler reached the end of its life.  Although the cooler can operate with as few as four (out of six) compressor elements, it was decided to switch operation to the redundant cooler.  On 11 August at 17:30 GMT the working cooler was switched off, and the redundant one was switched on.  Following this operation, an increase of temperature fluctuations in the 20\,K stage was observed.  The cause has been ascribed to the influence of liquid hydrogen remaining in the cold end of the inactive (previously operating) cooler.  These thermal fluctuations produced a measurable effect in the LFI data, but they propagate to the power spectrum at a level more than four orders of magnitude below
the CMB temperature signal \citep{planck2013-p02a} and have a negligible effect on the science data.  Furthermore, in February 2011 these fluctuations were reduced to a much lower level and have remained low ever since.

The 22~LFI radiometers have been extremely stable since the beginning of the observations \citep{planck2013-p02a}, with $1/f$  knee frequencies of order 50\,mHz and white noise levels unchanging within a few percent. After optimization during the calibration and performance verification phase, no changes to the bias of the front-end HEMT low-noise amplifiers and phase
switches were required throughout the nominal mission. 

The main disturbance to LFI data acquisition has been an occasional bit-flip change in the gain-setting circuit of the data acquisition electronics, probably due to cosmic ray hits \citep{planck2013-p02}.  Each of these events leads to the  loss of a fraction of a single ring for the affected detector.  The total level of data loss was extremely low, less than 0.12\,\% over the whole mission.

\subsubsection{HFI}
\label{sec:HFIOps}

HFI operations were extremely smooth.  The instrument parameters were not changed after being set during the calibration and performance verification phase. 

The satellite thermal environment had no major impact on HFI.  A drift of the temperature of the service vehicle module (SVM) due to the eccentricity of the Earth's orbit (Fig.~\ref{FigThermalEnvironment}) induced negligible changes of temperature of the HFI electronic chain.  Induced gain variations are of order 10$^{-4}$ per degree K.

The HFI dilution cooler \citep{planck2011-1.3} operated at the lowest available gas flow rate, giving a lifetime  twice the 15.5\,months of the nominal mission.  This was predicted to be possible following ground tests,  and demonstrates how representative of the flight environment these difficult tests were.

\begin{figure*}
   \centering
   \includegraphics[width=180mm]{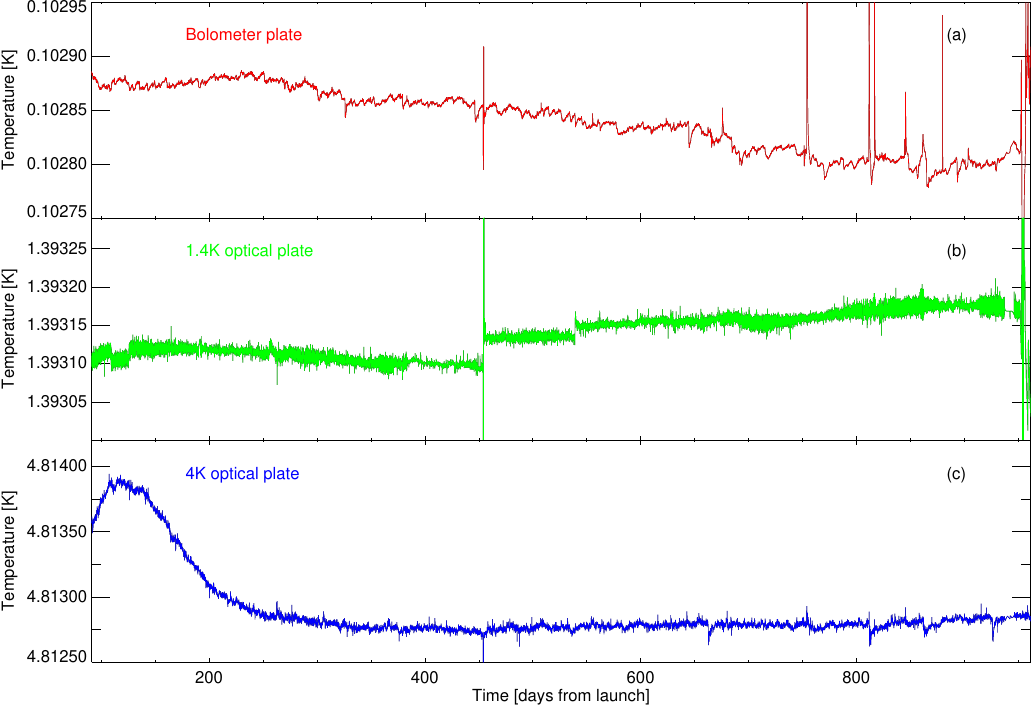} 
   \caption{Thermal stability of the HFI bolometer ({\it top\/}), 1.4\,K optical filter (\textit{middle}) and  4-K cooler reference load (\textit{bottom}) stages. The horizontal axis displays days since launch (the nominal mission begins on day 91). The sharp feature at Day 460 is due to the sorption cooler switchover.}
    \label{FigHFIStability}
\end{figure*}

The HFI cryogenic system remained impressively stable over the whole cryogenic mission.   Figure~\ref{FigHFIStability} shows the temperature of the three cold stages of the \HeJT\ and dilution coolers.  The temperature stability of the 1.6\,K and 4\,K plates, which support the feed horns, couple detectors to the telescope, and support the filters, was well within specifications and produced negligible effects on the scientific signals. The dilution cooler showed the  secular evolution of heat lift expected from the small drifts of the $^3$He and $^4$He flows as the pressure in the tanks decreased.  The proportional-integral-differential (PID) temperature regulation of the bolometer plate had a long time constant to avoid transferring cosmic-ray-induced glitches on the PID thermometers to the plate.  The main driver of the bolometer plate temperature drifts was the long-term change in the cosmic ray hit rate modulated by the Solar cycle, as described in \cite{planck2011-1.3} (see also Fig.~\ref{FigHFIStability}).  These very slow drifts did not induce any  significant direct systematic effect on the scientific signals.  Shorter-term temperature fluctuations of the bolometer plate driven by cosmic rays create steep low-frequency noise correlated between detectors.  This can be mostly removed using the measured temperatures, leaving a negligible residual at frequencies above the spin frequency of 0.016\,Hz.

The main effect of the cooling system on the scientific signals is an indirect one:  the very slow drift of the detector temperature over periods of weeks or months changes the amplitude of the modulated signal that shifts the science signal on the analogue-to-digital converters (ADCs) in each detector and thermometer electronic chain. The non-linearities of these devices, especially in the middle of their dynamic range, where most of the scientific signal is concentrated, lead to a systematic effect that can only be corrected empirically using the redundancies in its first order effect -- a gain change -- on the data processing (see Sect.~\ref{sec:HFITOI}).

Detector-to-detector cross-talk was checked in flight using Jupiter and strong glitches.  The level of cross-talk between detectors in different pixels is very low; however, the level of cross-talk between the two polarization sensitive bolometers (PSB) of a PSB pair is significant, in line with ground-based measurements.  For temperature-only analysis, this effect is negligible. 

Two of the bolometers, one at 143\,GHz and one at 545\,GHz, suffer heavily from ``random telegraphic signals" (RTS; \citealt{planck2013-p03}) and are not used.  Three other bolometers (two at 217\,GHz and one at 857\,GHz) exhibit short periods of RTS that are discarded; the periods of RTS span less than 10\,\% of the data from each of these detectors.

Cosmic rays induce short glitches in the scientific signal when they deposit energy either in the thermistor or on the bolometer grid.  They were observed in flight at the predicted rate with a decay time constant equal to the one measured during ground testing.  In addition, a different kind of glitch was observed, occurring in larger numbers but with lower amplitudes and long time constants; they are understood to be induced by cosmic ray hits on the silicon wafer of the bolometers \citep{planck2013-p03e}.   The different kinds of glitches observed in the HFI bolometers are described in detail in \cite{planck2013-p03e}.  High energy cosmic rays also induce secondary particle showers in the spacecraft and in the vicinity of the focal plane unit, contributing to correlated noise \citep{planck2013-p03}. 

A more detailed description of the performance of HFI is available in \citet{planck2013-p03}.

\subsubsection{Payload}
\label{sec:PaylPerf}

An early assessment of the flight performance of the \Planck\ payload (i.e., two instruments and telescope) was given in \citet[][LFI]{planck2011-1.4} and \citet[][HFI]{planck2011-1.5}, and summarized in \citet{planck2011-1.1}.  Updates based on the full nominal mission are given for LFI in \cite{planck2013-p02}, \cite{planck2013-p02a}, \cite{planck2013-p02d}, and \cite{planck2013-p02b}, and for HFI in \cite{planck2013-p03}, \cite{planck2013-p03c}, \cite{planck2013-p03f}, \cite{planck2013-p03d}, and \cite{planck2013-p03e}.

None of the LFI instrument performance parameters has changed significantly over time.  A complete analysis of systematic errors \citep{planck2013-p02a} shows that their combined effect is more than three orders of magnitude (in $\mu$K$^2$) below the CMB temperature signal throughout the measured angular power spectrum.  Similarly, the HFI performance in flight is very close to that measured on the ground, once the effects of cosmic rays are taken into account \citep{planck2013-p03e, planck2013-p03}.

Table~\ref{table:Instrument_performance} summarizes the performance of the \Planck\ payload for the nominal mission; it is well in line with, and in some cases exceeds, pre-launch expectations.

\begin{table*}[tmb]
\begingroup
\newdimen\tblskip \tblskip=5pt
\caption{\Planck\ performance parameters determined from flight data.} 
                            \label{table:Instrument_performance}
\nointerlineskip
\vskip -3mm
\footnotesize
\setbox\tablebox=\vbox{
    \newdimen\digitwidth
    \setbox0=\hbox{\rm 0}
    \digitwidth=\wd0
    \catcode`*=\active
    \def*{\kern\digitwidth}
    \newdimen\dpwidth
    \setbox0=\hbox{.}
    \dpwidth=\wd0
    \catcode`!=\active
    \def!{\kern\dpwidth}
\def\muKCMBs{\ifmmode \,\mu$K$_{\rm CMB}$\,s$^{1/2}\else \,$\mu$K$_{\rm CMB}$\,s$^{1/2}$\fi}
\halign{\hbox to 1.05in{#\leaderfil}\tabskip=2.0em&
      \hfil#\hfil\tabskip=3.0em&
      \hfil#\hfil\tabskip=3.0em&
      \hfil#\hfil\tabskip=1.0em&
      \hfil#\hfil\tabskip=3.0em&
      \hfil#\hfil\tabskip=0.1em&
      \hfil#\hfil\tabskip=0pt\cr
\noalign{\doubleline}
\omit&&&\multispan2\hfil S{\sc canning} B{\sc eam}$^{\rm c}$\hfil\cr
\noalign{\vskip -3pt}
\omit&&&\multispan2\hrulefill&\cr
\noalign{\vskip -18pt}
\omit&&&&&\multispan2\hfil N{\sc oise}$^{\rm d}$\hfil\cr
\noalign{\vskip 0pt}
\omit&&&&&\multispan2\hfil S{\sc ensitivity}\hfil\cr
\noalign{\vskip -1pt}
\omit&&$\nu_{\rm center}$$^{\rm b}$&FWHM&Ellipticity\cr
\noalign{\vskip -10pt}
\omit&&&&&\multispan2\hrulefill\cr
\omit\hfil C{\sc hannel}\hfil&$N_{\rm detectors}$$^{\rm a}$&[GHz]&[arcm]&&[\muKRJs]&[\muKCMBs]\cr
\noalign{\vskip 3pt\hrule\vskip 5pt}
*30\,GHz&*4&*28.4&33.16& 1.37&145.4*&148.5\cr
*44\,GHz&*6&*44.1&28.09& 1.25&164.8*&173.2\cr
*70\,GHz&12&*70.4&13.08& 1.27&133.9*&151.9\cr
100\,GHz&*8&100!*&*9.59& 1.21&*31.52&*41.3\cr
143\,GHz&11&143!*&*7.18& 1.04&*10.38&*17.4\cr
217\,GHz&12&217!*&*4.87& 1.22&**7.45&*23.8\cr
353\,GHz&12&353!*&*4.7*& 1.2*&**5.52&*78.8\cr
545\,GHz&*3&545!*&*4.73& 1.18&**2.66&0.026\rlap{$^{\rm d}$}\cr
857\,GHz&*4&857!*&*4.51& 1.38&**1.33&0.028\rlap{$^{\rm d}$}\cr
\noalign{\vskip 5pt\hrule\vskip 3pt}}}
\endPlancktablewide
\tablenote {{\rm a}} At 30, 44, and 70\,GHz, each {\it detector} is a linearly polarized radiometer, and there are two orthogonally polarized radiometers behind each horn.  Each radiometer has two diodes, both switched at high frequency between the sky and a blackbody load at $\sim4.5$\,K \citep{planck2011-1.4}.  At 100\,GHz and above, each {\it detector} is a bolometer \citep{planck2011-1.5}. Most of the bolometers are sensitive to polarization, in which case there are two orthogonally polarized detectors behind each horn.  Some of the detectors are spider-web bolometers (one per horn) sensitive to the total incident power. Two of the bolometers, one each at 143 and 545\,GHz, are heavily affected by random telegraphic signals (RTS; \citealt{planck2011-1.5}) and are not used.  Three other bolometers (two at 217\,GHz and one at 857\,GHz) exhibit short periods of RTS that are discarded. \par
\tablenote {{\rm b}} Effective (LFI) or Nominal (HFI) center frequency of the $N$ detectors at each frequency.\par
\tablenote {{\rm c}} Mean scanning-beam properties of the $N$ detectors at each frequency.  FWHM $\equiv$ FWHM of a circular Gaussian with the same volume.  Ellipticity gives the major-to-minor axis ratio for a best-fit elliptical Gaussian. In the case of HFI, the mean values quoted are the result of averaging the values of total-power and polarization-sensitive bolometers, weighted by the number of channels (not including those affected by RTS).  The actual point spread function of an unresolved object on the sky depends not only on the optical properties of the beam, but also on sampling and time domain filtering in signal processing, and the way the sky is scanned.  \par
\tablenote {{\rm d}} The noise level reached in 1\,s integration for the array of $N$ detectors, given the noise and integration time in the released maps, in both Rayleigh-Jeans and thermodynamic CMB temperature units for 30 to 353\,GHz, and in Rayleigh-Jeans and surface brightness units (M\,Jy\,sr\mo\,s$^{1/2}$) for 545 and 857\,GHz.  We note that for LFI the white noise level is within 1-2\,\% of these values. \par
\endgroup
\end{table*}

\subsection{Satellite Pointing}
\label{sec:Point}

The attitude of the satellite is computed on board from the star-tracker data and reprocessed daily on the ground.  The result is a daily attitude history file (AHF), which contains the filtered attitude of the three coordinate axes based on the star trackers every 0.125\,s during stable observations (``rings")  and every 0.25\,s during re-orientation slews.  The production of the AHF is the first step in the determination of the pointing of each detector on the sky (see Sect.~\ref{sec:FocPlaGeo}).

Early on it was realized that there were some problems with the pointing solutions.  First, the attitude determination during slews was poorer than during periods of stable pointing.  Second, the solutions were affected by thermoelastic deformations in the satellite driven by temperature variations resulting from imperfect thermal control loops, the sorption cooler, and the thermally-driven transfer of helium from one tank to another.

A significant effort was made to improve ground processing capability to address the above issues.  As a result, three different versions of the attitude history are now produced for the whole mission:

\begin{itemize}

\item The AHF, based on an optimized version of the initial (pre-launch) algorithm;

\item The Gyro-based attitude history hile (GHF), based on an algorithm that uses, in addition to the star tracker data, angular rate measurements derived from the on-board fibre-optic gyro\footnote{The \Planck\ fiber-optic gyro is a technology development instrument that flew on  \Planck\ as an opportunity experiment.  It was not initially planned to use its data for computation of the attitude over the whole mission.}. 

\item The Dynamical attitude history file (DHF), based on an algorithm that uses the star tracker data in conjunction with a dynamical model of the satellite, and which accounts for the existence of disturbances at known sorption cooler operation frequencies.

\end{itemize}

Both the GHF and DHF algorithms significantly improve the recovery of attitude during slews. Due to the deadlines involved, the optimised AHF algorithm has been used in the production of the 2013 release of \Planck\ products.  In the future we expect to use the improved algorithms, in particular enabling the use of the scientific data acquired during slews (9.2\,\% of the total).

The pointing characteristics at AHF level are summarized in Table~\ref{TabPointing}.

\begin{table}
\caption{Pointing performance over the nominal mission.}
\label{TabPointing} 
\vskip -8mm
\setbox\tablebox=\vbox{
\newdimen\digitwidth
\setbox0=\hbox{\rm 0}
\digitwidth=\wd0
\catcode`*=\active
\def*{\kern\digitwidth}
\newdimen\signwidth
\setbox0=\hbox{+}
\signwidth=\wd0
\catcode`!=\active
\def!{\kern\signwidth}
\newdimen\pointwidth
\setbox0=\hbox{.}
\pointwidth=\wd0
\catcode`@=\active
\def@{\kern\pointwidth}
\halign{\hbox to 2.2in{#\leaderfil}\tabskip 1.0em&
        \hfil#\hfil&
        \hfil#\hfil\tabskip 0pt\cr
\noalign{\vskip 3pt\hrule\vskip 1.5pt\hrule\vskip 5pt}
\omit\hfil Characteristic\hfil&  Median&  Std. dev.\cr
\noalign{\vskip 4pt\hrule\vskip 6pt}
Spin rate [deg\,s\mo]            & 6.00008& 0.00269\cr
\noalign{\vskip 4pt}
Small manoeuvre accuracy [arcsec]&   *5.1& 2.5\cr
\noalign{\vskip 4pt}
\omit Residual nutation amplitude\hfil\cr
\hglue 2em after manoeuvre  [arcsec]&*2.5& 1.2\cr
\noalign{\vskip 4pt}
\omit Drift rate during inertial\hfil\cr
\hglue 2em pointing [arcsec\,hr\mo]& 12.4& 1.7\cr
\noalign{\vskip 5pt\hrule\vskip 4pt}
}}
\endPlancktable
\end{table}

%============================================================

\section{Critical steps towards production of the \Planck\ maps}
\label{sec:DataProcessing}

\subsection{Overview and philosophy}
\label{sec:philosophy}

Realization of the potential for scientific discovery provided by \Planck's combination of sensitivity, angular resolution, and frequency coverage places great demands on methods of analysis, control of systematics, and the ability to demonstrate correctness of results.  Multiple techniques and methods, comparisons, tests, redundancies, and cross-checks are necessary beyond what has been required in previous experiments.  Among the most important are:

\result{Redundancy}Multiple detectors at each frequency provide the first level of redundancy, which allows many null tests.  As described in Sect.\ref{sec:ScanStrat}, the scanning strategy provides two more important levels of redundancy.  First, multiple passes over the same sky are made by each detector at each position of the spin axis (``rings'').  Differences of the data between halves of one ring (whether first and second half or interleaved data points) provide a good measure of the actual noise of a given detector, because the sky signal subtracts out to high accuracy.  (Maps of ``half-ring'' data are referred to as ``half-ring maps''.)  Second, null tests on Surveys and on one-year intervals reveal important characteristics of the data.  

\result{Comparison of LFI and HFI}The two instruments and the systematics that affect them are fundamentally different, but they scan the same sky in the same way.  Especially at frequencies near the diffuse foreground minimum, where the CMB signal dominates over much of the sky, comparison of results from the two provides one of the most powerful demonstrations of data quality and correctness ever available in a single CMB experiment.

\result{Multiplicity of methods}Multiple, independent methods have been developed for all important steps in the processing and analysis of the data.  Comparison of results from independent methods provides both a powerful test for bugs and essential insight into the effects of different techniques.  Diffuse component separation provides a good example (Sect.~\ref{subsec:CMBmapNG}; \citealt{planck2013-p06}).

\result{Simulations}Simulations are used in four important ways.  First, simulations quantify the effects of systematics.  We simulate data with a systematic effect included, process those data the same way we process the sky data, and measure residuals. Second, simulations validate and verify tools used to measure instrument characteristics from the data.  We simulate data with known instrument characteristics, apply the tools used on the sky data to measure the characteristics, and verify the accuracy of their recovery.  Third, simulations validate and verify data analysis algorithms and their implementations.  We simulate data with known science inputs (cosmology and foregrounds) and instrument characteristics (beams, bandpasses, noise), apply the analysis tools used on the sky data, and verify the accuracy of recovered inputs. Fourth, simulations support analysis of the sky data.  We generate massive Monte Carlo simulation sets of the CMB and noise, and pass them through the analyses used on the sky data to quantify uncertainties and correct biases.
The first two uses are instrument-specific; distinct pipelines have been developed and employed by LFI and HFI.  The last two uses require consistent simulations of both instruments in tandem.  Furthermore, the Monte Carlo simulation-sets are the most computationally intensive part of the \Planck\ data analysis and require large computational capacity and capability.

\subsection{Simulations}
\label{sec:FFP}

We simulate time-ordered information (TOI) for the full focal plane (FFP) for the nominal mission.  Each FFP simulation comprises a single ``fiducial'' realization (CMB, astrophysical foregrounds, and noise), together with separate Monte Carlo (MC) realizations of the CMB and noise.  The first \Planck\ cosmology results were supported primarily by the sixth FFP simulation-set, hereafter FFP6. The first five FFP realizations were less comprehensive and were primarily used for validation and verification of the \Planck\ analysis codes and for cross-validation of the Data Processing Centres (DPCs) and FFP simulation pipelines. 

To mimic the sky data as closely as possible, FFP6 used the actual pointing, data flags, detector bandpasses, beams, and noise properties of the nominal mission.  For the fiducial realization, maps were made of the total observation (CMB, foregrounds, and noise) at each frequency for the nominal mission period,  using the Planck Sky Model \citep{delabrouille2012}.  In addition, maps were made of each component separately, of subsets of detectors at each frequency, and of half-ring and single Survey subsets of the data. The noise and CMB Monte Carlo realization-sets also included both all and subsets of detectors at each frequency, and full and half-ring data sets for each detector combination.  With about 125 maps per realization and 1000 realizations of both the noise and CMB, FFP6 totals some 250,000 maps --- by far the largest simulation set ever fielded in support of a CMB mission.

\subsection{Timeline processing}

\subsubsection{LFI}
\label{sec:LFITOI}

The processing of LFI data (Planck Collaboration II--V 2013) is divided into three levels.  Level~1  retrieves information from telemetry packets and auxiliary data received each day from the Mission Operation Center (MOC), and transforms the scientific time-ordered information (TOI) and housekeeping (H/K) data into a form that is manageable by the Level~2  scientific pipeline.  

The Level~1 steps are:

\begin{itemize}

\item uncompress the retrieved packets;

\item de-quantize and de-mix the uncompressed packets to retrieve the original signal in analogue-to-digital units (ADU);

\item transform ADU data into volts; and 

\item time stamp each sample.

\end{itemize}

\noindent The Level 1 software has not changed since the start of the mission.  Detailed information is given in \citet{planck2011-1.6} and \citet{planck2013-p02}.  

Level~2 processes scientific and H/K information into data products.  The highly stable behaviour of the LFI radiometers means that very few corrections are required in the data processing at either TOI or map level.  The main Level~2 steps are:

\begin{itemize}

\item Build the reduced instrument model (RIMO) that contains all the main instrumental characteristics (beam size, spectral response, white noise etc). 

\item Remove spurious effects at the diode level.  Small electrical disturbances, synchronous with the 1-Hz on-board clock, are removed from the 44\,GHz data streams.  For some channels (in particular LFI25M-01) non-linear behaviour of the ADC is corrected by analyzing the white noise level of the total power component.  No corrections are applied to compensate for thermal fluctuations in the 4\,K, 20\,K, and 300\,K stages of the instrument, since H/K monitoring and instrument thermal modelling confirm that their effect is below significance. 

\item Compute and apply the gain modulation factor to minimize the $1/f$ noise.  The LFI timelines are produced by taking differences between the signals from the sky and from internal blackbody reference loads cooled to about 4.5\,K. Radiometer balance is optimized by introducing a gain modulation factor, typically stable to 0.04\,\% throughout the mission, which greatly reduces $1/f$ noise and improves immunity from a wide class of systematic effects. 

\item Combine the diodes to remove a small anti-correlated component in the white noise. 

\item Identify and flag periods of time containing anomalous fluctuations in the signal.  Fewer than 1\,\% of the data acquired during the nominal mission are flagged. 

\item Compute the corresponding detector pointing for each sample based on auxiliary data and the reconstructed focal plane geometry (Sect.~\ref{sec:FocPlaGeo}).

\item Calibrate the scientific timelines in physical units ($K_{\rm CMB}$), fitting the dipole convolved with a $4\pi$ representation of the beam (Sect.~\ref{sec:GainCal}).

\item Combine the calibrated TOIs into maps at each frequency (Sect.~\ref{sec:lfi_maps}).

\end{itemize}

Level 3 then collects instrument-specific Level 2 outputs (from both HFI and LFI) and derives various scientific products as maps of separated astrophysical components.

\subsubsection{HFI}
\label{sec:HFITOI}

Following Level~1 processing similar to that of LFI, the HFI data pipeline consists of TOI processing, followed by map making and calibration (Planck Collaboration VI--X 2013).

The HFI processing pipeline steps are:

\begin{itemize}

\item Demodulate, as required by the AC square-wave polarization bias of the bolometers. 

\item Flag and remove cosmic-ray-induced glitches, including the long-time-constant tails of glitches induced in the silicon wafer.  More than 95\,\%  of the acquired samples are affected by glitches.  Glitch templates constructed from averages are fitted and subtracted from the timelines; the fast part of each glitch is rejected.  The fraction of time-ordered data rejected due to glitches is 16.5\,\%\footnote{varying from 10 to 26.7\,\%   depending on the bolometer.} when averaged over the nominal mission. 

\item Correct for the slow drift of the bolometer response induced by the bolometer plate temperature variation
described in Sect.~\ref{sec:HFIOps}.  A baseline drift estimated from the signal from the dark bolometers on the same plate (smoothed over 60\,s) is removed from each timeline. 

\item Deconvolve the bolometer complex time response (analyzed in detail in \citet{planck2013-p03c}).  

\item Remove the narrow lines induced by electromagnetic interference from the \HeJT\ cooler, exploiting the fact that the cooler is synchronized with the HFI readout and operates at a harmonic of the sampling rate. 

\item Analyze the statistics of the time-scale of pointing periods and discard anomalous ones (less than 1\,\% of the data are discarded). 

\end{itemize}

Apparent gain variations seen when comparing identical pointing circles one year apart actually originate in non-linearities in the ADCs of the bolometer readout system.  Lengthy on-board measurements of the non-linear properties of the ADCs have been carried after the end of 0.1-K operations, and algorithms to correct for these non-linearities have been developed.  The electromagnetic interference from the \HeJT\ cooler described above  induces voltages in the readout circuits before digitization by the ADC.  That interference itself is localized in frequency and therefore easy to remove; however, it makes it more difficult to estimate the ADC non-linearity correction accurately for the detectors most affected by it.  The ADC non-linearity correction is still under development and has not been applied to the data in this 2013 release.  Instead, a calibration scheme (see Sect.~{\ref{sec:GainCal} and \citealt{planck2013-p03f}) that estimates a varying gain corrects very well the first order effects of the ADC non-linearity.  A full correction will be implemented for the release in 2014 of the polarization data, for which  higher order effects are not negligible.

\subsection{Beams}
\label{sec:Opt}

As described in \citet{planck2013-p02d}, the main beam parameters of the LFI detectors and the geometry of the focal plane were determined using  Jupiter as a source.  By combining four Jupiter transits (around days 170, 415, 578, and 812)  the beam shapes were measured down to $-20$\,dB from peak at 30 and 44\,GHz, and $-25$\,dB at 70\,GHz. The FWHM of the beams is determined with a typical uncertainty of 0.3\,\% at 30 and 44\,GHz, and 0.2\,\% at 70\,GHz.  The alignment of the focal plane and the location of each detector's phase centre were determined by varying their values in a GRASP\footnote{developed by TICRA, http://www.ticra.com/} physical optics model to minimize the difference between model co-polar and cross-polar patterns and the measurements.  To estimate the uncertainties in the determination of the in-flight beam models, a set of optical models representative of the measured LFI scanning beams$^7$ was found, using GRASP to randomly distort the wavefront error of the physical model of telescope and detectors, then rejecting those distorted models whose predicted patterns fell outside the error envelope of the measured ones.

Sidelobe pick-up by the LFI of the CMB dipole and diffuse Galactic emission \citep{planck2013-p02a} is clearly seen at the $\sim10\muK$ level in odd-even Survey difference maps (which enhance the effects of sidelobes) at 30\,GHz.  This  contamination was fitted to a model that incorporates the radiometer bandpass and the optical response variation across the band.  The modelled contamination was then removed.  Residual straylight effects in the maps are estimated to be less than $\sim 2\muK$ at all frequencies.

The in-flight scanning  beams\footnote{The term ``scanning beam" refers to the angular response of a single detector to a compact source, including the optical beam and (for HFI) the effects of time domain filtering.  In the case of HFI, a Fourier filter deconvolves the bolometer/electronics time response and lowpass-filters the data.  In the case of LFI, the sampling tends to smear signal in the time domain.} of HFI \citep{planck2013-p03c} were measured using observations of Mars.  Observations of Saturn and Jupiter are used to estimate the near sidelobes and other residuals. The HFI bolometers have a complex time response, characterized by multiple time constants.  To obtain a compact scanning beam, this time behaviour must be deconvolved from the measured timelines.  The deconvolution algorithm is iterative, allowing an  estimate of the parameters of the bolometer transfer function, and forcing the resulting scanning beams to be more compact.  A spline representation of the beams is used, allowing capture of the near-sidelobe structure down to about -40\,dB from the peak.  Stacking of multiple crossings of Saturn and Jupiter allows us to obtain high SNR maps of these near sidelobes, and to quantify the level of unmodelled effects.  The stacked data  at 353\,GHz show the presence of skirts in the pattern close to the main beam due to diffraction at the edge of the secondary mirror. These skirts could not be measured accurately at lower frequencies with the 2013 data and were not included in the beam representation (Section~\ref{sec:BeamRep}). Instead, upper limits to their contribution to the solid angle at each frequency were estimated \citep{planck2013-p03c} to be in the range 0.2 to 0.4\,\%, and were included in the uncertainty budget for the gain calibration and transfer functions.

At the three highest frequencies, the stacked planet maps  also clearly reveal  sharp ``grating lobes" due to dimpling of the telescope reflector surfaces into the cells of the internal honeycomb structure \citep{tauber2010b}.  The amplitude of these lobes is larger than predicted, possibly indicating mechanical changes in the reflectors after launch; however, the total power contained in these lobes remains negligible in terms of impact on the scientific data, and therefore they are ignored in the scanning beam model. 

The uncertainties in the estimation of the HFI scanning beams and other systematic effects in the maps are determined at window function level, using realistic Monte-Carlo simulations that include pointing effects, detector noise, and measurement effects.  Additional estimates are made of the effect of planet emission variability, beam colour corrections, and more.  The total uncertainties in the scanning beam solid angles are under 0.5\,\% for the CMB channels. 

Sidelobe pick-up by the HFI due to spillover past the primary reflector \citep{tauber2010b}, is clearly seen in Survey difference maps at 545 and 857\,GHz \citep{planck2013-pip88} at times when the central part of the Galactic plane is aligned with the elongated far sidelobe.  GRASP models are fit to odd-even Survey difference maps to estimate sidelobe levels for each detector.  These levels are highly variable between the 857\,GHz detectors  and not in agreement with levels predicted by GRASP; this difference may plausibly be caused by deviations of the as-built horns from the design.   A model of the primary reflector spillover signal can then be removed from the time-ordered data before mapmaking.  Being close to the spin axis, these signals are largely unmodulated by the spinning motion, and are mostly removed by the destriping map-making code (this is the case for both instruments).  

The HFI far sidelobe signals and zodiacal light can be removed at TOI level in the same pipeline.  Two sets of maps are released in 2013 \citet{planck2013-p03}.  In the ``default'' set, far sidelobes and zodiacal light are not removed.  In the second set, far sidelobes and zodiacal light are removed.

\subsection{Focal Plane Geometry and Pointing}
\label{sec:FocPlaGeo}

The focal plane geometry\footnote{That is, the relative location on the sky of the multiple detectors.} of LFI was determined independently for each Jupiter crossing \citep{planck2013-p02d}. The solutions for the first and second and for the third and fourth crossings agree to 2\arcs; however, a shift of $\sim$15 arcsecs (largely in the in-scan direction) is found between the two pairs. For this reason, the focal plane geometry is assumed constant over time, with the exception of a single jump on day 540. 

The focal plane geometry of the HFI detectors was also measured using planet observations \citep{planck2013-p03,planck2013-p03c}. The relative location of individual detectors differs from the ground prediction typically by 1\arcm, mainly in the in-scan direction, indicating some de-alignment of the HFI focal plane or of the telescope in flight.  The high SNR available on Jupiter allows us to estimate pointing ``errors" on a 1-minute timescale; these measurements show the presence of thermo-elastic deformations of the star tracker mounting structure that are well correlated with a known on-board thermal control cycle.  This specific cycle was changed on OD\,540, leading to a reduction in this ``error" from 3\arcs\ to 1\arcs. These small high-frequency effects are not taken into account at the present time; however, larger (up to 15\arcs) slow pointing variations  are observed with time scales of order 100\,days using measurements of bright compact radio sources. The HFI focal plane geometry variation with time is corrected for this trend, leaving an estimated total pointing reconstruction error of a few arcseconds rms. 

The time-dependent pointing direction of each detector in \Planck\ is based on the filtered star tracker data provided in the daily Attitude History Files (see Sect.~\ref{sec:Point}), corrected for three effects: 1)~changes in the inertial tensor of the satellite resulting from fluid depletion; 2)~time-varying thermoelastic deformations of the structure between the star tracker and the telescope boresight; and 3)~stellar aberration. The measured focal plane geometry is then used to estimate the corrected pointing timestream for each detector.

Time-stamping of all detector samples is synchronized to the satellite's Central Time Reference, also used by the star trackers. The stability of this reference has been shown to have negligible impact on \Planck's science products, when compared to other sources \citep{planck2013-p28}.

\subsection{Calibration}
\label{sec:GainCal}

The photometric (gain) calibrations of LFI and HFI are achieved by comparing the measured data against the expected signal from the Solar and orbital dipoles.  The Solar dipole parameters are assumed to be those determined by WMAP7 \citep{jarosik2010}, which during the initial period of \Planck\ analysis provided a convenient, stable, and accurate reference.  In the future, a \Planck-determined dipole will replace the WMAP7 dipole.   The orbital dipole is determined from the measured satellite orbital motion, so in principle it can provide an independent absolute calibration.

In the LFI pipeline \citep{planck2013-p02b}, a full $4\pi$ beam is used to calculate the expected dipole signal.  An iterative fit outside a Galactic mask plus low-frequency filtering yields a single calibration constant for each pointing period (or ring).  For 44 and 70\,GHz, an adaptive smoothing function removes the effect of low-frequency noise stripes from the time series of ring gains. The 30\,GHz data are more affected by sidelobe signals; for this channel it has been found more effective to estimate the relative changes in the ring gains over time by using the total power variations of the 4-K reference loads measured by each radiometer.  This relative estimation is carried out over three different time periods (to account for significant temperature changes in the 20\,K and 4\,K stages due to operational events).  The estimated gains are applied to the acquired data in the time domain.  The Solar and orbital dipole signals are then removed from the time-ordered data before proceeding into mapmaking. 

At the present stage, the overall uncertainty in the LFI calibration relative to the \WMAP\ Solar dipole is of order 0.4\,\%, dominated by the beam uncertainty over the band (0.1--0.4\,\% depending on the frequency), sidelobe convolution ($\sim0.2\,\%$), and residual systematics (0.1--0.2\,\%). There is a 0.25\,\% uncertainty in the \WMAP\ dipole itself \citep{hinshaw2009}, which we conservatively add linearly to the LFI relative calibration uncertainty to get the overall calibration uncertainties given in Table \ref{tab:accuracyBudgetResult}. 

Preliminary results on the orbital dipole as measured by LFI are in agreement with WMAP at around the 1$\sigma$ level. This indicates good consistency of LFI and WMAP calibrations at $\ell = 1$. The inter-channel calibration discussed below allows us to extrapolate the independent measurement of the Solar dipole by the LFI to the HFI CMB channels.

In the HFI pipeline \citep{planck2013-p03f}, a pencil beam is assumed when comparing data to the expected dipole signal.  In this process, the contribution of the far sidelobes due to spillover past the primary reflector (see Sect.~\ref{sec:Opt}) is neglected.  Its contribution to the Solar dipole signal is estimated to be $\sim$0.13\,\% after destriping, and affects only large angular scales ($\ell <$ 10).  Near-side-lobe corrections are more substantial (0.2--0.4\,\%, see Sect.~\ref{sec:Opt}) but have not yet been implemented. These effects are small compared to other corrections not yet made, such as the ADC non-linearity for HFI (see Sect.~\ref{sec:HFITOI}) or beam uncertainties over the band for LFI \citep{planck2013-p02b}.

Fitting for the HFI gains is done at ring level, including a spatial model of Galactic emission based on HFI maps, but excluding the Galactic plane.  It yields three parameters per ring: a dipole gain, a Galactic gain, and an offset.  An initial gain model consists of a fixed gain averaged over a contiguous set of 4000 rings, where the ring-to-ring dispersion is less than 1\,\%.  An iterative scheme is then applied to the 100--217\,GHz channels that fits for relative variations of the gain over the whole mission.  This scheme is not applied to the 353\,GHz channel because intra-pixel emission gradients cause instability in the iterations.  Remaining residuals in the calibrated rings are estimated by comparisons between detectors, and are of order 0.3\,\% for 100--217\,GHz and 1\,\% at 353\,GHz.  Once the HFI rings are calibrated, they are converted to maps using a destriping algorithm (see Sect.~\ref{sub:hfi_maps}), and the WMAP7 dipole is removed from the maps.

The current calibration scheme of HFI has been checked against an independent algorithm based only on the orbital dipole  \citep{2011A&A...534A..88T}. Relative variations are typically 0.1\,\% and always smaller than 0.2\,\% except for a systematic bias of 0.5--1\,\% believed to be due to the (currently uncorrected) ADC non-linearities (Sect.~\ref{sec:HFITOI}), which affect the two methods differently.  The total gain calibration accuracy, as evaluated from the ring-to-ring variability, overestimates the real accuracy since part of the variability is being corrected. 

Calibrated maps can be examined for signatures of calibration errors. The presence of Solar dipole residuals  provides limits to the quality of the LFI and HFI calibration processes with respect to their common reference, i.e., the \WMAP\ Solar dipole. In the 2013 CMB-calibrated maps (Section~\ref{sec:FreqMaps}) , the residual dipole amplitudes are of order 0.3\,\% for the HFI CMB channels \citep{planck2013-p03f} and 0.2\,\% to 0.3\,\% for LFI \citep{planck2013-p02}. 
These residual levels do not constrain calibration uncertainties at high multipoles, which are due to transfer function errors at multipoles larger than $\ell$=1. In HFI, the transfer function is constrained by measurement of the scanning beams on planets, which are representative of the transfer function only to the extent that uncertainties due to the deconvolution from the long time constants do not dominate. These uncertainties are estimated to be less than 0.5\,\%, thus comparable to those due to near-sidelobe effects.

A reliable estimate of the final calibration accuracy can also be obtained by measuring  the relative calibration between individual detectors and between frequency bands directly on the CMB dipole and anisotropies.  Comparisons \citep{planck2013-p08} show that the HFI 100 and 217\,GHz channels are within 0.3\,\%  of the 143\,GHz channel and 70\,GHz is within 0.5\,\%  (see also Section~\ref{sec:Consis} and Fig.~\ref{HFIDPCFig35}). A detailed analysis of possible systematic effects accounting for the estimated difference between LFI and HFI channels shows that uncorrected near-side-lobe effects in the HFI channels can account for most of it \citep{planck2013-p01a}.

To summarize, the relative calibration accuracy of all \Planck\ channels between 44 and 217\,GHz is of order $0.6\,\%$, and if we include the 30 and 353\,GHz channels the relative accuracy remains within $1\,\%$.  The absolute calibration is  limited by the accuracy of the CMB dipole determination (an additional uncertainty of 0.25\,\%), leading to an absolute calibration accuracy of the \Planck\ CMB channels of order 0.8\,\%. Significant improvements are expected for the next release, when orbital dipole analysis and a more accurate accounting of beam solid angle will be fully implemented for both instruments. By themselves, near-sidelobe effects in HFI channels should contribute a common-mode improvement of 0.2 to 0.4\,\% to the total. 

The calibration philosophy for the two submillimetre channels of HFI (545 and 857\,GHz) has changed from that reported in \citet{planck2011-1.7}.  The original approach compared HFI and {\it COBE}/FIRAS \citep{1999ApJ...512..511M} maps at high Galactic latitude.  Detailed investigation revealed a number of inconsistencies \citep{planck2013-p03f}, and the calibration scheme at 545 and 857\,GHz is now based on fitting measurements of the flux density of Uranus and Neptune to planetary emission models, which have an absolute (relative) accuracy of about 5\,\% (2\,\%). This process is described in detail in \citet{planck2013-p03f}. The overall estimated accuracy of the gain calibration with this method is estimated to be 10\,\%.

%============================================================

\section{Frequency Maps}
\label{sec:FreqMaps}

\Planck\ uses HEALPix \citep{gorski2005} as its basic representation scheme for maps.

\subsection{Beam representation}
\label{sec:BeamRep}

As described in Sect.~\ref{sec:Opt}, the LFI scanning beams are represented by GRASP models fitted to observations of Jupiter, while the HFI scanning beams are represented by B-spline surfaces fitted to observations of Mars.  The \Planck\ frequency maps are constructed from many detectors, which sample each pixel at different scan angles and multiple times.  The integrated angular response function at map level, the ``effective beam'' at a given pixel, can be quite different from the scanning beam and varies significantly across the sky.  Effective beams are computed for each pixel and frequency using the {\tt FEBeCoP} algorithm \citep{mitra2010}, which calculates the effective beam at a given pixel by computing the real space average of the scanning beam over all observed crossing angles at that position.   Table \ref{table:stat-BS} gives some statistics of the variation across the sky of parameters of the effective beams, and Fig.~\ref{FigFebecopMaps} shows these variations graphically for 100\,GHz.  The effective beams include pixelization effects (essentially the HEALpix pixelization window function).

\begin{table}[tmb]
\begingroup
\newdimen\tblskip \tblskip=5pt
\caption{Statistics of effective beam parameters: FWHM, ellipticity, and solid angle}                          
\label{table:stat-BS} 
\nointerlineskip
\vskip -3mm
\footnotesize
\setbox\tablebox=\vbox{
   \newdimen\digitwidth 
   \setbox0=\hbox{\rm 0} 
   \digitwidth=\wd0 
   \catcode`*=\active 
   \def*{\kern\digitwidth}
   \newdimen\dpwidth 
   \setbox0=\hbox{.} 
   \dpwidth=\wd0 
   \catcode`!=\active 
   \def!{\kern\dpwidth}
\halign{\hbox to 1.8cm{#\leaderfil}\tabskip 1.3em&
     \hfil#\hfil \tabskip 1em&
     \hfil#\hfil \tabskip 1em&
     \hfil#\hfil \tabskip 0pt\cr 
\noalign{\doubleline}
\omit& FWHM$^a$&& $\Omega$\cr
\omit\hfil Band\hfil&[arcmin]& Ellipticity&[arcmin$^2$]\cr
\noalign{\vskip 3pt\hrule\vskip 5pt}
 30& $32.239\pm0.013$& $1.320\pm0.031$& $1189.51*\pm*0.84*$\cr
44&  $27.01*\pm0.55$*& $1.034\pm0.033$& *$833!***\pm32!***$\cr
70&  $13.252\pm0.033$& $1.223\pm0.026$& *$200.7**\pm*1.0$**\cr                  
100& *$9.651\pm0.014$& $1.186\pm0.023$& *$105.778\pm*0.311$\cr
143& *$7.248\pm0.015$& $1.036\pm0.009$& **$59.954\pm*0.246$\cr
217& *$4.990\pm0.025$& $1.177\pm0.030$& **$28.447\pm*0.271$\cr
353& *$4.818\pm0.024$& $1.147\pm0.028$& **$26.714\pm*0.250$\cr
545& *$4.682\pm0.044$& $1.161\pm0.036$& **$26.535\pm*0.339$\cr
857& *$4.325\pm0.055$& $1.393\pm0.076$& **$24.244\pm*0.193$\cr
\noalign{\vskip 3pt\hrule\vskip 5pt}
}
}
\endPlancktable 
\tablenote a {Mean of  best-fit Gaussians to the effective beams.} \par
\endgroup
\end{table}

\begin{figure}
   \centering
\includegraphics[width=88mm]{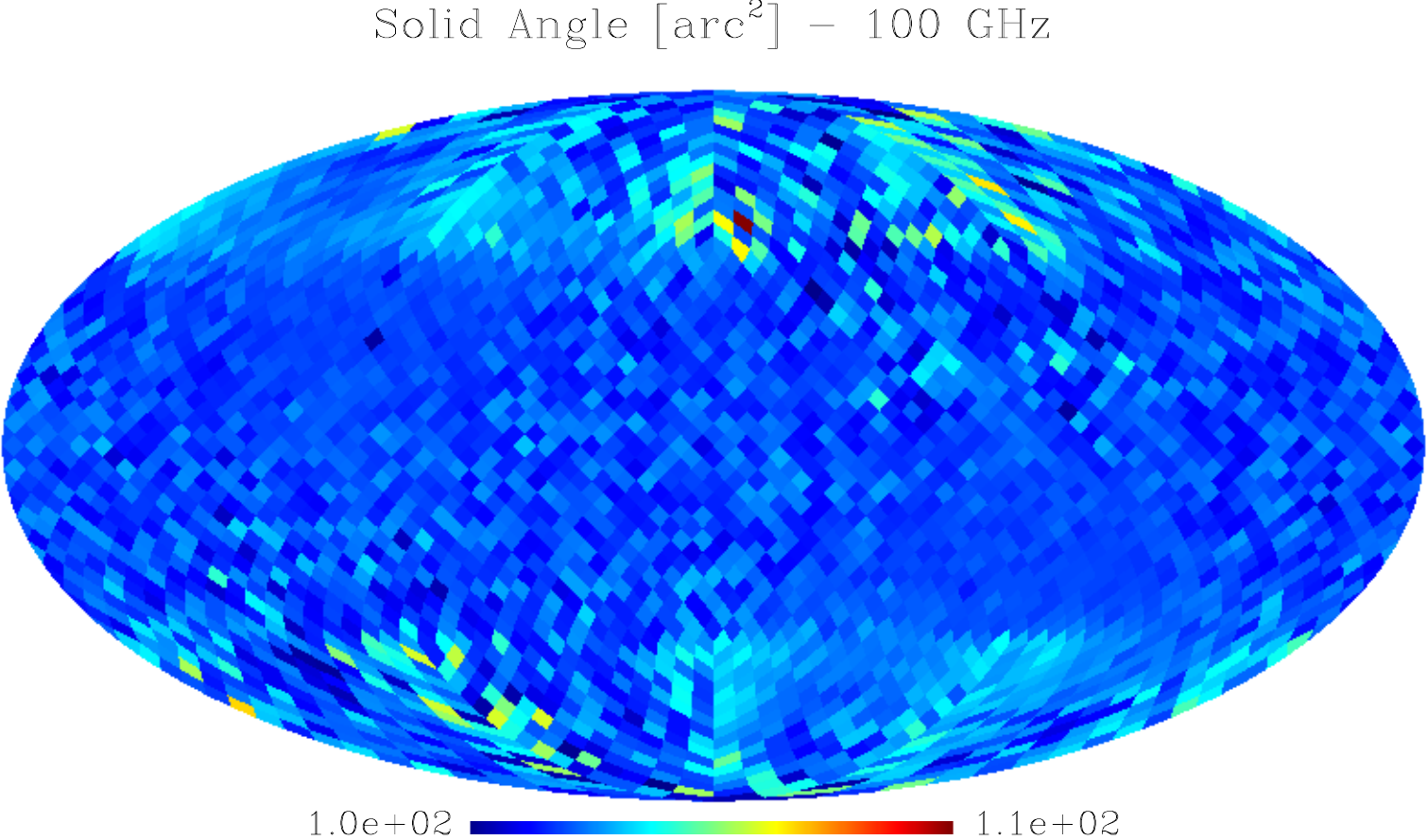}
\vskip 10pt 
\includegraphics[width=88mm]{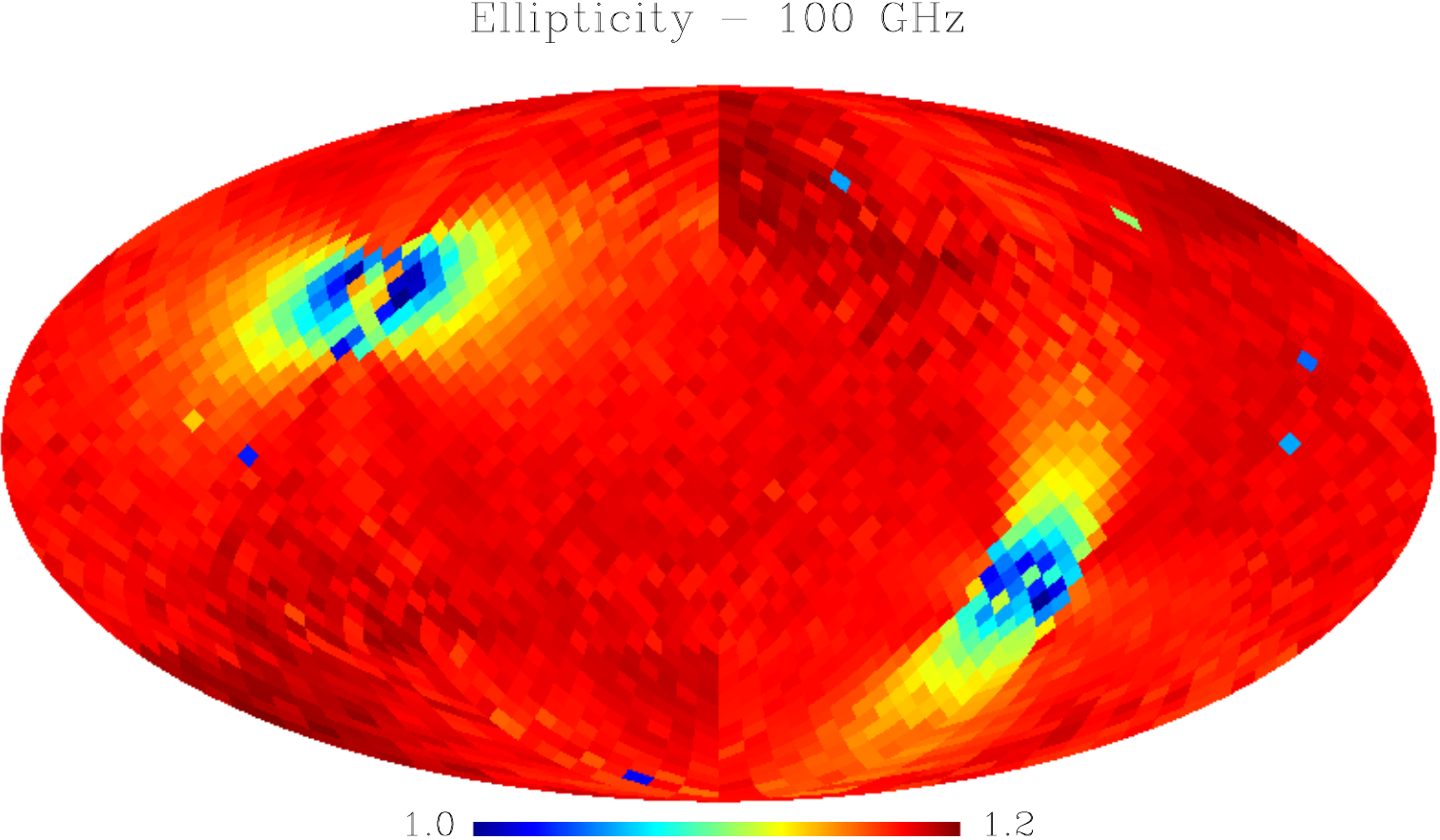}
   \caption{Distribution across the sky of the solid angle (top) and ellipticity of the effective beams at 100\,GHz, which is typical of all channels.}
              \label{FigFebecopMaps}
    \end{figure}

For LFI, the effective beam window function is calculated by {\tt FEBeCoP} using an ensemble of signal-only simulations convolved with the effective beams.  For HFI, it is calculated using the {\tt quickbeam} harmonic space effective beam code \citep{planck2013-p03c}.  

To estimate the uncertainties of the effective beams, the ensemble of allowed LFI GRASP models (Sect.~\ref{sec:Opt}) is propagated through {\tt FEBeCoP}.  For HFI, {\tt quickbeam} is used to propagate an ensemble of simulated Mars observations to harmonic space.  The total uncertainties in the effective beam window function (in $B_\ell^2$ units) at $\ell = 600$ are 2\,\% at 30\,GHz and 1.5\,\% at 44\,GHz.  At $\ell=100$, they are 0.7\,\%, 0.5\,\%, 0.2\,\%, and 0.2\,\% for 70, 100, 143, and 217\,GHz respectively \citep{planck2013-p02d, planck2013-p03c}.

\subsection{Mapmaking}
\label{sec:mapmaking}

\subsubsection{LFI }
\label{sec:lfi_maps}

Maps of the LFI channels are made using the {\tt Madam} destriping code \citep{keihanen2010}, from calibrated TOI of each radiometer and the corresponding pointing data in the form of the Euler angles $(\theta,\phi,\psi)$. {\tt Madam} is a polarized, maximum-likelihood code.  The noise is modelled as white noise plus a set of offsets, or baselines.  The algorithm estimates the amplitudes of the baselines, subtracts them from the TOI, and bins the result into maps of the Stokes parameters $T$, $Q$, and $U$.  The resulting $T$ maps are shown in Fig.~\ref{FigFreqMaps}.

\begin{sidewaysfigure*}
\centering
\vspace*{18cm}~\\
\includegraphics[width=0.95\textwidth]{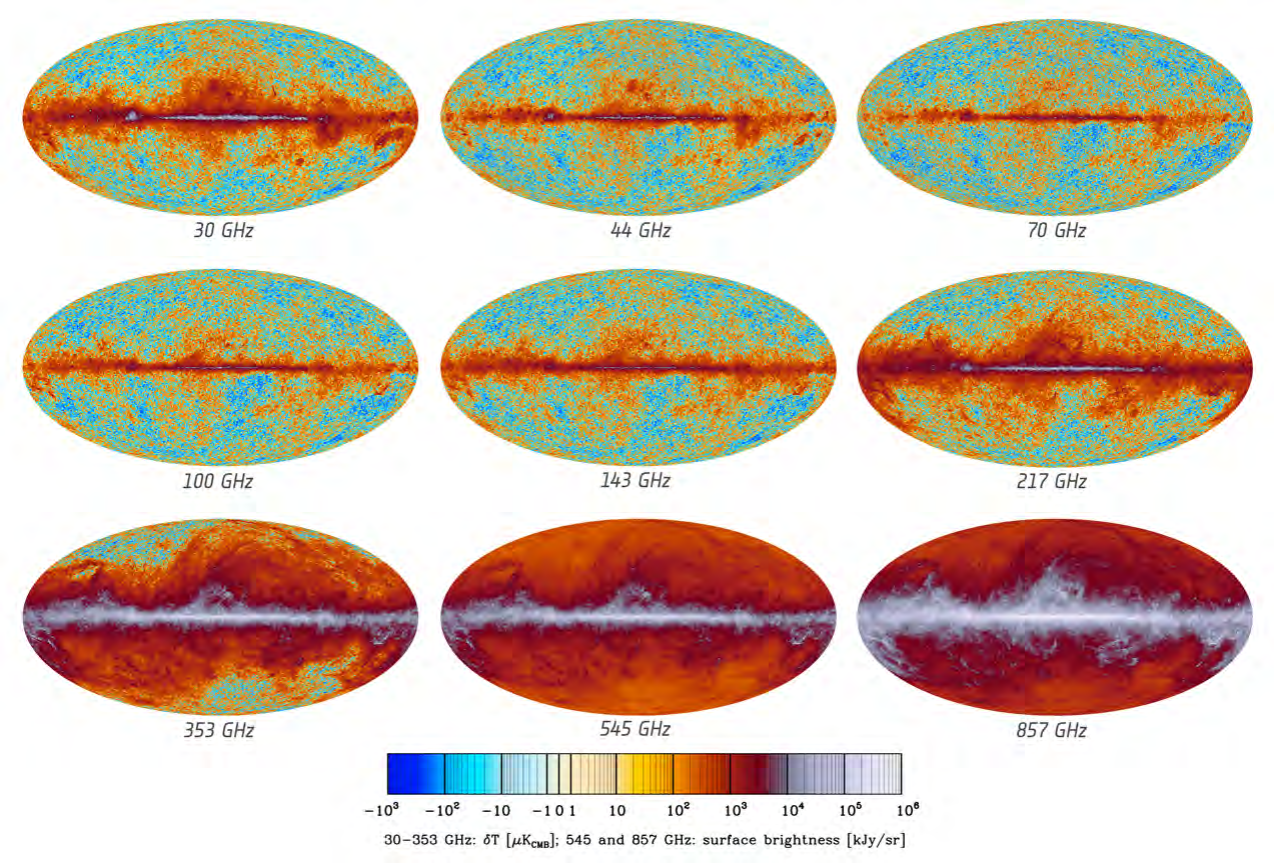} 
\caption{ The nine \Planck\ frequency maps  show the broad frequency response of the individual channels. The color scale, based on inversion of the function $y = 10^x - 10^{-x}$, is tailored to show the full dynamic range of the maps. 
 }
\label{FigFreqMaps}
\end{sidewaysfigure*}

A key parameter in the {\tt Madam} algorithm is the baseline length, the time interval over which the baseline approximation to low-frequency noise is applied.  We choose baseline lengths corresponding to an integer number of samples (33, 47, and 79 at 30, 44, and 70\,GHz, respectively) such that the total integration time over the baseline corresponds approximately to one second.  This selection is based on a compromise between computational load and map quality, and we find that shortening the baselines below one second has practically no effect on the residual noise. 

{\tt Madam} requires an estimate of the noise covariance matrix.  In general, we use a white noise covariance matrix.  The pipeline allows the use of different user-defined weighting schemes.  The maps being released are made using the horn-uniform weighting scheme with
\begin{equation}
C_w^{-1} = \frac{2}{\sigma_M^2 + \sigma_S^2}\, ,
\end{equation}
where $\sigma_M$ and $\sigma_S$ are the white noise levels of the Main and Side radiometers of a given horn,  weighted equally.

Half-ring maps (Sect.~\ref{sec:philosophy}) are made in the same way.  A time-weighted difference between the first-half and second-half ring maps captures the noise properties directly from the data of noise components whose frequency is $f\gsim 2/20\,{\rm min} = 1.7$\,mHz, i.e., half the duration of the pointing period.

The monopoles in \Planck\ maps are unconstrained, and it is conventional to adjust them such that they yield plausible values of the diffuse Galactic foreground signals. For LFI, the monopoles are estimated \citep{planck2013-p02b} in a 1\deg-diameter patch at high galactic latitudes, first subtracting the CMB (using the  independent component analysis estimate described in Sec. \ref{subsec:CMBmapNG}),   
and then carrying out a multi-frequency foreground fit using both \Planck\ and \WMAP\ channels. The offsets in each map are then adjusted to attain the fitted foreground level.

\subsubsection{HFI}
\label{sub:hfi_maps}

Maps of the HFI channels are made by projecting processed HEALPix rings built from the TOIs (Sect.~\ref{sec:HFITOI}) onto a HEALPix map \citep{planck2013-p03, planck2013-p28}.  Specifically, maps of individual rings are created by averaging filtered and baseline-subtracted TOIs into HEALPix pixels.  These ring maps are used in the photometric calibration of each detector (Sect.~\ref{sec:GainCal}).  Calibrated ring maps are combined via a least-squares destriping procedure \citep{planck2013-p03f} that estimates a constant offset per ring.  Maps are made for individual detectors, subsets of detectors at each frequency, and all the detectors at a given frequency.  Half-ring maps are made using the first and second halves of all rings, to monitor the statistical and systematic noise properties of the maps.

Because of \Planck's rotation and orbit, contributions to the TOI from far-sidelobe pickup (mostly from the Galaxy at high frequencies) and of the zodiacal light do not project onto fixed sky coordinates.  They are, however, a significant contaminant at 353\,GHz and above, which must be estimated and removed (\citealt{planck2013-pip88}; \citealt{planck2013-p28}).  In this release we provide two sets of HFI maps:

\begin{itemize}

\item A default set of maps from which neither far-sidelobe straylight nor zodiacal emission has been removed. These default maps are the ones that we use for the extraction of the CMB for the low-$\ell$ likelihood and to search for non-Gaussianity. The rationale for not removing zodiacal emission in the frequency maps is that its removal there produces artifacts during component separation \citep{planck2013-p06}.  For CMB extraction, it is more effective to let the the component separation method (Sect.~\ref{subsec:CMBmapNG}) remove zodiacal emission. 

\item A second set of maps from which estimated far-sidelobe straylight and zodiacal emission have been removed \citep{planck2013-pip88}.  The zodiacal emission is estimated by fitting to the \textit{COBE} emission model \citep{kelsall1998}, and subtracted from the TOI before mapmaking.  Zodiacal emission is removed at all frequencies.  Far-sidelobe emission is estimated and removed at 545 and 857\,GHz only.  This is the set of maps that should be used for work related to diffuse foregrounds.

\end{itemize}

The 2013 HFI maps contain significant Galactic CO emission.  Specific component separation pipelines yield separate estimates of it (Sect.~\ref{sec:COmaps} and \citealt{planck2013-p03a}) optimized for different scientific uses.

The HFI frequency maps contain an offset that arises from two different components, the diffuse interstellar medium and the cosmic infrared background.  The offset level due to the diffuse interstellar medium is estimated by correlating the HFI maps with a map of the column density of neutral hydrogen.  This offset (reported in Table \ref{tab:accuracyBudgetResult}) should be removed from the released maps before carrying out scientific analysis of Galactic emission.  The mean level contributed by the cosmic infrared background at each frequency is estimated by means of an empirical model that fits the current data.  For analysis of total emission, the CIB level (\citealt{planck2013-p03f}) must be added to the released maps after readjustment for the interstellar medium offset described above. 

The resulting HFI maps are shown in Fig.~\ref{FigFreqMaps}.

\subsection{Map Units}
\label{sec:MapUnits}

Broad-band detection instruments like those on \Planck\ measure radiative power through a filter characterized by its bandpass.  As described in Sect.~\ref{sec:GainCal}, \Planck\ is calibrated to the CMB dipole except for the two highest frequency channels (545 and 857\,GHz), for which planets are used.  If a target source has the same spectral energy distribution (SED) within the bandpass as the calibrator, the true brightness is the ratio of the response to the target and the response to the calibrator.  In practice only CMB anisotropies satisfy this condition; other sources of radiation have different SEDs.  The raw brightness values in the maps cannot then be taken as monochromatic values; rather, they represent the integral of the detected power weighted by the bandpass.

To allow comparison with other measurements and with models, the brightness of the CMB-calibrated channels is given as differential CMB temperature, $\Delta T_\mathrm{CMB} = \Delta I_\nu / (dB_\nu/dT)_{T_0}$, where $B_\nu$ is the Planck function, $T_0 = 2.7255$\,K \citep{fixsen2009}, and $\nu$ is a specified reference frequency for each channel\footnote{Our definition of $\Delta T_\mathrm{CMB}$ is linearized, and deviates significantly from the true variation in the equivalent blackbody temperature in the brightest regions of the 217 and 353\,GHz maps.}.  The units are thermodynamic temperature,  K$_{\rm CMB}$.

The 545 and 857\,GHz channel maps are instead given in intensity units (MJy\,sr\mo), assuming the reference SED $I_\nu = I_0\times (\nu_0 / \nu)$ (used previously by {\it IRAS\/} and {\it COBE}-DIRBE).  For all \Planck\ channels, the intensity (or flux density for unresolved sources) is attached to a choice of both reference frequency and assumed SED. 

Each foreground observed by \Planck\ has a different SED (power law, modified black body, SZ distortion, CO lines).  To evaluate intensities for these SEDs, e.g., for component separation, we provide unit conversions and colour corrections for each band, where the corrected values are such that the power integrated in the spectral bandpass and throughput is equal to the measured power.  These are described for LFI in \citet{planck2013-p02b}, which tabulates conversion from the CMB fluctuation SED to power-laws with various indices, and for HFI in \citet{planck2013-p03d}, which gives conversions between the two standard HFI SEDs (CMB fluctuation and {\it IRAS\/} standard), and also the Compton $y_{\rm sz}$ parameter.  In addition, a unit conversion and colour correction software tool ({\tt UcCC}) covering all \Planck\ bands is provided as part of the data release and described in \citet{planck2013-p03d}. Users are cautioned to read the detailed descriptions carefully, as in general a sequence of steps is required to convert from the units and assumed SED of the map calibration to the those appropriate for a given foreground.  Our colour conversions are uniformly cast as a multiplicative correction, yielding the brightness at the standard band reference frequency for the required SED.  It would also have been possible in most cases to quote an ``effective frequency'' at which the numerical value of the map brightness applies to the required SED, but this is less practical in general and cannot be applied at all to line emission. 
The effective frequencies for the LFI detectors in Table \ref{table:Instrument_performance} correspond to the SED for the CMB.

\begin{table*}[tmb]
\begingroup
\newdimen\tblskip \tblskip=5pt
\caption{\label{tab:accuracyBudgetOverview} Contributors to uncertainties at map level}
\nointerlineskip
\vskip -6mm
\tiny
\setbox\tablebox=\vbox{
   \newdimen\digitwidth 
   \setbox0=\hbox{\rm 0} 
   \digitwidth=\wd0 
   \catcode`*=\active 
   \def*{\kern\digitwidth}
   \newdimen\signwidth 
   \setbox0=\hbox{+} 
   \signwidth=\wd0 
   \catcode`!=\active 
   \def!{\kern\signwidth}
\halign{
\hbox to 4.5cm{#\leaderfil}\tabskip 2.0em&
\hfil #\hfil&
#\hfil&
#\hfil\tabskip=0pt\cr
\noalign{\doubleline}
\omit&&\multispan2\hfil Method used to assess uncertainty\hfil\cr
\noalign{\vskip -2pt}
\omit&&\multispan2\hrulefill\cr
\omit\hfil Uncertainty\hfil&\omit\hfil Applies to\hfil&\omit\hfil LFI\hfil&\omit\hfil HFI\hfil\cr
\noalign{\vskip 3pt\hrule\vskip 5pt}
\noalign{\vskip 3pt}
Gain calibration standard& All sky&  \WMAP\ dipole& 100--353\,GHz: \WMAP\ dipole\cr
\omit&&& 545--857\,GHz: Planet model\cr
Zero level& All sky& Galactic cosecant model& Galactic zero: correlation to HI\cr
\omit&& Comparison with \WMAP\&CIB: empirical model\cr
Beam uncertainty& All sky& GRASP models via Febecop& Beam MC realizations via \tt{Quickbeam}\cr
Color corrections& non-CMB emission& Comparison of ground/flight bandpass leakages& Ground measurements\cr
Beam Color corrections& non-CMB emission& GRASP models& GRASP models\cr
\noalign{\vskip 8pt}
Residual systematics& All sky& Null tests& Null tests\cr
\noalign{\vskip 5pt\hrule\vskip 3pt}}}
\endPlancktablewide 
\endgroup
\end{table*}

\begin{table*}[tmb]
\begingroup
\newdimen\tblskip \tblskip=5pt
\caption{\label{tab:accuracyBudgetResult} Properties of the \Planck\ maps$^{\rm a}$ }
\nointerlineskip
\vskip -3mm
\tiny
\setbox\tablebox=\vbox{
   \newdimen\digitwidth 
   \setbox0=\hbox{\rm 0} 
   \digitwidth=\wd0 
   \catcode`*=\active 
   \def*{\kern\digitwidth}
   \newdimen\signwidth 
   \setbox0=\hbox{+} 
   \signwidth=\wd0 
   \catcode`!=\active 
   \def!{\kern\signwidth}
\halign{
\hbox to 5.7 cm{#\leaderfil}\tabskip 1.6em&
\hfil #\hfil&
\hfil #\hfil\tabskip=1.25em&
\hfil #\hfil&
\hfil #\hfil&
\hfil #\hfil&
\hfil #\hfil&
\hfil #\hfil&
\hfil #\hfil&
\hfil #\hfil&
\hfil #\hfil\tabskip=0pt\cr
\noalign{\doubleline}
\omit&&\multispan9\hfil Frequency [GHz]\hfil\cr
\noalign{\vskip -2pt}
\omit&&\multispan9\hrulefill\cr
\omit\hfil Property\hfil&
\omit\hfil Applies to\hfil&
\omit\hfil 30\hfil&
\omit\hfil 44\hfil&
\omit\hfil 70\hfil&
\omit\hfil 100\hfil&
\omit\hfil 143\hfil&
\omit\hfil 217\hfil&
\omit\hfil 353\hfil&
\omit\hfil 545\hfil&
\omit\hfil 857\hfil\cr
\noalign{\vskip 3pt\hrule\vskip 5pt}
\noalign{\vskip 3pt}
Effective frequency [GHz]& Mean& 28.4& 44.1& 70.4 & 100& 143& 217& 353& 545& 857\cr
Noise rms per pixel [$\mu$K$_{\rm CMB}$]& Median& 9.2& 12.5& 23.2 &11& 6& 12& 43&\dots&\dots\cr
\phantom{Noise rms per pixel} [MJy\,sr\mo]& Median&\dots&\dots&\dots&\dots&\dots&\dots&\dots&0.0149& 0.0155\cr
Combined gain calibration uncertainty$^{\rm b}$ [\%]& All sky& 0.82& 0.55 & 0.62& 0.5& 0.5& 0.5& 
   1.2& 10& 10\cr
Zero level$^{\rm c}$ [MJy\,sr\mo]& All sky& 0& 0& 0&0.0047& 0.0136& 0.0384& 0.0885& 0.1065&  0.1470\cr
Zero level uncertainty [$\mu\mathrm{K}_\mathrm{CMB}$]& All sky& $\pm 2.23$& $\pm 0.78$& $\pm 0.64$&\dots&\dots&
   \dots&\dots&\dots&\dots\cr
\phantom{Zero level uncertainty} [MJy\,sr\mo]& All sky&\dots&\dots&\dots& $\pm 0.0008$& $\pm 0.001$& $\pm 0.0024$&
    $\pm 0.0067$& $\pm 0.0165$& $\pm 0.0147$\cr
Colour correction uncertainty [\%]&Notes d,e& $0.1\beta$& 0.3$\beta$& 0.2$\beta$ &
   0.11$\Delta\alpha$&0.031$\Delta\alpha$&0.007$\Delta\alpha$&0.006$\Delta\alpha$&
   0.020$\Delta\alpha$&0.048$\Delta\alpha$\cr 
Beam colour correction uncertainty [\%]&Notes d,e& 0.5& 0.1& 0.3&$<$0.3& $<$0.3&
   $<$0.3& $<$0.5&$<$2.0& $<$1.0\cr
\noalign{\vskip 3pt}
\noalign{\vskip 5pt\hrule\vskip 3pt}}}
\endPlancktablewide
\tablenote {{\rm a}} The HFI default maps do not include removal of zodiacal emission.\par
\tablenote {{\rm b}} Includes the absolute uncertainty (0.25\,\%) of the calibration standard used, which is the CMB dipole estimated by WMAP7 \citep{hinshaw2009}.\par
\tablenote {{\rm c}} A zero level has been removed from the LFI maps ($-300.84$, $-22.83$, and $-28.09\,~\mu\mathrm{K}_\mathrm{CMB}$ at 30, 44, and 70\,GHz), but not from the HFI maps. The value given in this table corresponds to an estimated zero level of Galactic emission \citep{planck2013-p03f} in the maps, which include zodiacal emission.  For total emission studies, the level contributed by the cosmic infrared background must be added \citep{planck2013-p03f}.\par
\tablenote {{\rm d}} $\beta$ is the temperature spectral index of the source (\citealt{planck2013-p02d}) and $\Delta\alpha$ is the difference in source spectral index from $-1$ (\citealt{planck2013-p03d}).  For $\alpha=-1$, $\nu I_{\nu} = \hbox{constant}$ following the IRAS convention.  No uncertainties are assumed on $\alpha$ and $\beta$.\par
\tablenote {{\rm e}} For the HFI channels, we show the upper limit in solid angle change due to color correction from a planet spectrum source (roughly $\nu^2$) to {\it IRAS} convention ($\nu^{-1}$). \par
\endgroup
\end{table*}

\subsection{Map characterization}
\label{sec:MapChar}

Null tests are a powerful way to evaluate the quality of LFI and HFI maps.  Among these are half-ring difference maps (which capture noise properties), and Survey-to-Survey differences (which capture different types of systematic signals).  Simulation of known systematics is also a viable way to validate the effects seen in the real data, especially in Survey-to-Survey differences.  Comparison of angular cross-power spectra of maps made with individual detectors within a frequency band, and of maps at different frequencies, is used to give confidence in the results.  Many such tests have been implemented, as described in \citet{planck2013-p02}, \citet{planck2013-p02a}, and \citet{planck2013-p02b} for LFI,  and \citet{planck2013-p03}, \citet{planck2013-p03c}, \citet{planck2013-p03f}, and \citet{planck2013-p03d} for HFI.   Table~\ref{tab:accuracyBudgetOverview} summarises the sources contributing to uncertainties at map level.  Table~\ref{tab:accuracyBudgetResult} gives the actual uncertainty levels for each map.  For residual systematic levels, see \citet{planck2013-p02a} (LFI) and \citet{planck2013-p03} (HFI).

\subsection{Consistency tests}
\label{sec:Consis}

One of the key design features of \Planck\ is that it contains two separate instruments, subject to independent calibration and systematic effects. The simple fact that they observe the same CMB anisotropies in nearly adjacent frequency bands, and that they do so with high signal-to-noise ratio, provides a powerful cross-check on data quality. However, the very high accuracy ($\sim$0.1\,\%) that is the aim of \Planck\  also implies that every minute difference in how the CMB anisotropies are observed must be taken into account when comparing data from LFI and HFI. This applies particularly to instrumental issues (beam shapes, noise levels) and residual foreground signals. Similar considerations apply when comparing \Planck\  data to data from other experiments, e.g. WMAP.

Figure~\ref{FigConsistB2} shows a map-level comparison between 70 and 100\,GHz, the closest frequencies between the two instruments. The CMB structures at high Galactic latitude disappear in the difference made in $K_{\rm cmb}$ units as shown by the uniform background of noise. The deep nulling of the CMB anisotropy signal directly achieved by this straightforward differencing demonstrates that the \Planck\ maps are free from serious large- to intermediate-scale imperfections. It also reveals in an immediate and interesting manner the foreground residuals.

\begin{figure*}
   \centering
\includegraphics[width=180mm]{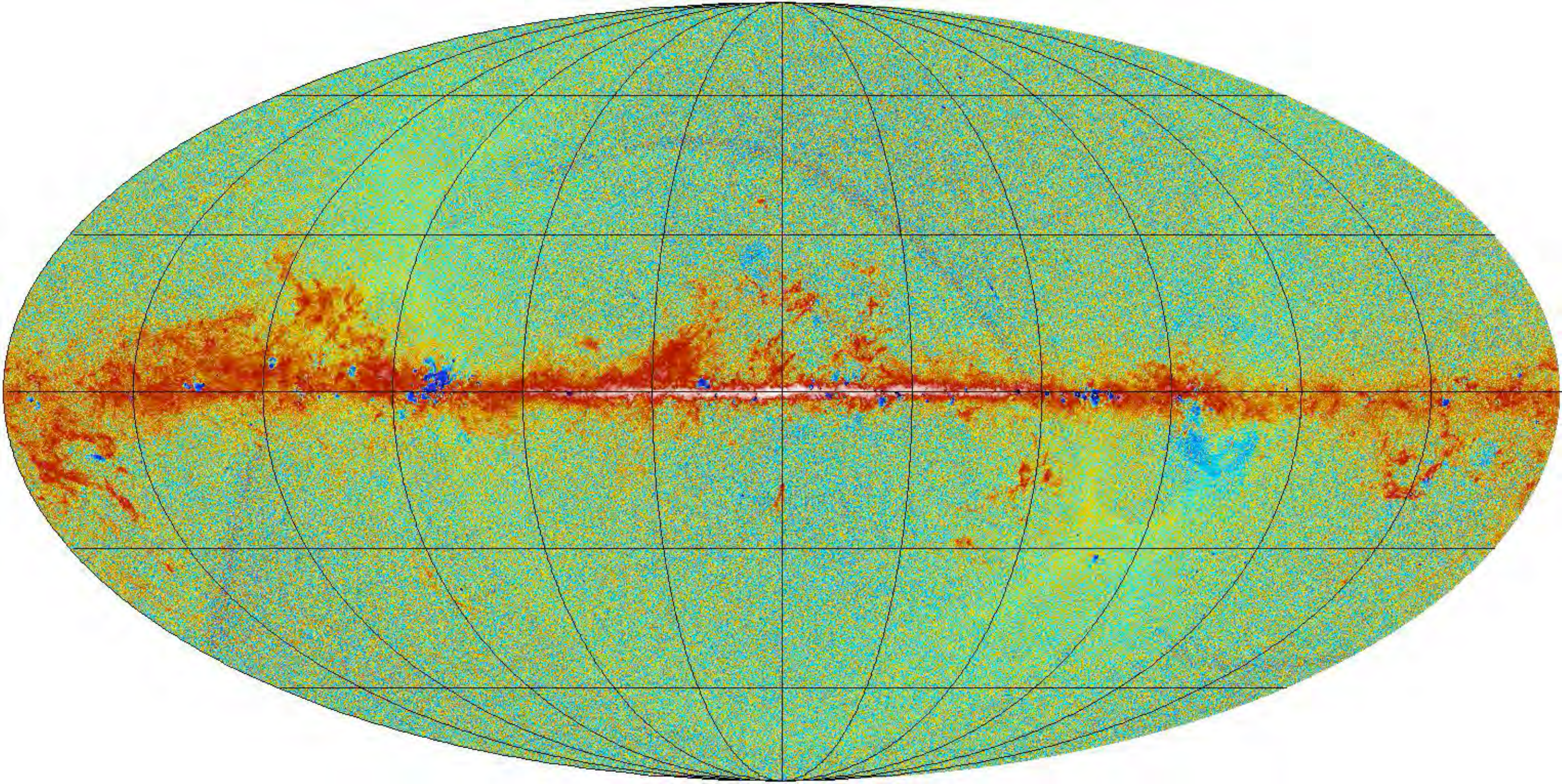}
\includegraphics[width=100mm]{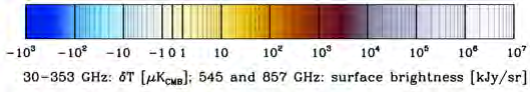}
   \caption{Example map constructed for nulling CMB anisotropies by differencing the 100\,GHz and 70\,GHz channels. A good fraction of the Galactic emission which stands out at low latitudes arises from CO in the 100\,GHz channel (see Section~\ref{sec:COmaps}).  The overall impression of green, a colour not used in the colour bar, is due to the interaction between noise, the colour scale, and the display resolution.  Positive and negative swings between pixels in the 70\,GHz noise map pick up reds and blues ``far'' from zero, which when displayed at less than full-pixel resolution give green.}
   \label{FigConsistB2}
\end{figure*}

Spectral analysis allows a more quantitative assessment. 
The result of a detailed comparison between neighbouring maps at 70, 100, and 143\,GHz is shown in Figure~\ref{FigConsisSpec},  checking consistency between LFI and HFI, and agreement between the main CMB channels overall. In doing this comparison, residual foregrounds cannot be ignored (see Section~\ref{SummaryFGs}, Figures \ref{FigFgContour} and \ref{FigFgPS}) and are corrected for channel-by-channel.  Beam effects also have to be taken into account for such spectral comparisons. Since the data release of March 2013, strong evidence has arisen that a small fraction ($\sim$0.2\,\%) of the HFI beams' solid angle was not taken into account in the window functions (Section~\ref{sec:Opt}), and a correction for this has been included in Figure~\ref{FigConsisSpec}.  See \citet{planck2013-p03c} and \citet{planck2013-p01a} for a complete description.

\begin{figure}
   \centering
   \includegraphics[width=88mm]{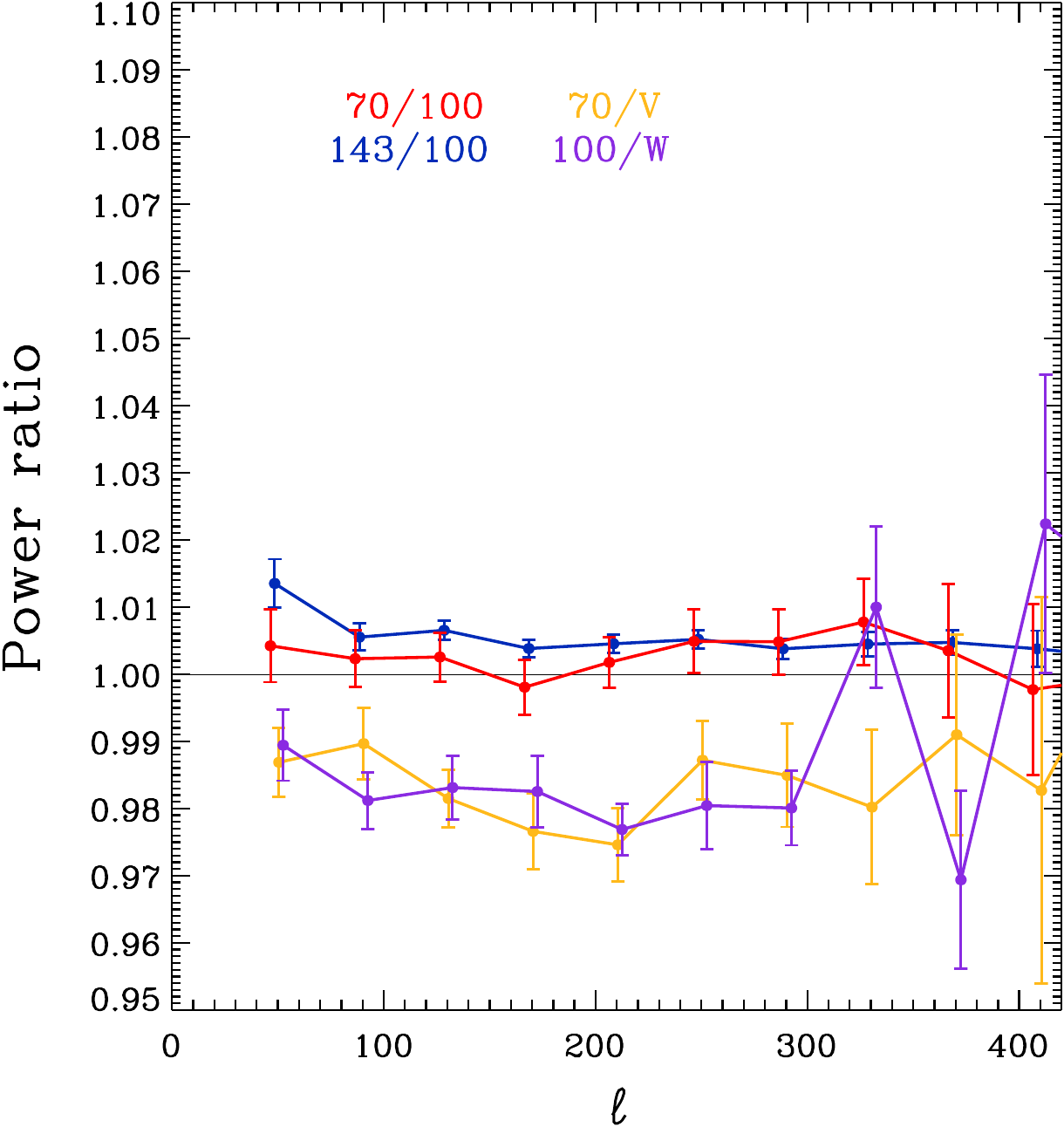}
   \caption{Ratios of power spectra of \planck\ and \WMAP\ maps, using a joint mask ($f_{sky}=56.7\,\%$), and including corrections for both \Planck\ beams and for \Planck\ and \WMAP\  discrete source residuals. (70, 100, 143) and (V, W) refer to the corresponding \Planck\ and \WMAP\ channels.}
   \label{FigConsisSpec}
\end{figure}

Figure~\ref{FigConsisSpec} demonstrates that, once residual foregrounds and known beam effects are corrected for, spectral consistency between the LFI 70\,GHz and HFI 100\,GHz maps is achieved at a level of 0.43\,\% , and between the HFI 100 and 143\,GHz channels at a level of 0.46\,\% (the values quoted are averages between $\ell = 70$ and 390). These consistency levels are within the uncertainties assumed for the 2013 scientific analysis.

Additional checks can be made. For example, Fig.~\ref{HFIDPCFig35} \citep{planck2013-p03} compares CMB anisotropies in all \Planck\ LFI and HFI channels  (44, 70, 100, 143, 217, 353 GHz) following component separation at the power spectrum level and using 143 GHz as the reference. The three main HFI CMB channels agree within 0.3\,\% of each other. The recalibration coefficients for the bands 70 to 217\,GHz are all within 0.4\,\%;  44 and 353\,GHz are at the 2\,\% level.  

\begin{figure}
   \centering
   \includegraphics[width=88mm]{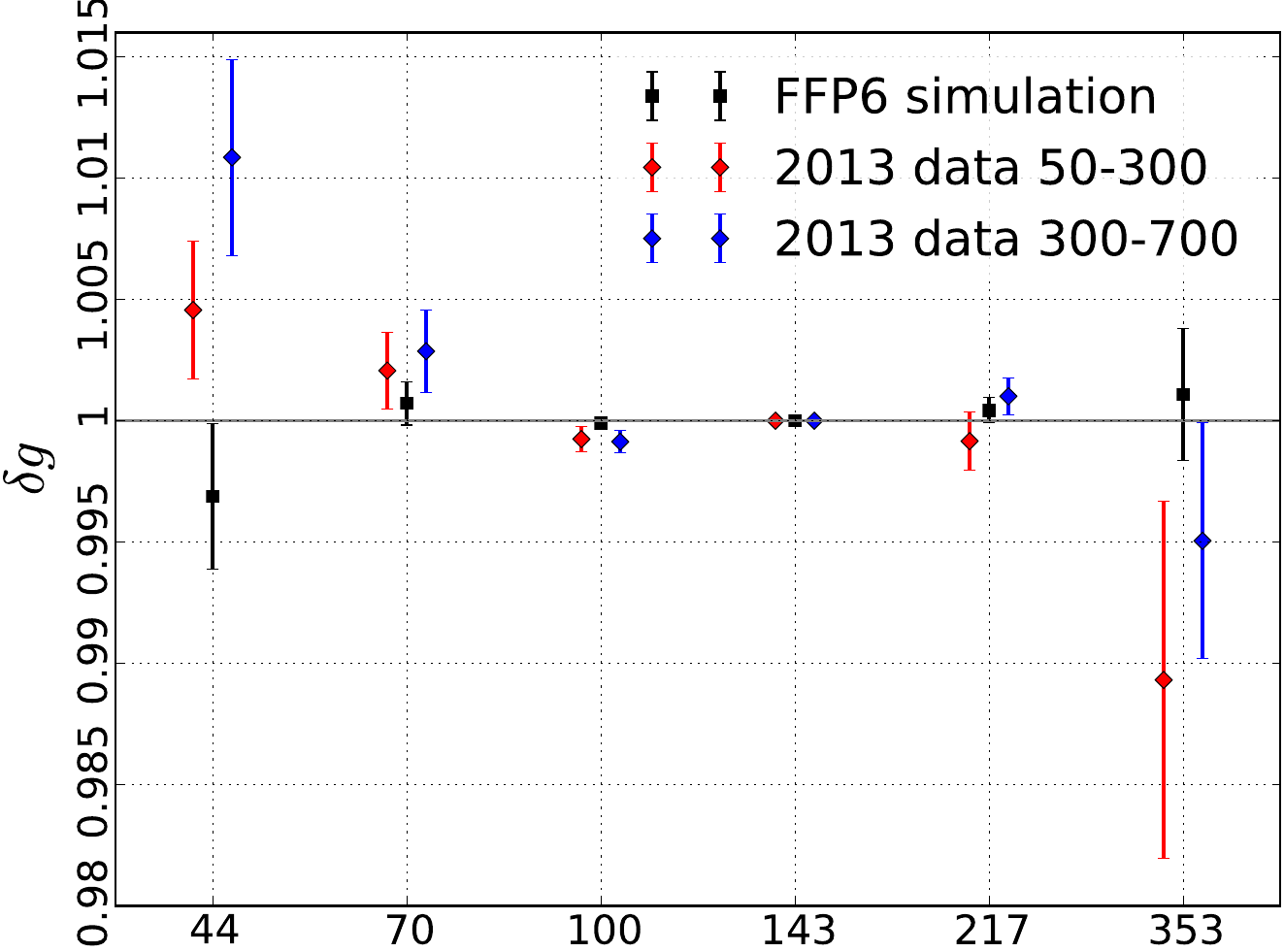}
   \caption{Recalibration factor maximizing the CMB consistency in simulations (black) and in the data considering different multipole ranges (red and blue), at each \Planck\ frequency (in GHz on the horizontal axis). This is Figure~35 from \citep{planck2013-p03}.} 
   \label{HFIDPCFig35}
\end{figure}

An independent approach to consistency is based on the likelihood analysis described in \cite{planck2013-p08} and \cite{planck2013-p11} (see also Section \ref{sec:CMBLike}), which works on cross-spectra of individual detector sets and solves simultaneously for calibration, foreground, and beam parameters. We have verified that such an approach achieves similar levels of consistency between LFI and HFI channels as those illustrated in Figure~\ref{FigConsisSpec}, once differences in sky coverage, noise properties, and foreground modelling are taken into account.

A difference of 0.4--0.5\,\% between the power spectra of LFI and HFI adjacent channels is within the uncertainties assumed for the 2013 scientific analysis \citep{planck2013-p11}.  A specific analysis described in \citet{planck2013-p01a} has verified that fractional percent shifts in the overall amplitude of power spectra result in fractional $\sigma$ shifts in cosmological parameters, except for the cosmological amplitude $A_{\rm s}$ and parameters related to it, as expected.

We can extend the spectral comparisons of neighbouring \Planck\ channels to power spectra of CMB maps from \WMAP.  Figure~\ref{FigConsisSpec} shows a significant discrepancy of $\sim$1.7-2.0\,\% in the $\ell$ range 70 to 390 in the power ratio of  the closest \Planck\ and \WMAP\ channels.   Additional checks based on component-separated CMB maps confirm these findings. Fig.~\ref{fig:TTspectrum}, taken from \citet{planck2013-p08}, shows that the \WMAP\ spectrum lies systematically above the \Planck\ spectra, with the difference being of order $20\muK^2$ at $\ell < 25$, and possibly rising slowly with $\ell$ (see \cite{planck2013-p01a}).  At higher multipoles, the comparison between \WMAP\ and \Planck\ is discussed in Appendix~A of \cite{planck2013-p11}, and shows that a multiplicative factor of 0.974 applied to the \WMAP\ V+W spectrum brings it into excellent
 point-by-point agreement with the \Planck\ 100$\times$100\,GHz spectrum over the range $50 < \ell < 400$.

\begin{figure*}
\begin{centering}
\includegraphics[width=1\textwidth]{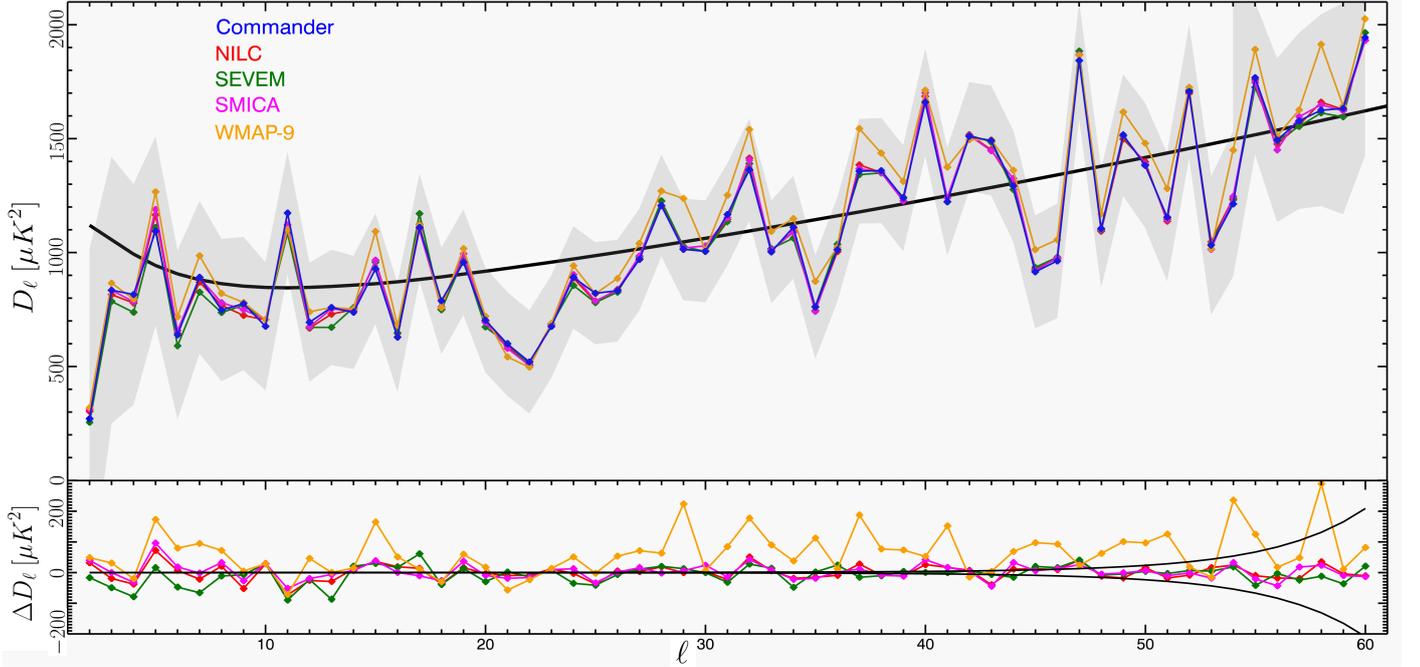}
\par\end{centering}
\caption{{\it Top\/}: temperature power spectra evaluated from downgraded \Planck\ maps, estimated with  {\tt Commander}, {\tt NILC}, {\tt SEVEM}, and {\tt SMICA}, and the 9-year \WMAP\ ILC map, using the {\tt Bolpol} quadratic estimator. The grey shaded area indicates the $1\,\sigma$ Fisher errors.  The solid line is the \Planck\ \LCDM\ best fit model. {\it Bottom\/}: Difference between the other power spectra in the top panel and {\tt Commander}.  The black lines show the expected $1\,\sigma$ uncertainty due to (regularization) noise.  This is Fig.~34 of \cite{planck2013-p08}.} 
\label{fig:TTspectrum}
\end{figure*}

The ratios of \Planck\ and \WMAP\ spectra represent a 1.5--$2\sigma$ difference from unity, based on the uncertainties in absolute calibration determined for \Planck\ and \WMAP\ \citet{planck2013-p01a}, somewhat larger than expected.\footnote{We note, however, that the difference is of the same order as the change in calibration reported by the \WMAP\  team between the 3rd and 5th year releases; in fact \Planck\  is quite consistent with the earlier \WMAP\  calibration.}.  As the primary calibration reference used by \Planck\ in the 2013 results is the \WMAP\ Solar dipole, the inconsistency between \Planck\ and \WMAP\ is unlikely to be the result of simple calibration factors.  Reinforcing this conclusion is the fact that the intercalibration comparison given in Fig.~35 of \citet{planck2013-p03} for CMB anisotropies shows agreement between channels to better than 0.5\,\%  over the range 70--217\,GHz, and 1\,\% over all channels from 44 to 353\,GHz, using 143\,GHz as a reference.  Problems with transfer functions are more likely to be the cause. The larger deviations at higher multipoles in the \Planck\ intercalibration comparison just referred to also point towards transfer function problems.  At present, we do not have an explanation for the $\sim 2\,\%$ calibration difference between WMAP and Planck. The differences between \WMAP\ and \Planck\ are primarily multiplicative in the power spectra, and so have little impact on cosmological parameters other than on the amplitude of the primordial spectrum $A_{\rm s}$ and directly related derived parameters.

\section{CMB Science Products}
\label{sec:SciProds}

With \Planck\ we observe the millimetre- and submillimetre-wave sky in greater detail than previously possible.  Component separation --- the process of separating the observed sky emission into its constituent astrophysical sources --- is therefore a central part of our data analysis, a necessary step in reaching the mission goal of measuring the primary CMB temperature anisotropies to a precision limited mainly by uncertainty in foreground subtraction, as well as in producing maps of foreground components for astrophysical studies.  We apply a variety of component separation techniques, some developed specifically for the \Planck\ analysis.
	
Multiple Galactic and extragalactic emission mechanisms contribute to the observed sky over the \Planck\ frequency range.  Synchrotron radiation from relativistic electrons spiraling in the Galactic magnetic field dominates at the lowest frequencies (30 and 44\,GHz), falling in brightness temperature $T$ as $\nu^\alpha$ with $\alpha\approx -3$.  Free-free emission from ionized interstellar gas and \ion{H}{ii} regions is also prominent, decreasing with a power-law index $\alpha=-2.15$.  Anomalous microwave emission (AME), almost certainly due to rotating dust grains spun up by photons in the interstellar radiation field and collisions, has a spectrum that peaks somewhere around 30\,GHz and falls rapidly with frequency through the lower \Planck\ bands.  Thermal emission from dust grains heated near 20\,K is the dominant Galactic emission at frequencies above 70\,GHz, rising with frequency according to a greybody spectrum with emissivity $\epsilon \propto \nu^{1.5{\rm -}2}$.  In addition, we observe line emission from CO at 100, 217, and 353\,GHz.  
	
In the lowest-foreground half of the sky at high Galactic latitudes, foregrounds in \Planck's ``CMB channels'' (70--217\,GHz) are dominated by dust in the Milky Way at $\ell < 50$, and by extragalactic radio (low frequencies) or infrared (217\,GHz) sources at $\ell > 200$.  Some of these sources are seen as individual objects by \Planck, while most form an unresolved background. In the case of infrared sources, this background is the CIB, which has a spectrum close to that of Galactic thermal dust.  There is also a contribution from secondary CMB anisotropies, notably the Sunyaev-Zeldovich effect produced by galaxy clusters.  The strongest clusters are detected as individual compact sources, but the weakest contribute to an unresolved Sunyaev-Zeldovich foreground. 

The extraction of cosmological information from the \Planck\ data follows two main paths: the search for non-Gaussianity and other signatures of statistical order greater than two, which can only be found in the map of the CMB; and the estimation of the parameters of models of the Universe, which can be determined from the angular power spectrum, a complete statistical description of the sky up to order two.   For the path that requires maps, diffuse foregrounds must be removed at map level.  Unresolved source residuals, however, can only be removed statistically in the power spectra, as \Planck\ itself can determine nothing about the  location on the sky of such residuals.  It is also possible to separate diffuse astrophysical foregrounds at high-$\ell$ at the power spectrum level over a fraction of the sky. 

For the 2013 \Planck\ results, we use a component-separated map of the anisotropies (see Sect.~\ref{subsec:CMBmapNG}, \citealt{planck2013-p06}) for non-Gaussianity analysis and other higher-order statistics.  For parameter estimation, we use a likelihood code \citep{planck2013-p08} based on a component-separated CMB map at $\ell<50$ and on a self-consistent set of parameters of physically-motivated foreground models determined simultaneously with the best-fit CMB model (see Sect.~\ref{sec:CMBLike}) at $\ell\ge50$.

The generation of the \Planck\ CMB-science products is diagramed in Fig.~\ref{FigCMBProducts}. 

\begin{figure*}
   \centering
   \includegraphics[width=180mm]{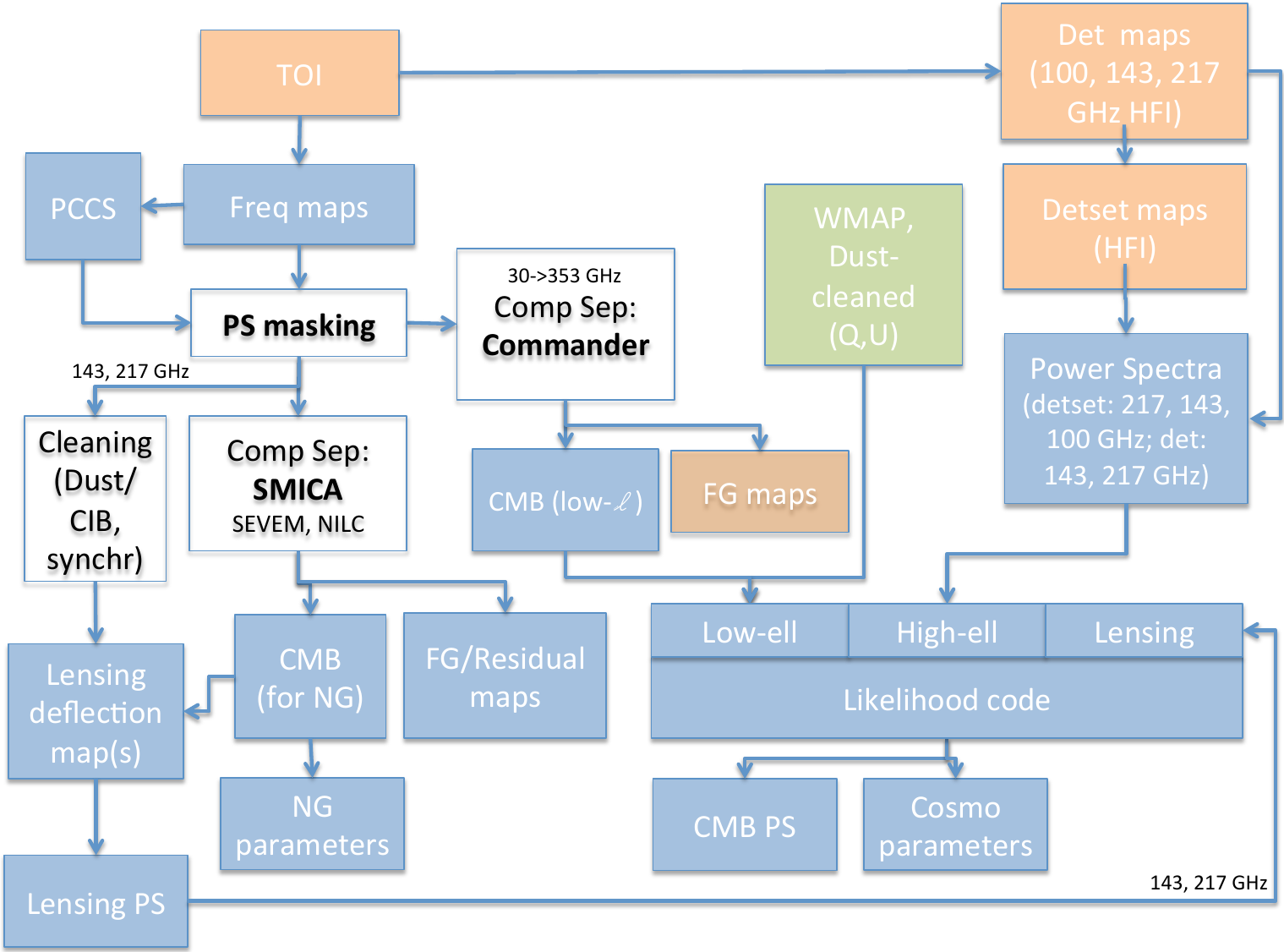}
   \caption{Diagram of the generation of CMB-science products being delivered by \Planck, in blue. Products in green are external, and products in orange are not being delivered in the current release.  Each product delivered is accompanied by specific data characterizing it (not shown on the diagram). This diagram does not include other data used for parameter estimation, either from \Planck\ itself (lensing, CIB, SZ) or from other CMB experiments (SPT, ACT, etc). }
    \label{FigCMBProducts}
    \end{figure*}

\subsection{CMB map extraction}
\label{subsec:CMBmapNG}

Our approach to component separation for \Planck, and more specifically to extraction of a CMB map, is described in detail in \citet{planck2013-p06}.  We cannot extract maps independently for all diffuse components (CMB, synchrotron, free-free, AME, dust, CIB, SZ) using \Planck\ data alone, as the number of parameters needed to describe them exceeds the number of frequency channels.  However, by treating some or all of the foregrounds in combination, we can extract the CMB itself quite effectively.  Four different methods were developed and optimized to do this using \Planck\ maps alone: {\tt SMICA} (independent component analysis of power spectra, \citealt{delabrouille2003}; \citealt{cardoso2008}); {\tt NILC} (needlet-based internal linear combination, \citealt{delabrouille2009}); {\tt Commander-Ruler} (pixel-based parameter and template fitting  with Gibbs sampling, \citealt{eriksen2006}; \citealt{eriksen2008}); and {\tt SEVEM} (template fitting, \citealt{fernandez2012}). %

All four codes were tested and characterized on the FFP6 simulations of \Planck\ data (see Sect.~\ref{sec:FFP}).  Based on performance in simulations and on statistical tests conducted on the \Planck\ data, {\tt SMICA} was selected to extract the high-resolution CMB map used for non-Gaussianity and higher-order statistics, while {\tt Commander} (run without the {\tt Ruler} extension) was selected to extract the low-resolution CMB map used to construct the low-$\ell$ likelihood.  The {\tt SMICA} high-resolution map is used in a wide variety of analyses presented in this release \citep{planck2013-p09,planck2013-p09a,planck2013-p19,planck2013-p20,planck2013-p12,planck2013-p14}.  Although performance details vary somewhat from method to method \citep{planck2013-p06} and some methods are preferred for specific purposes, all four methods yield CMB maps suitable for cosmological analysis.  Moreover, the use of multiple methods giving consistent results provides important cross-validation, and demonstrates the robustness of the CMB map obtained by \Planck.   We therefore release all four maps, to give users a grasp of both the uncertainties and the robustness associated with these methods.

The SMICA map in Fig.~\ref{FigCMBMap} estimates the CMB over about 97\,\% of the sky; the remaining area is replaced with a constrained Gaussian realization.   It has an angular resolution of 5\arcm, but its harmonic content is cut off for $\ell> 4000$.  In the pixel domain, the noise has an average RMS of about 17\muK\ (for the cutoff at $\ell=4000$), but its distribution is highly inhomogeneous (Fig.~\ref{FigCMBMapRMS}).

\begin{figure*}
  \centering
  \includegraphics[width=\textwidth]{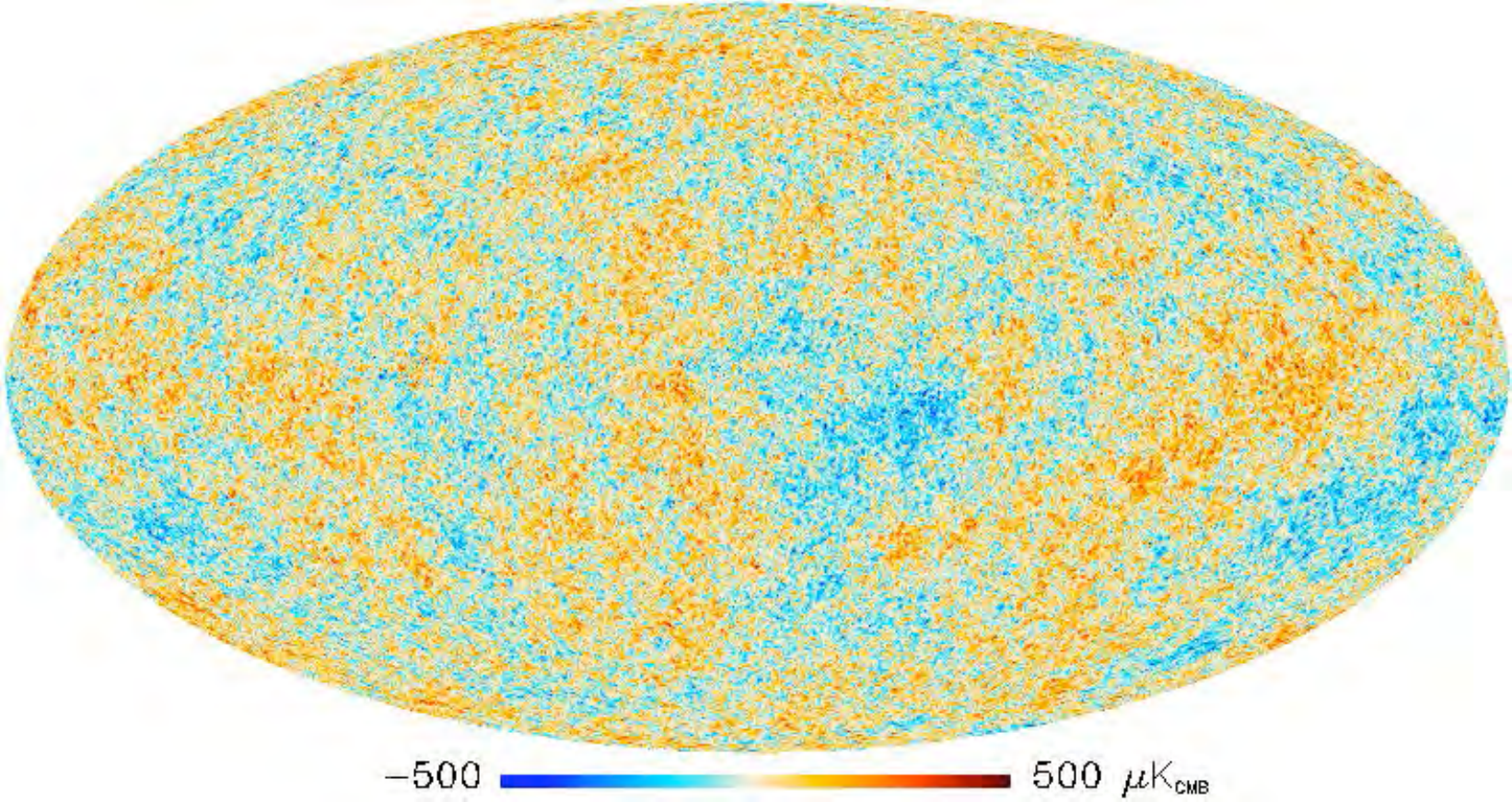}
  \caption{The SMICA CMB map, with 3\,\% of the sky replaced by a constrained Gaussian realization.  For the non-Gaussianity analysis (Sect.~\ref{sec:CMBNG} and  \citealt{planck2013-p09a}), 73\,\% of the sky was used. Apart from filling of the blanked pixels, this is the same map as shown in Fig. 1 of \cite{planck2013-p06}. }
  \label{FigCMBMap}
\end{figure*}

\begin{figure}[!ht]
\hspace{1em}
  \centering
  \includegraphics[width=\columnwidth]{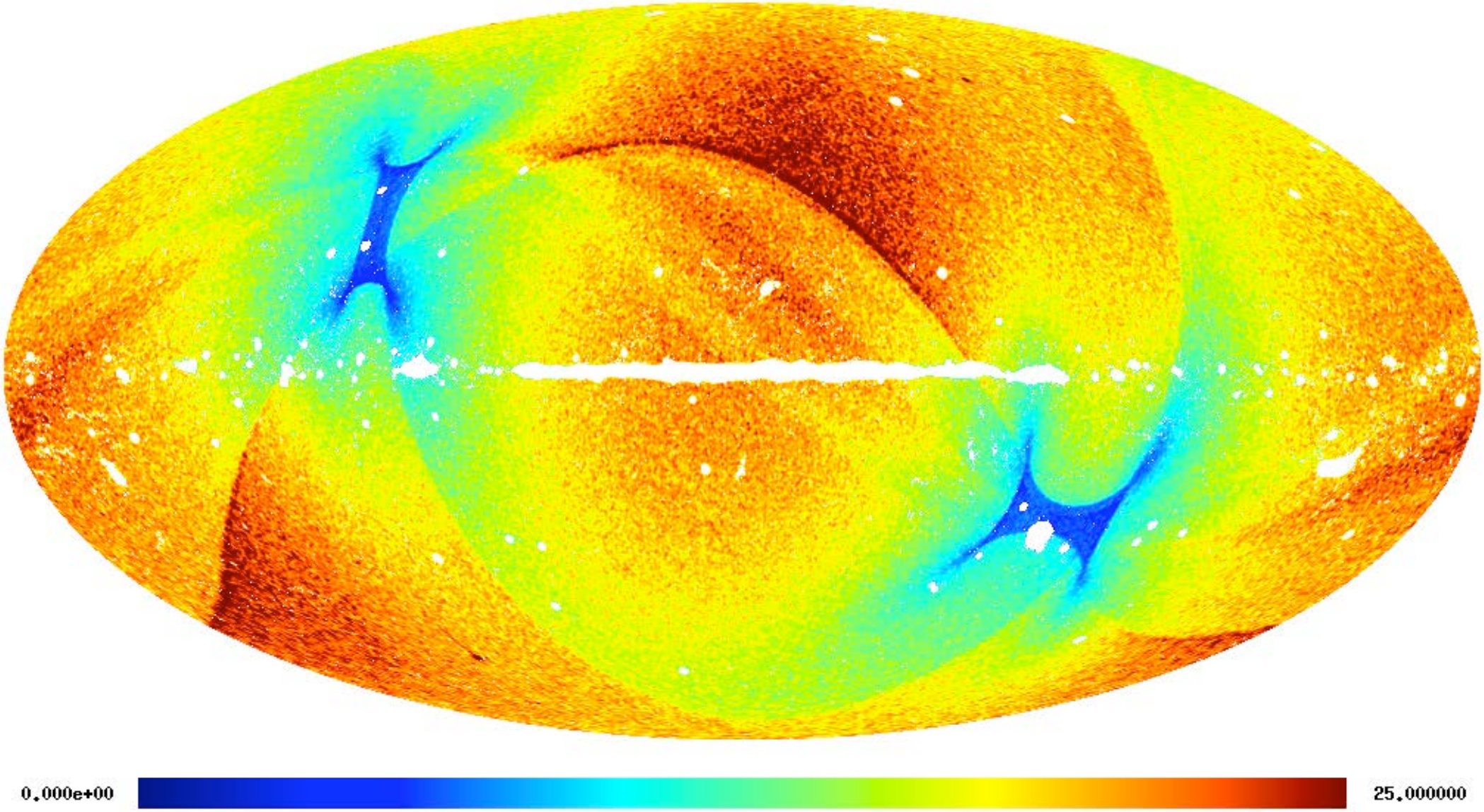}
  \caption{Spatial distribution of the noise RMS on a color scale of 25\muK\ for the SMICA CMB map, from the noise map obtained by running SMICA through the half-ring maps and taking the half-difference. The average RMS noise is 17\muK.  SMICA does not produce CMB values in the blanked pixels.  They are replaced by a constrained Gaussian realization.   }
  \label{FigCMBMapRMS}
\end{figure}

Figure~\ref{FigCMBNoiseSpec} illustrates the SNR reached by \Planck\ for the CMB signal.  It shows the angular power spectrum of the SMICA map and the associated half-ring noise, and their difference (both raw and smoothed) after beam correction.  The latter noise-corrected spectrum shows the CMB spectrum plus any remaining contamination.  Seven acoustic peaks are visible, and the SNR reaches unity (for single multipoles) at $\ell\sim 1700$.

\begin{figure}[!ht]
\hspace{1em}
  \centering
  \includegraphics[width=\columnwidth]{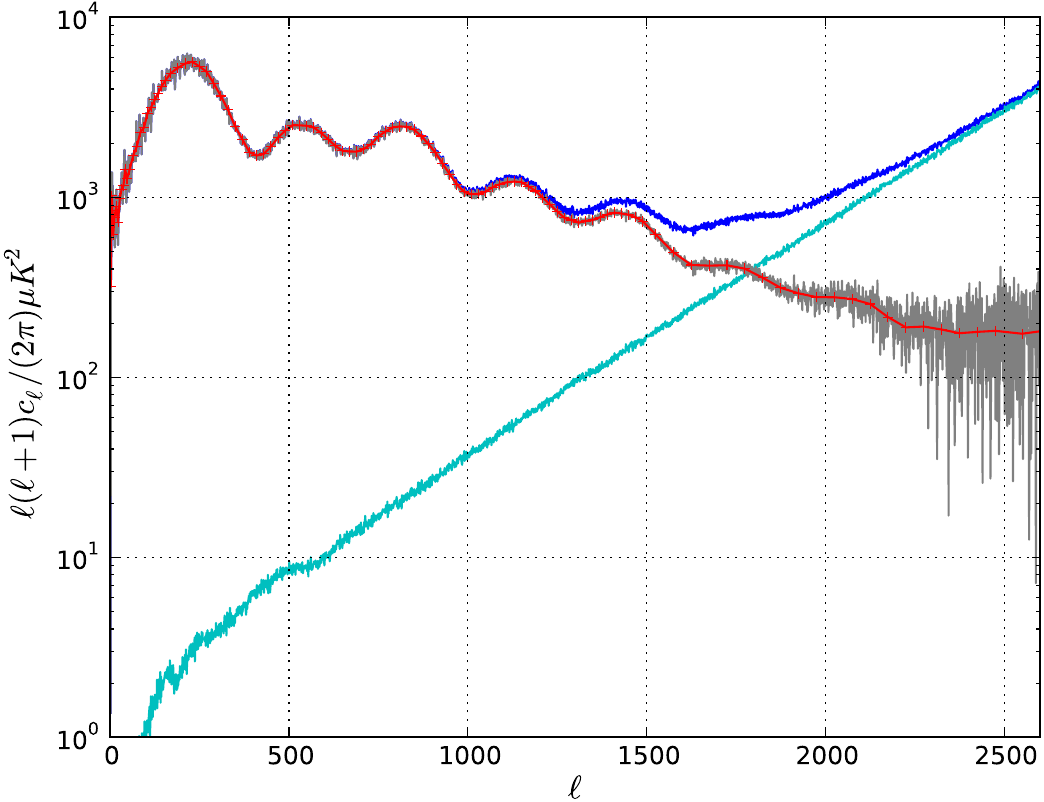}
  \caption{Angular spectra for the SMICA CMB products, evaluated over the confidence mask, and after removing the beam window function: spectrum of the CMB map (dark blue), spectrum of the noise in that map from the half-rings (magenta), their difference
(grey), and a binned version of it (red). }
  \label{FigCMBNoiseSpec}
\end{figure}

All four methods yield a set of ``residual'' maps that contain astrophysical foregrounds and other sources of noise.  As noted previously, the number of constraints provided by the \Planck\ data is less than even the minimal number of parameters that could describe all of the physically meaningful foreground components individually, so that without ancillary information we cannot separate all of the components individually.  Nonetheless, we release the residual maps for analysis in conjunction with the extracted CMB maps.

Additional maps based on \Planck\ data have been produced and subjected to the same characterization as the four maps described above \citep{planck2013-p06}. They are:

\begin {itemize}

\item the low-resolution ($\sim$1$\deg$) CMB map produced by {\tt Commander} and used as input for the low-$\ell$  part of the \Planck\ likelihood code.   The component separation incorporates physically-motivated parametric foreground models. In contrast to the other schemes developed to extract the CMB (Sect.~\ref{subsec:CMBmapNG}), it provides direct samples of the  likelihood posterior and rigorous propagation of uncertainties.  The Commander CMB map is not ideal for non-Gaussianity studies, due to its lower angular resolution, but it is good for the low-$\ell$ likelihood, which does not require high resolution.   This map is packaged into the input data required by the likelihood code.

\item a set of astrophysical components (see Sect.~\ref{sec:CommFGs} and \citealt{planck2013-p06}) at 7\arcm\ resolution extracted using {\tt Commander-Ruler} along with the CMB map described above.  The algorithm was optimised for recovery of astrophysical foregrounds as well as CMB extraction.  Nonetheless, it performs comparably well to the other CMB maps up to $\ell \sim 1500$ in terms of power spectrum estimations and extraction of cosmological parameters, as is shown in \citet{planck2013-p06} and \citet{planck2013-p08}.

\end{itemize}

It is well known that the CMB dipole is caused by our motion through the sea of background photons, with the velocity precisely measured \citep[e.g.,][]{hinshaw2009}, and in fact used to calibrate \Planck.  The 2013 results from \Planck\ do {\it not\/} include a new measurement of our velocity---that will follow in 2014.  However, the all-sky coverage, sensitivity, and angular resolution of \Planck\ enable two other tests of our motion, which were not possible up until now, namely, the aberration of the CMB sky and the dipole modulation of the anisotropies \citep{planck2013-pipaberration}.  These effects can be thought of as a strengthening and shrinking of the anisotropy pattern in the direction we are moving, and a weakening and enlarging of the pattern in the opposite direction.  We measure these effects using a quadratic estimator operating on the covariance matrix of the CMB fluctuations, which is essentially the same approach that we use for gravitational lensing \citep{planck2013-p12}.  We find a combined significance of these Doppler boosting effects above the $4\,\sigma$ level, with an amplitude and direction consistent with expectation.

\subsection{CMB Lensing products}
\label{sec:CMBLensing}

The CMB fluctuations measured by \Planck\ are perturbed by gravitational lensing, primarily by the structure of the Universe on very large scales (near the peak of the matter power spectrum at $300$\ Mpc comoving) at relatively high redshifts (with a kernel peaking at $z \sim 2$).  Lensing blurs the primary CMB fluctuations, slightly washing out the acoustic peaks of the CMB power spectrum \citep{planck2013-p08,planck2013-p11}.  Lensing also introduces several distinct non-Gaussian statistical signatures, which are studied in detail in \cite{planck2013-p12}.  The deflections caused by lensing on such large scales are weak, with an RMS of $2\parcm5$, and their effect may be represented as a remapping by the gradient of a lensing potential $\phi(\hat{n})$ as
\begin{equation}
T( \hat{n} ) = \tilde{T}( \hat{n} )( \hat{n} + \nabla \phi (\hat{n}) ),
\end{equation}
where $\hat{n}$ is the direction vector and $\tilde{T}$ is the unlensed CMB.  In \cite{planck2013-p12}, we reconstruct a map of the lensing potential $\phi(\hat{n})$, as well as estimates of its power spectrum $C_L^{\phi\phi}$.  Although noisy, the \Planck\ lensing potential map represents a projected measurement of all matter back to the last scattering surface, with considerable statistical power.  Figure~\ref{fig:xphimap} shows the \Planck\ lensing map, and Fig.~\ref{fig:phispec} shows the lensing power spectrum. 

As a tracer of the large scale gravitational potential, the \Planck\ lensing map is significantly correlated with other tracers of large scale structure.  We show several representative examples of such correlations in \cite{planck2013-p12}, including the NVSS quasar catalog \citep{NVSS1998}, the MaxBCG cluster catalog \citep{Koester:2007bg}, luminous red galaxies from SDSS \citep{Ross:2011cz}, and a survey of infrared sources from the WISE satellite \citep{Wright:2010qw}.  The strength of the correlations between the \Planck\ lensing map and such tracers provides a fairly direct measure of how they trace dark matter; from our measurement of the lensing potential, the \Planck\ maps provide a mass survey of the intermediate redshift Universe, in addition to a survey of the primary CMB temperature and polarization anisotropies.

\begin{figure}[!ht]
\includegraphics[width=88mm]{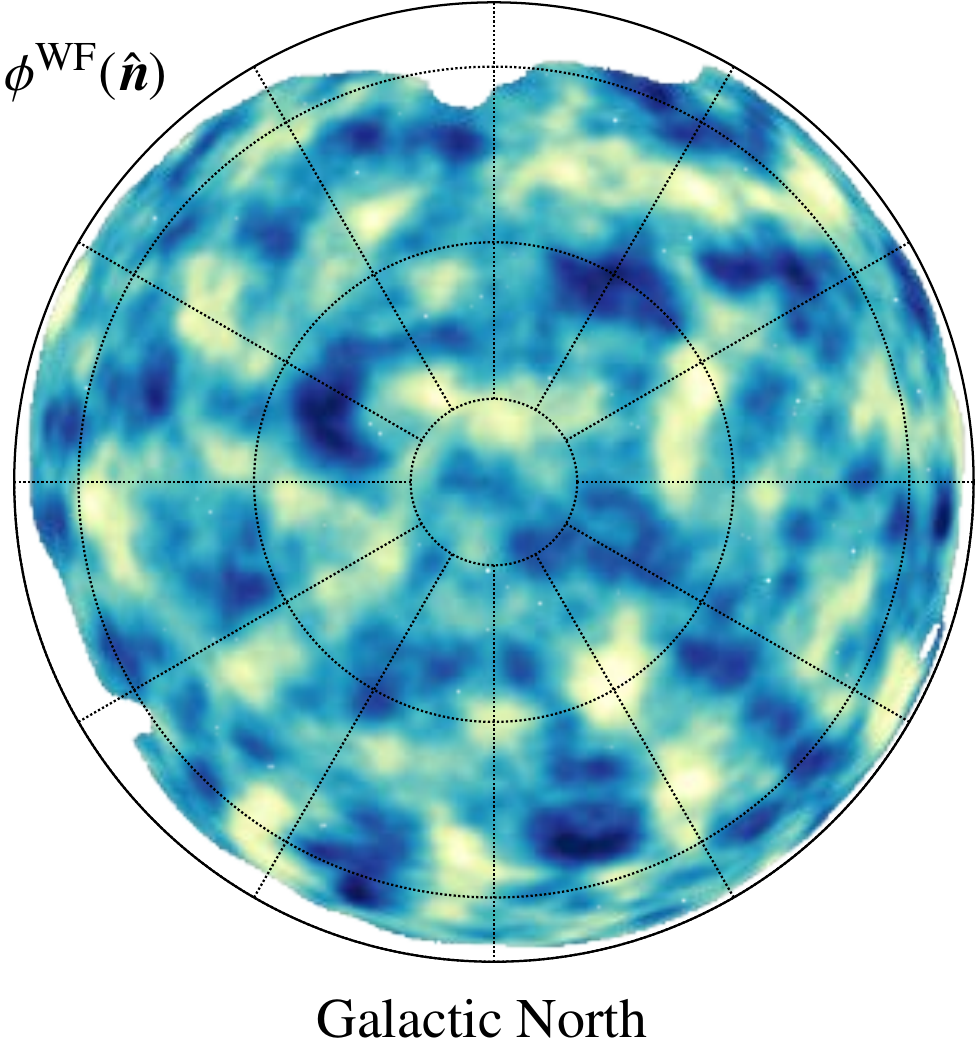}
\vglue 3mm
\includegraphics[width=88mm]{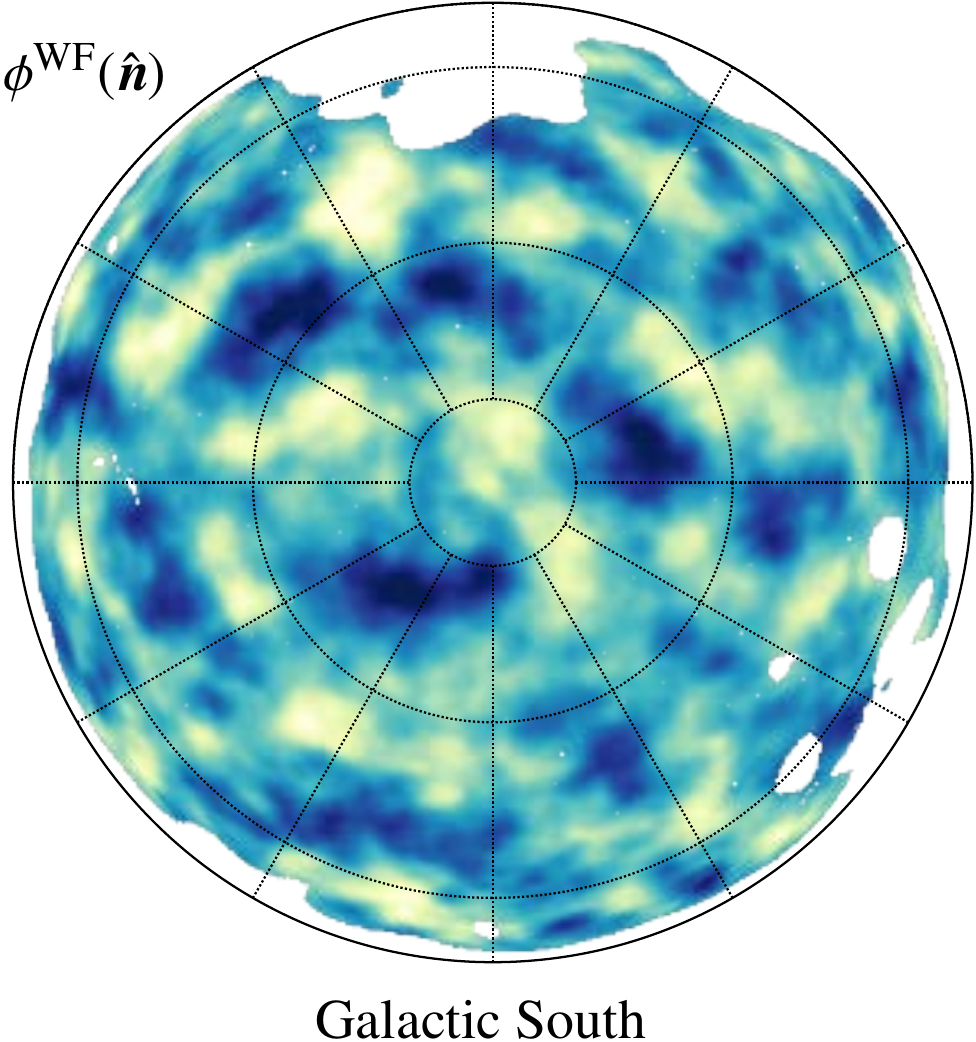} 
\caption{Wiener-filtered lensing potential estimate, in Galactic coordinates using orthographic projection \citep{planck2013-p12}.  The reconstruction was bandpass filtered to \mbox{$L \in [10, 2048]$}.  Note that the lensing reconstruction, while highly statistically significant, is still noise dominated for every individual mode, and is at best ${\rm SNR} \approx 0.7$ around $L=30$.}
\label{fig:xphimap}
\end{figure}

\begin{figure*}
\begin{center}
\begin{overpic}[width=180mm]{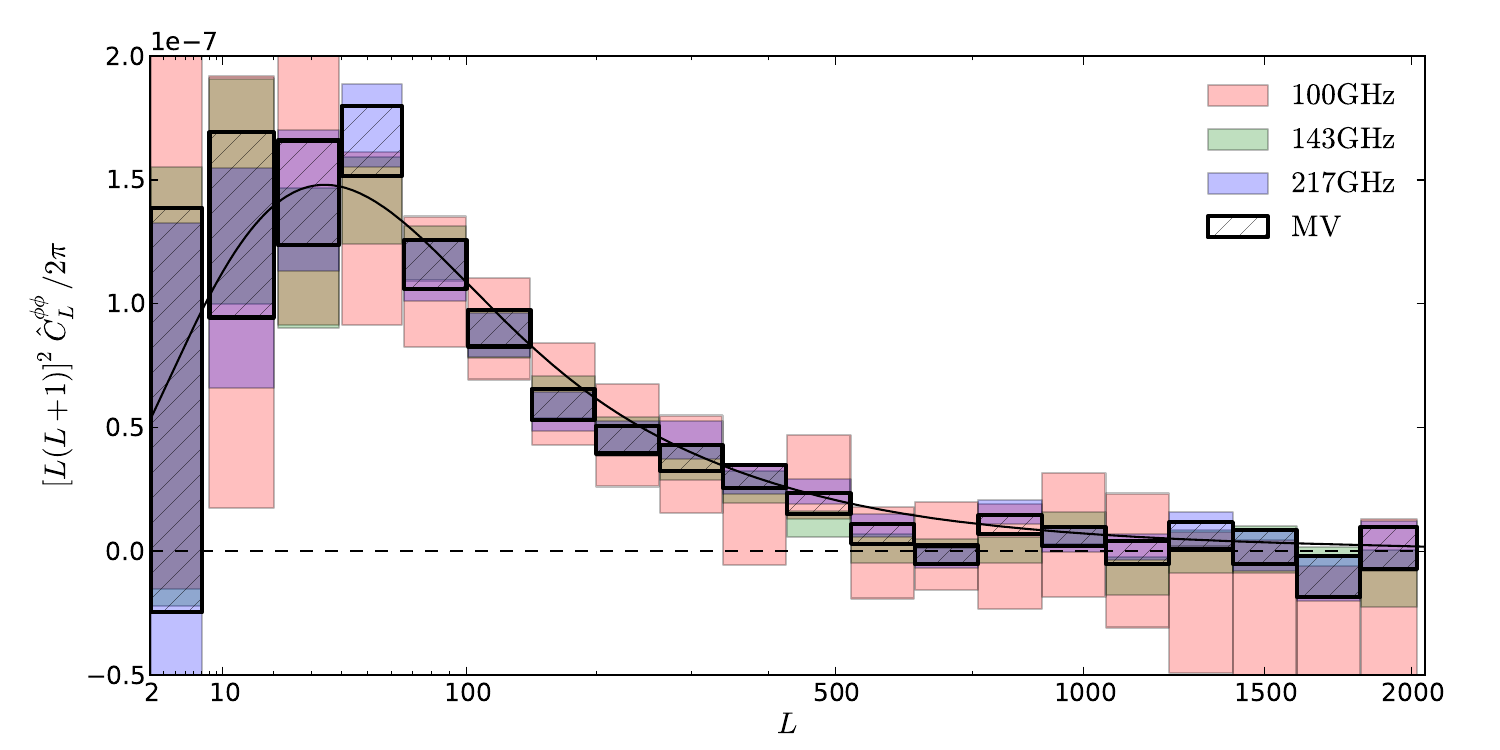}
\put(42, -1.5){ \small{Lensing Multipole $L$} }
\put(42, 48.5){ \small{Angular Scale [deg.]} }
\end{overpic}
\end{center}
\caption{Fiducial lensing power spectrum estimates based on the 100, 143, and 217\,GHz frequency reconstructions, as well as the minimum-variance reconstruction that forms the basis for the \planck\ lensing likelihood  \citep{planck2013-p12}.}
\label{fig:phispec}
\end{figure*}

\subsection{Likelihood code}
\label{sec:Like}

\subsubsection{CMB likelihood}
\label{sec:CMBLike}	

We construct a hybrid likelihood for the \Planck\ temperature data using an exact likelihood at large scales ($\ell<50$) and a pseudo-C$_{\ell}$ power spectrum at smaller scales ($50\le\ell<2500$). This follows similar analyses in, e.g., \citet{spergel2007}. The likelihood is described fully in \citep{planck2013-p08}; here we summarize its main features.

On large scales, the distribution for the angular power spectrum cannot be assumed to be a multivariate Gaussian, and the Galactic contamination is most significant. We use the $30<\nu<353$\,GHz maps from LFI and HFI to separate Galactic foregrounds. This procedure (Sects.~\ref{sec:SciProds} and \ref{subsec:CMBmapNG}) uses a Gibbs-sampling method to estimate the CMB map and the probability distribution of its power spectrum, $p(C_\ell|d)$, for $\ell<50$, using the cleanest 87\,\% of the sky.  We supplement this `low-$\ell$' temperature likelihood with the pixel-based polarization likelihood at large scales ($\ell<23$) from the \WMAP\ 9-year data release (Bennett et al. 2012).  The \WMAP\/9 data must be corrected for the dust contamination, for which we use the \WMAP\ procedure.  However, we have checked that switching to a correction based on the 353\,GHz \Planck\ polarization data changes the parameters extracted from the likelihood by less than $1 \sigma$. 

At smaller scales, $50<\ell<2500$, we compute the power spectra of the multi-frequency \Planck\ temperature maps and their associated covariance matrices using the 100, 143, and 217\,GHz channels, and cross-spectra between these channels\footnote{Interband calibration uncertainties with respect to 143\,GHz have been estimated by comparing directly the cross spectra and found to be within 2.4 and $3.4\times10^{-3}$ for 100 and 217\,GHz, respectively.}. Given the limited frequency range used in this part of the analysis, the Galaxy is conservatively masked to avoid contamination by Galactic dust, retaining 58\,\% of the sky at 100\,GHz and 37\,\% at 143 and 217\,GHz.

Bright extragalactic ``point'' sources detected in the frequency range 100 to 353\,GHz are also masked.  Even after masking, extragalactic (including thermal and kinetic Sunyaev-Zeldovich) sources contribute significantly to the power spectra at the smallest angular scales probed by \Planck.  We model this extra power as the sum of multiple emission components, following similar analyses in previous CMB experiments (e.g., \citealt{2013arXiv1301.0824S}; \citealt{2011ApJ...739...52D}; \citealt{2012ApJ...755...70R}; \citealt{2013arXiv1301.0776D}).

In this `high-$\ell$' \Planck\ likelihood, we model the total theoretical power as the sum of the lensed CMB and a set of parameterized foreground emission spectra.  Parameters are introduced describing the beam uncertainties. We simultaneously marginalize over all the  ``nuisance" parameters when estimating cosmological model parameters.  This approach is designed to allow easy combination with data from the ACT and SPT experiments, which measure the millimetre-wave spectra from scales of $\ell\sim 200$ to $\ell<10000$.

Unresolved extragalactic point sources are modelled as a Poisson power spectrum with one amplitude per frequency.  The power spectra of the anisotropies associated with the correlated infrared background and the thermal and kinetic Sunyaev-Zeldovich effect are described by four amplitudes and power law indices and one correlation coefficient.  In the mask used for the cosmology analysis, the CIB dominates at 217\,GHz over Galactic dust and extragalactic foregrounds above $\ell\sim 500$. At higher multipoles (above 2500 at 217\,GHz and 4000 at 143\,GHz), the Poisson part dominates over the CMB.  The foreground parameter values recovered in the  likelihood analysis \citep{planck2013-p08} are all compatible with cuurent knowledge of source counts, and with the new \Planck\ determination of the cosmic infrared background \citep{planck2013-pip56} and the thermal Sunyev-Zeldovich effect \citep{planck2013-p05b}, taking into account the rather large distribution of uncertainties for these weak foregrounds at the highest multipoles.

Removing the best-fitting extragalactic foreground model and combining multiple frequencies, we obtain the CMB temperature power spectrum shown in Fig.~\ref{fig:cmb_planck}.  \Planck\ measures the first seven acoustic peaks to high precision. For comparison, we show in Fig.~\ref{fig:cmb_smica} a power spectrum estimated from the SMICA CMB map discussed in Sect.~\ref{subsec:CMBmapNG}.

\begin{figure*}
\includegraphics[width=180mm]{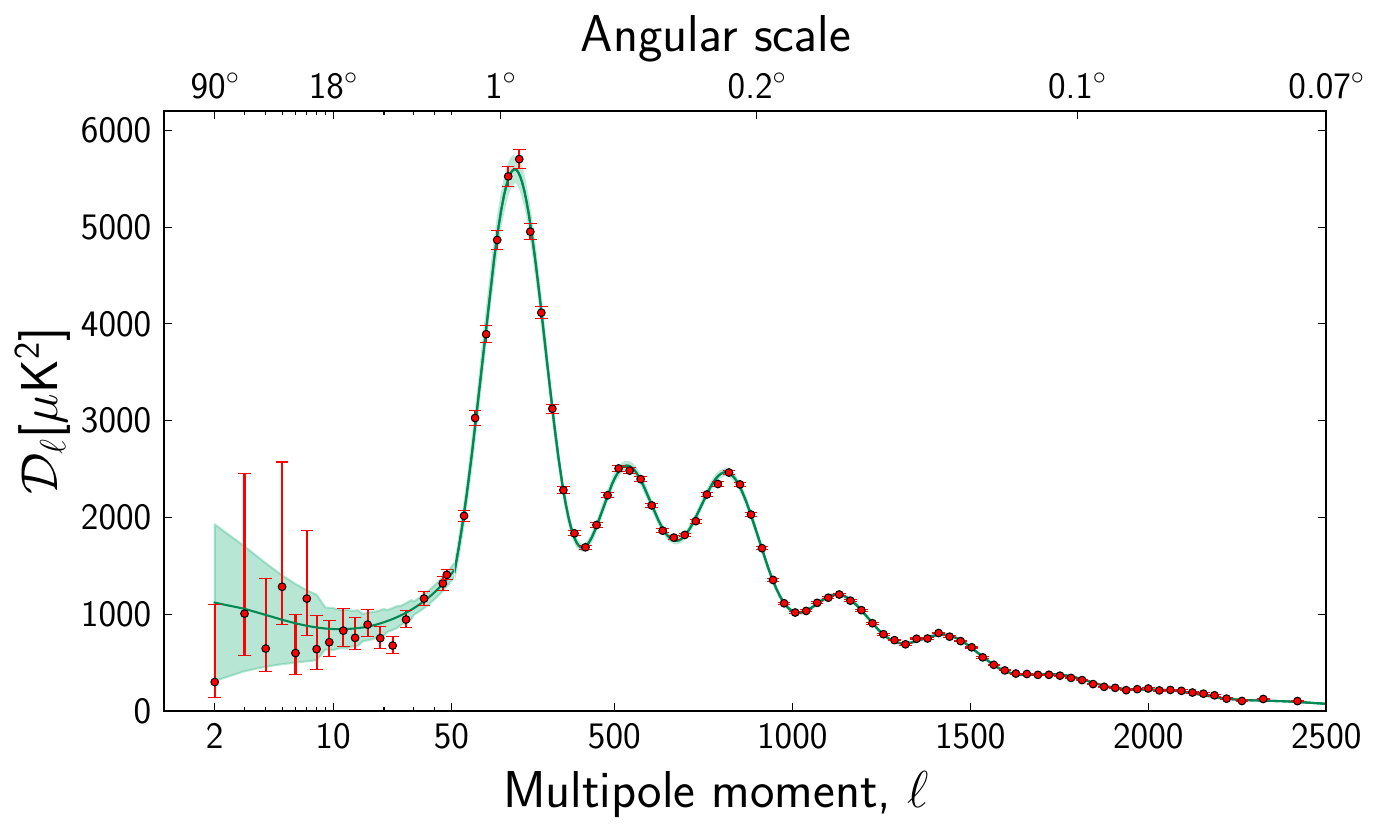}
\caption{Temperature angular power spectrum of the primary CMB from \Planck, showing a precise measurement of seven acoustic peaks that are well- fitted by a six-parameter $\Lambda$CDM model (the model plotted is the one labelled [Planck+WP+highL] in \citealt{planck2013-p11}). The shaded area around the best-fit curve represents cosmic/sample variance, including the sky cut used.  The error bars on individual points also include cosmic variance.  The horizontal axis is logarithmic up to $\ell = 50$, and linear beyond.  The vertical scale is $\ell(\ell+ 1)C_l/2\pi$. The measured spectrum shown here is exactly the same as the one shown in Fig. 1 of \citet{planck2013-p11}, but it has been rebinned to show better the low-$\ell$ region. }
\label{fig:cmb_planck}
\end{figure*}

To test the robustness of the \Planck\ power spectrum, we perform null tests between different detectors within a frequency band, between different Surveys, and between frequency bands. To test the likelihood formalism, we perform a suite of tests modifying aspects including the foreground modeling, beam treatment, and angular range considered.  We check that they have minimal effect on cosmological parameters, and also check that the same results are obtained using two independent power spectrum pipelines \citep{planck2013-p08}. 

The current version of the \Planck\ likelihood software is made available with the 2013 data release, together with the multi-frequency power spectra, the best-fitting CMB power spectrum, and the maps and masks used to construct the power spectrum and likelihood.

\begin{figure}
\includegraphics[width=0.5\textwidth]{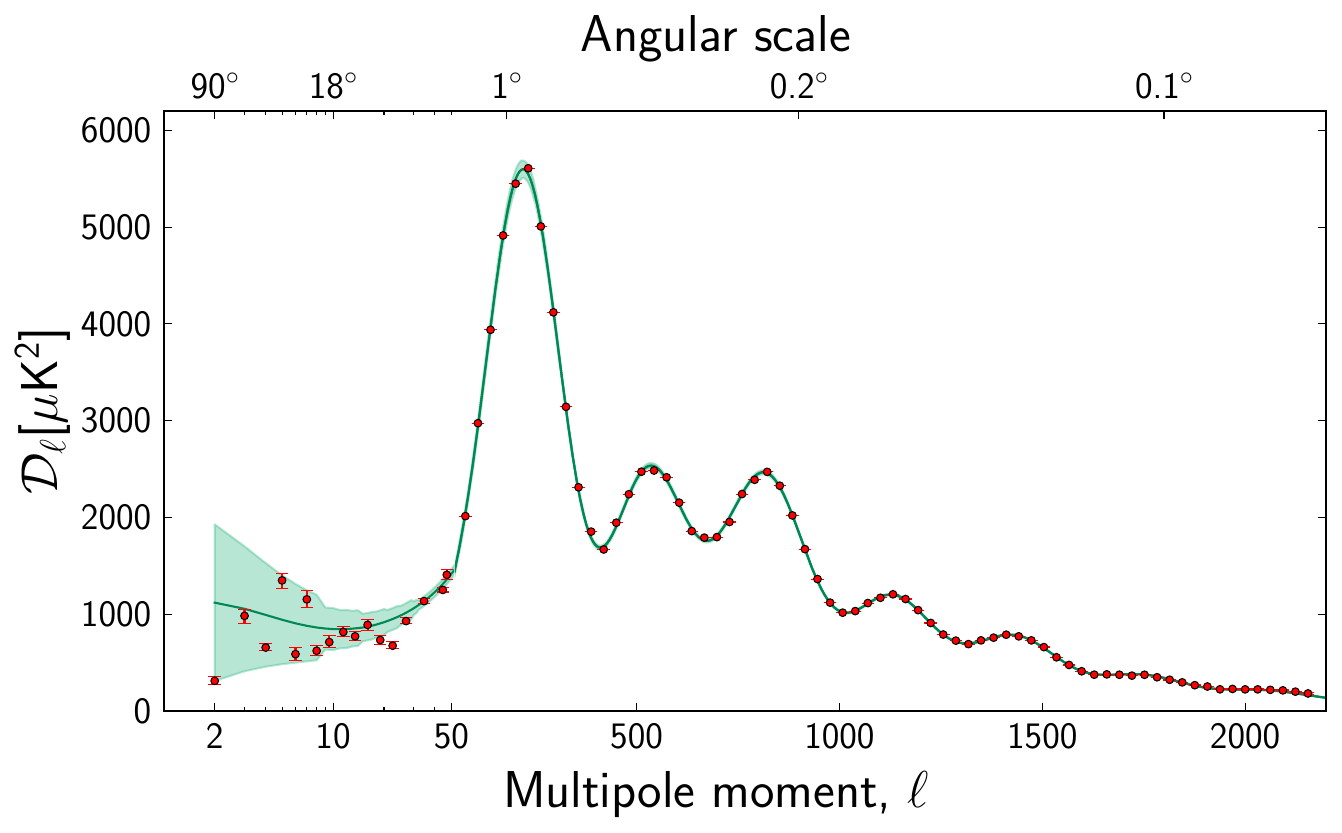}
\caption{The temperature angular power spectrum of the CMB, estimated from the SMICA \Planck\ map. The model plotted is the one labelled [Planck+WP+highL] in \citet{planck2013-p11}.  The shaded area around the best-fit curve represents cosmic variance, including the sky cut used.  The error bars on individual points do not include cosmic variance.  The horizontal axis is logarithmic up to $\ell = 50$, and linear beyond. The vertical scale is $\ell(\ell+ 1)C_l/2\pi$. The binning scheme is the same as in Fig.~\ref{fig:cmb_planck}. }
\label{fig:cmb_smica}
\end{figure}

\subsubsection{Lensing likelihood}
\label{sec:LensLike}

From the measurement of the lensing power spectrum described in Sect.~\ref{sec:CMBLike} and plotted in Fig.~\ref{fig:phispec}, we construct a simple Gaussian likelihood in eight bins of $C_L^{\phi\phi}$ between $40 \le L \le 400$. The bin size ($\Delta L = 45$) is such that we maintain some parameter leverage from the power spectrum shape information, while reducing the covariance between bins enough to neglect it. We analytically marginalize over the beams, diffuse point sources, and first order bias uncertainty and include them in the covariance. The cosmological uncertainty on the normalization is accounted for by a first-order correction.  Our power spectrum measurement constrains the lensing potential power spectrum to a precision of $\pm4\,\%$, corresponding to a $2\,\%$ constraint on the overall amplitude of matter fluctuations ($\sigma_8$).  The construction of the lensing likelihood is described in \cite{planck2013-p12}, and its cosmological implications are discussed in detail in \cite{planck2013-p11}.

\section{Astrophysical products}
\label{sec:AstroProds}

The generation of the \Planck\ astrophysical products is outlined in Fig.~\ref{FigFGProducts}. 

\begin{figure*}
\centering
\includegraphics[width=180mm]{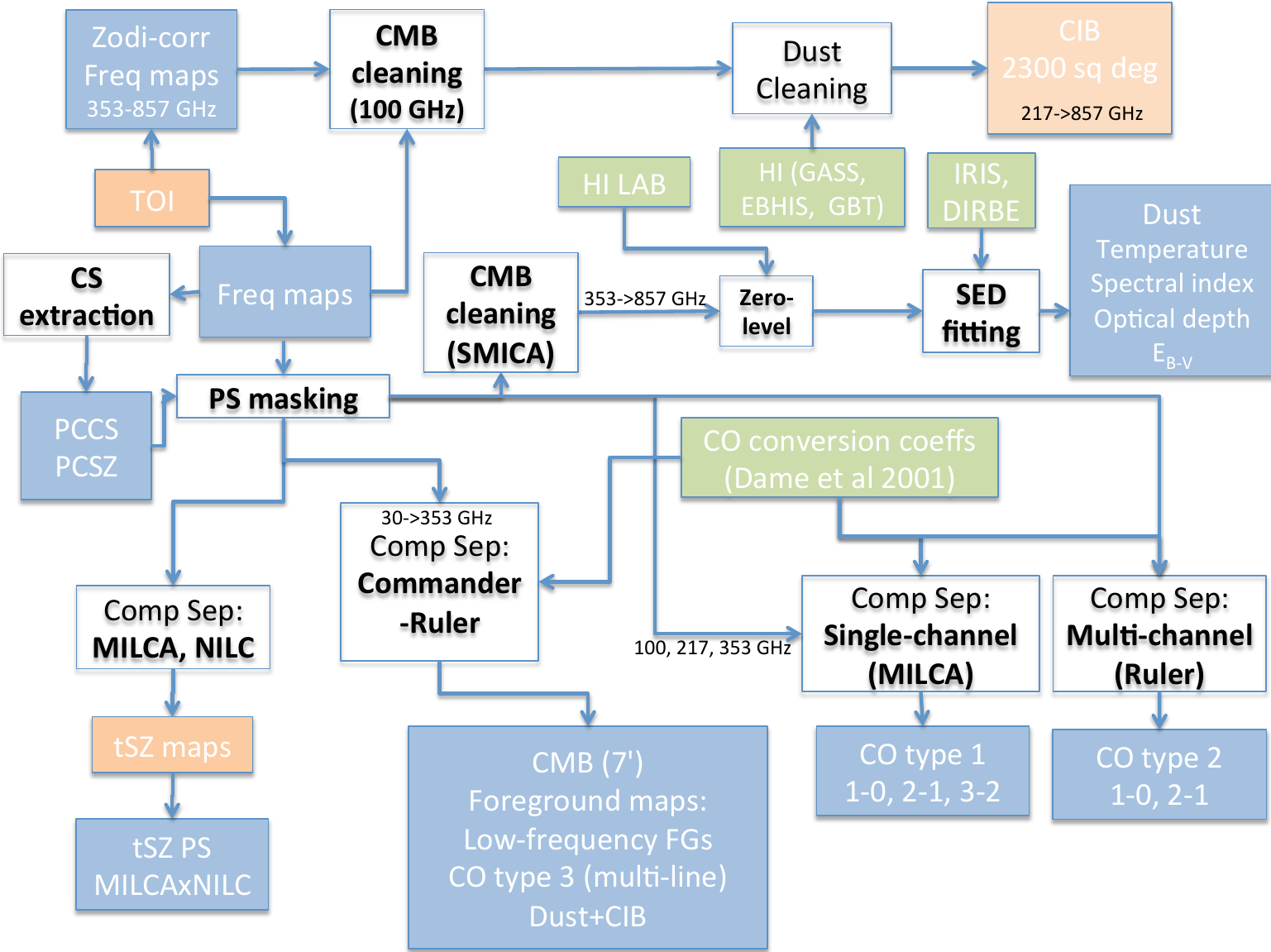}
\caption{Outline of the generation of astrophysical products being delivered by \Planck, in blue. Products in green are external; products in orange are not being delivered in the current release. Each product delivered is accompanied by specific data characterizing it (not shown on the diagram).  The CIB analysis uses maps corrected for zodiacal emission beween 353 and 857\,GHz;  for simplicity this is not reflected in the diagram.}
\label{FigFGProducts}
\end{figure*}

\subsection{The \Planck\ catalogues}
\label{sec:PCCS}

Many types of compact sources are found in the \Planck\ data, including quasars, radio galaxies, infrared galaxies, clusters of galaxies, \ion{H}{ii} regions, and young star-forming regions.  The \Planck\ Early Release Compact Source Catalogue (ERCSC), published in January 2011 \citep{planck2011-6.2}, included lists of compact sources extracted independently at each frequency, an early list of sources detected via the Sunyaev-Zeldovich effect (the ESZ), and a list of Galactic cold cores selected by temperature.  The \Planck\ Catalogue of Compact Sources (PCCS; \citealt{planck2013-p05}) and the \Planck\ Catalogue of Sunyaev-Zeldovich sources  (PSZ; \citealt{planck2013-p05a}) significantly expand the ERCSC and the ESZ.

\subsubsection{Main catalogue}
 
The PCCS contains sources detected over the entire sky, including both Galactic and extragalactic objects.  The PCCS differs from the ERCSC in its extraction philosophy: whereas the ERCSC emphasized high reliability suitable for follow-up especially with the short-lived {\it Herschel} telescope, the PCCS emphasizes completeness. Additional data, different selection processes, and improvements in calibration and mapmaking (Sects.~\ref{sec:GainCal} and \ref{sec:mapmaking}) result in greater depth and more sources compared to the  ERCSC.

Sources were extracted from the 2013 frequency maps (Sect.~\ref{sec:FreqMaps}) using a Mexican Hat Wavelet algorithm \citep{2006MNRAS.370.2047L, 2006MNRAS.369.1603G}.  The MTXF algorithm \citep{herranz2008, herranz2009} was used for validation and characterization.  

Most sources were observed at least three times during the 15.5\,month observing period.  The source selection for the PCCS is made on the basis of SNR. However, the  properties of the background in the maps vary substantially depending on frequency and part of the sky.  Up to 217\,GHz, the CMB is the dominant source of confusion at high Galactic latitudes. At higher frequencies, confusion from Galactic foregrounds dominates the noise at low Galactic latitudes, while fluctuations in the cosmic infrared background dominate at high Galactic latitudes.  The SNR has therefore been adapted to the different frequencies.  Specifically, we use two detection thresholds at frequencies above 353\,GHz, one in the brightest 52\,\% of the sky (called the ``Galactic zone''), and a different one in the cleanest 48\,\% of the sky (called the ``extragalactic zone'').  This strategy ensures interesting depth and good reliability in the extragalactic zone, but also high reliability in the Galactic zone.  The actual thresholds are listed in Table \ref{tab:pccs}.

\begin{table*}
\begingroup
\newdimen\tblskip \tblskip=5pt
\caption{characteristics of the Planck Compact Source Catalogue (PCCS).}
\label{tab:pccs}
\nointerlineskip \vskip -3mm \footnotesize
\setbox\tablebox=\vbox{
\newdimen\digitwidth
\setbox0=\hbox{\rm 0}
\digitwidth=\wd0
\catcode`*=\active
\def*{\kern\digitwidth}
\newdimen\signwidth
\setbox0=\hbox{+}
\signwidth=\wd0
\catcode`!=\active
\def!{\kern\signwidth}
\halign{\hbox to 2.0in{#\leaderfil}\tabskip 2.2em &
 \hfil#\hfil& 
 \hfil#\hfil& 
 \hfil#\hfil& 
 \hfil#\hfil& 
 \hfil#\hfil& 
 \hfil#\hfil& 
 \hfil#\hfil& 
 \hfil#\hfil& 
 \hfil#\hfil\tabskip=0pt\cr
\noalign{\doubleline}
\omit&\multispan9\hfil Channel [GHz]\hfil\cr
\noalign{\vskip -1pt}
\omit&\multispan9\hrulefill\cr
\noalign{\vskip 2pt}
\omit\hfil Characteristic\hfil& 30& 44& 70& 100& 143& 217& 353& 545& 857\cr
\noalign{\vskip 3pt\hrule\vskip 5pt}
{\bf Frequency} [GHz]& 28.4& 44.1& 70.4& 100.0& 143.0& 217.0& 353.0& 545.0& 857.0\cr
\noalign{\vskip 3pt}
{\bf Wavelength} [$\mu$m]& 10561& 6807& 4260& 3000& 2098& 1382& 850& 550& 350\cr
\noalign{\vskip 3pt}
{\bf Beam FWHM}$^{\rm a}$ [arcmin]& 32.38& 27.10& 13.30& 9.65& 7.25& 4.99& 4.82& 4.68& 4.33\cr
\noalign{\vskip 3pt}
\omit\bf SNR threshold\hfil\cr
\noalign{\vskip 3pt}
\hglue 2em Full sky& 4.0& 4.0& 4.0& 4.6& 4.7& 4.8&\ldots&\ldots&\ldots\cr
\hglue 2em Extragactic zone$^{\rm b}$&\ldots&\ldots&\ldots&\ldots&\ldots&\ldots& 4.9& 4.7& 4.9\cr
\hglue 2em Galactic zone$^{\rm b}$&\ldots&\ldots&\ldots&\ldots&\ldots&\ldots& 6.0& 7.0& 7.0\cr
\noalign{\vskip 3pt}
\omit\bf Number of sources\hfil\cr
\noalign{\vskip 3pt}
\hglue 2em Full sky&    1256& 731& 939& 3850& 5675& 16070& 13613& 16933& 24381\cr
\hglue 2em $|b|>30\deg$&*572& 258& 332& *845& 1051& *1901& *1862& *3738& *7536\cr
\noalign{\vskip 3pt}
\omit\bf Flux densities\hfil\cr
\noalign{\vskip 3pt}
\hglue 2em Minimum$^{\rm c}$ [mJy]&   461& *825& 566& 266& 169& 149& 289& 457& 658\cr
\hglue 2em 90\,\% completeness [mJy]& 575& 1047& 776& 300& 190& 180& 330& 570& 680\cr
\hglue 2em Uncertainty [mJy]&         109& *198& 149& *61& *38& *35& *69& 118& 166\cr
\noalign{\vskip 3pt}
{\bf Position uncertainty}$^{\rm d}$ [arcmin]& 1.8& 2.1& 1.4& 1.0& 0.7& 0.7& 0.8& 0.5& 0.4\cr
\noalign{\vskip 5pt\hrule\vskip 3pt}}}
\endPlancktablewide
\tablenote {{\rm a}} {\tt FEBeCoP} band-averaged effective beam: $\hbox{FWHM}_{\rm eff}= \sqrt{\frac{\Omega_{\rm eff}}{2\pi}8\log{2}}$, where $\Omega_{\rm eff}$ is the {\tt FEBeCoP} band-averaged effective solid angle (see \citealt{planck2013-p02d} and \citealt{planck2013-p03c} for a full description of the \Planck\ beams). This table shows the exact values that were adopted for the PCCS.  In constructing the PCCS, we used a value of the effective FWHM for the LFI channels that is slightly different (by $\ll1$\,\%) from the final values specified in \citet{planck2013-p02d} paper. A  correction will be made in later versions of the catalogue.\par
\tablenote {{\rm b}} See text.\par
\tablenote {{\rm c}} Minimum flux density of the catalogue at $|b|>30\deg$, after excluding the faintest 10\,\% of sources.\par
\tablenote {{\rm d}} Positional uncertainty derived by comparison with PACO sample up to 353\,GHz and with Herschel samples in the other channels.\par
\endgroup

\end{table*}

Because the properties of the sky vary so widely from low to high frequencies, the PCCS contains more than one estimate of the flux density of each source. The choice of the most accurate measure to use depends on frequency and foreground surface brightness as well as the solid angle subtended by the source: these choices are discussed in detail in \citet{planck2013-p05}.

The PCCS has been subject to both external and internal validation.  At the three lowest frequencies, it is possible to validate most source identifications, completeness, reliability, positional accuracy and in some cases flux-density accuracy using external data sets, particularly large-area radio surveys. Such ``external validation" was undertaken using the following catalogues and surveys: 1)~the full sky NEWPS catalogue, based on \WMAP\ results \citep{LopezCaniego2007,2009MNRAS.392..733M};  2)~the southern hemisphere AT20G catalogue at 20\,GHz \citep{2010MNRAS.402.2403M}; and 3)~the northern hemisphere (where no large-area, high-frequency survey covering the frequency range of AT20G is available)  CRATES catalogue \citep{2007ApJS..171...61H}.  These catalogues have similar frequency coverage and source density as the PCCS.

The higher (HFI) frequency channels have been validated through an internal Monte-Carlo quality assessment process that injects simulated sources into both real and simulated maps.  For each channel, the quality of the detection, photometry, and astrometry are calculated for multiple detection codes. The results are summarized by the completeness and reliability of the catalogue. Completeness is a function of the intrinsic flux density, the selection threshold applied to the detection (SNR), and location.  The reliability of extragalactic sources is a function only of the detection SNR. The reliability of sources detected within cirrus clouds is relatively lower because of the higher probability to detect fluctuations of the structure of the diffuse interstellar medium rather than actual individual sources. The quality of photometry and astrometry is assessed through direct comparison of detected position and flux density parameters.  Comparisons have also been performed with ACT (Gralla and members of the ACT team, in preparation), {\it Herschel}-SPIRE \citep{2010A&A...518L...3G}, and H-ATLAS \citep{2010PASP..122..499E}, as discussed in \citet{planck2013-p05}. 

Table \ref{tab:pccs} summarizes the characteristics of the PCCS.  The sources detected by \Planck\ are dominated in number at frequencies up to 217\,GHz by radio galaxies (synchrotron emission), and at frequencies above 217\,GHz by infrared galaxies (thermal dust emission), in agreement with previous findings \citep{planck2011-1.10,planck2011-1.10sup,planck2011-6.1,planck2012-VII} based on the ERCSC. %; Planck Collaboration VII 2013).          %,planck2012-VII}. %the last ref breaks hyperref!
The large spectral range covered by LFI and HFI gives a unique view of the two populations and their relative weight as a function of frequency, e.g., through the evolution of the spectral indices \citep{planck2013-p05}. 

 Many low-frequency extragalactic sources are variable; however, considering typical variability timescales and amplitudes, the temporal observing pattern of \Planck, and the sensitivity of the PCCS, a clear detection of variability is likely only for the brightest sources. Recent efforts in this direction can be found in \cite{Chen2013, Kurinsky2013}.  Similar work along these lines has not yet been attempted with the PCCS, but will be pursued with the full-mission dataset in the future.

\subsubsection{Cluster catalogue}
\label{sec:PSZ}

The PSZ \citep{planck2013-p05a} is deeper and six times larger than the ESZ \citep{planck2013-p05a}.  It includes 1227~sources found by three  SZ-detection algorithms down to a SNR of 4.5, distributed over 83.7\,\% of the sky (Table \ref{tab:valid_sum}, Fig.~\ref{fig:sz_dist}).  The  SZ detections were validated using existing X-ray, optical, and near-infrared data, and with a multi-frequency follow-up programme. A total of 861~SZ detections are associated with previously known or newly confirmed clusters, of which 178 are new \planck-discovered clusters. The remaining 366 cluster candidates have not yet been followed up, and are divided into three classes according to their estimated reliability, i.e., the probability that they are real clusters.  Only 142 are in the lowest reliability class.

The information derived from the validation of the \Planck\ SZ detections and included in the released catalogue, in particular the SZ-based mass estimate, provides high value to the catalogue, and will make it a reference for studies of cluster physics. Considering that only a small fraction of the new \Planck\ cluster candidates have been followed up with other observatories to date, it will also motivate multi-wavelength follow-up efforts.

\begin{table*}
\begingroup
\newdimen\tblskip \tblskip=5pt
\caption{Summary of the  classification of PSZ sources, based on external validation and confirmation from follow-up observations.  Previously known clusters can be found in the catalogues indicated. Confirmations from follow-up do not include the observations performed by the \Planck\ collaboration to measure missing redshifts of known clusters. Confirmation from archival data covers X-ray data from Chandra, {\it XMM}, and ROSAT PSPC pointed observations only.  References to individual source catalogues mentioned in this Table can be found in \cite{planck2013-p05a}. In a number of cases, NED and SIMBAD were used to obtain supporting information such as redshifts, and the original references are listed in \cite{planck2013-p05a}.}
\label{tab:valid_sum}
\nointerlineskip \vskip -3mm \footnotesize
\setbox\tablebox=\vbox{
\newdimen\digitwidth
\setbox0=\hbox{\rm 0}
\digitwidth=\wd0
\catcode`*=\active
\def*{\kern\digitwidth}
\newdimen\signwidth
\setbox0=\hbox{+}
\signwidth=\wd0
\catcode`!=\active
\def!{\kern\signwidth}
\halign{\hbox to 4.5cm{#\leaderfil}\tabskip=2em&
   \hfil#\hfil& 
   \hfil#\hfil\tabskip=0pt& 
   #\hfil\tabskip=0pt\cr
\noalign{\doubleline}
\omit\hfil Category\hfil&$N$&&\omit\hfil Comment\hfil\cr
\noalign{\vskip 3pt\hrule\vskip 3pt}
Previously known&*683&$\Biggl\{$&$\vcenter{\hbox{472\quad \hbox to 0.5in{X-ray:\strut\hfil}   MCXC meta-catalogue\hfill}
                                           \hbox{182\quad \hbox to 0.5in{Optical:\strut\hfil} Abell, Zwicky, SDSS\hfill}
                                           \hbox{*16\quad \hbox to 0.5in{SZ:\strut\hfil}      SPT, ACT\hfill}
                                           \hbox{*13\quad \hbox to 0.5in{Misc:\strut\hfil}    NED \& SIMBAD\hfill}}$\cr
\noalign{\vskip 15pt}
New confirmed&*178&&\hbox{***\quad XMM, ENO, WFI, NTT, AMI, SDSS}\cr
\noalign{\vskip 15pt}
New candidate&*366&   $\Biggl\{$&$\vcenter{\hbox{*54\quad High reliability}
                                           \hbox{170\quad Medium reliability}
                                           \hbox{142\quad Low reliability}}$\cr
\noalign{\vskip 15pt}
\omit&\hrulefill\cr
\bf Total \Planck\ SZ catalogue&\bf 1227\cr
\noalign{\vskip 5pt\hrule\vskip 3pt}}}
\endPlancktable
\endgroup
\end{table*}

Using an extended sub-sample of the \Planck\ SZ clusters with high-quality XMM-Newton data, the scaling relation between SZ and X-ray properties has been reassessed and updated \citep{planck2013-p05a}.  With better quality data and thus higher precision, we show excellent agreement between SZ and X-ray measurements of the intra-cluster gas properties. The mean of $Y_{500}$ to $Y_{\mathrm{X}}$ is very well constrained with a precision of $2.5\,\%$, $\log(Y_{SZ}/Y_X)=-0.027\pm0.010$.

To date, a total of 813 \Planck\ clusters have measured redshifts, ranging from $z=0.01$ to nearly 1, with two thirds of the clusters lying below $z=0.3$. For the clusters with redshifts, we have used the Compton $Y$ measure to estimate masses, which range between $\sim$0.1 and $1.6\times 10^{15}$\,\Msolar. 

Except at low redshifts, the \Planck\ cluster distribution exhibits a nearly redshift-independent mass limit.  Owing to its nearly
mass-limited selection function and its all-sky observations, \Planck\ detects new clusters in a region of the mass-redshift plane that is
sparsely populated by the RASS catalogues \citep[e.g.,][]{boe00,boe04,bur07,ebe07}. Furthermore, \Planck\ has the unique capability of detecting the most massive clusters, $M\ge5\times 10^{14}$\,\Msolar, at high redshifts, $z\ge 0.5$.  Such clusters, in the exponential tail of the cluster mass function, are the best clusters for cosmological studies. 

The \Planck\ catalogue of SZ sources serves to define samples for cosmological studies. A first step in this direction consists of the selection of a sub-sample consisting of 189 clusters detected above a SNR of 7 and with measured redshifts (see 
Sect.~\ref{sec:SZsci}), which has been used to constrain cosmological parameters \citep{planck2013-p15}.

\begin{figure}[t]
\begin{center}
\includegraphics[width=0.5\textwidth]{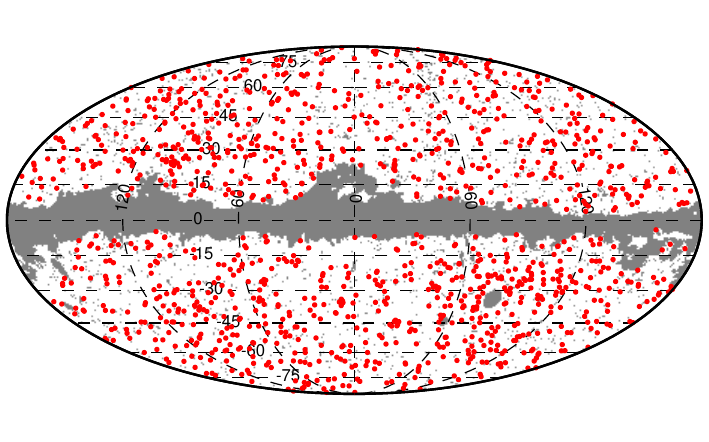}
\caption{ Sky distribution of the {1227} clusters and candidates (red dots), in a Mollweide projection with the Galactic plane horizontal and centered at zero longitude. The small grey dots show the positions of masked point sources, and the grey shading shows the mask used to exclude the Magellanic clouds and the Galactic plane. The mask covers 16.3\,\%\ of the sky.}
\label{fig:sz_dist}
\end{center}
\end{figure}

\subsection{Diffuse emission}
\label{sec:DiffFGs}

Eight types of diffuse foreground have been identified that must be removed or controlled for CMB analysis: dust thermal emission; dust anomalous emission (from rotating small grains); three CO rotational lines; free-free emission; synchrotron emission; the CIB not fully correlated between frequencies; Sunyaev-Zeldovich secondary CMB distortions; and the background of unresolved radio sources.  Some of these have been independently extracted from the \Planck\ maps, in some cases with the help of external information (e.g., ancillary maps tracing specific astrophysical components, or prior knowledge of the spectral energy distribution of the power spectrum),  to yield astrophysically meaningful foregrounds.  We describe first the astrophysical foregrounds that result from CMB-directed component separation, which are combinations of physically distinct components, and then describe several physical foregrounds extracted with the help of additional data or specialized techniques.

\subsubsection{Foregrounds from CMB component separation}
\label{sec:CommFGs}

\Planck's wide frequency range allows us to use component separation techniques based on \Planck\ data alone to derive tight constraints on several astrophysical components in addition to the primary CMB fluctuations.  In \citet{planck2013-p06}, we present individual maps of: 1)~a combined high-frequency component accounting for Galactic thermal dust emission and fluctuations in the CIB; 2)~Galactic carbon monoxide (CO {\sc Type~3}, see \ref{sec:COmaps}); and 3)~a combined low-frequency component accounting for synchrotron, free-free, and anomalous microwave emission (AME, almost certainly emission from microscopic spinning dust grains; Fig. \ref{FigCommanderProds}).  Only \Planck\ frequencies between 30 and 353\,GHz are used, as the systematics of 545 and 857\,GHz are less well understood than those of the lower ones, the dust signal is already strongly dominant at 353\,GHz, and higher frequencies may include emission from higher-temperature dust.

The astrophysical components are derived by Bayesian parameter estimation, in which an explicit parametric model is fitted to the raw observations within the bounds of physically motivated priors.  This process is implemented in two stages,  referred to as ``Commander" and ``Commander-Ruler" respectively.  In the first, the frequency maps are smoothed to a common resolution of 40\arcm\ FWHM, pixelized at $N_{\rm side}=256$), and all model parameters (spectral parameters and signal amplitudes) are fitted jointly by the CMB Gibbs sampler Commander.  The CMB samples produced by this code form the basis of the low-$\ell$ \Planck\ CMB temperature likelihood, as described in \citet{planck2013-p08}.  In the second stage, the spectral parameters from the low-resolution fit are formally upgraded to $N_{\rm side}=2048$, and full-resolution CMB and thermal dust amplitudes are estimated using a generalized least-squares fit \citep{planck2013-p06}.  The low-resolution foreground components are limited by the angular resolution of the lowest frequencies, and the products from the low-resolution stage are therefore retained for these.

The thermal dust emission is modelled as a one-component greybody with free emissivity, $\beta_{\rm dust}$, and temperature, $T_{\rm d}$, per pixel.  Since we only include frequencies up to 353\,GHz here, the dust temperature is largely unconstrained in our fits, and we therefore adopt a tight prior around the commonly accepted mean value of $T_{\rm d} = 18 \pm 0.05$\,\hbox{K}.  The only reason it is not fixed completely at 18\,K is to allow for modelling errors near the Galactic center.  The dust emissivity prior is set to $\beta_{\rm d} = 1.5\pm0.3$, where the mean is once again set by a dedicated MCMC run.  Because the CIB is a statistically isotropic signal, it can be well-approximated by a dominant monopole plus a small spatially varying fluctuation, analogous to the CMB itself.  Further, as shown by Planck Collaboration XXX (2013), the CIB frequency spectrum follows very nearly a one-component greybody function with similar parameters to those of the Galactic thermal dust component.  The current model therefore accounts for the CIB component without introducing an additional and dedicated CIB parameter, simply by first subtracting off a best-fit monopole at each frequency, and, second, through the free dust parameters (amplitude and spectral parameters) for each pixel.  The dust amplitude map shown in Fig. \ref{FigCommanderProds} therefore contains both Galactic thermal dust and extragalactic CIB fluctuations.  The CIB fluctuations are strongly sub-dominant everywhere on the sky except in the very cleanest regions.

The CO component is modelled in terms of a mean amplitude per pixel at 100\,GHz, which is then extrapolated to 217 and 353\,GHz through a spatially constant overall factor per frequency called a ``line ratio.''  To minimize parameter degeneracies, the default line ratios are estimated in a dedicated preliminary run using only the pixels with the highest CO-to-thermal-dust ratio (0.5\,\% of the sky), and holding the dust and synchrotron spectral indices spatially constant. This gives line ratios of 0.60 at 217\,GHz and 0.30 at 353\,GHz, in excellent agreement with those derived from the ``{\sc Type-2}'' analysis of \citet[Sect.~8.2.4]{planck2013-p03a}.  The resulting ``{\sc Type~3}'' CO map maximizes the SNR, and it can be seen as a discovery map for new potential CO clouds; however; it combines the information about the single line transition into a total intensity one.

The low-frequency component is modelled as a straight power-law in intensity units, with a free spectral index per pixel.  We adopt a prior of $\beta=-3 \pm 0.3$ for the low-frequency spectral index; this is mostly relevant only at high Galactic latitudes where the SNR is low and the dominant foreground component is expected to be synchrotron emission.  In the signal-dominated AME and free-free regions at low latitudes, the data are sufficiently strong that the prior becomes irrelevant.  Figure~\ref{FigCommanderProds} shows the resulting component maps; Fig.~\ref{FigCommanderSpectra} shows their frequency dependence at high latitudes.

\begin{figure*}
\centering
\includegraphics[width=180mm]{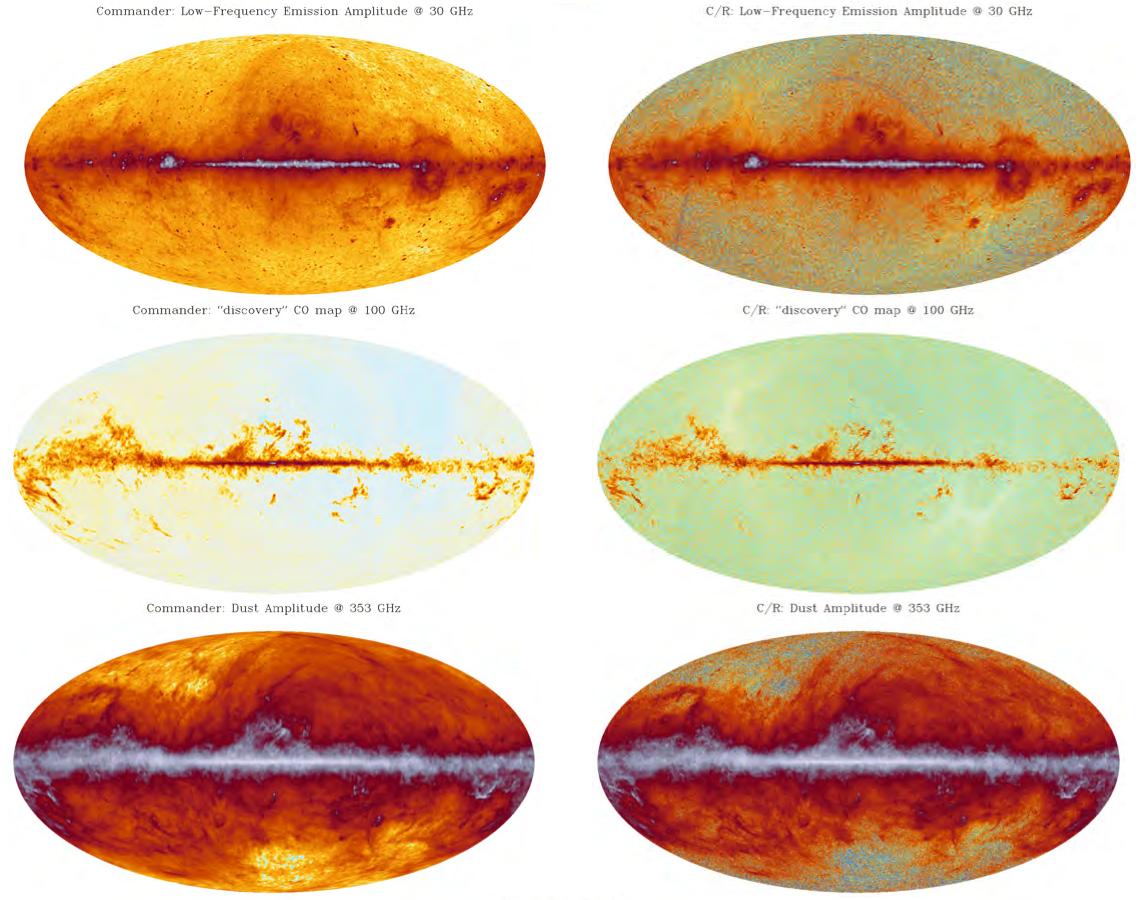} \\ 
\includegraphics[width=100mm]{Figs/KMG_component_separation_maps_colorbar.pdf} 
\caption{Foreground maps produced by {\tt Commander} (left, resolution 1\deg) and by {\tt Commander-Ruler} (right, resolution 7\arcm).  See Sect.~\ref{sec:CommFGs} for details.  {\it Top\/}: amplitude of low-frequency foregrounds (synchrotron, free-free, and anomalous microwave emission) at 30\,GHz . {\it Middle\/}: integrated intensity of CO {\sc Type~3}. {\it Bottom\/}: amplitude of high-frequency foregrounds (dust thermal emission and the cosmic infrared background) at 353\,GHz . }
\label{FigCommanderProds}
\end{figure*}

\begin{figure*}
\centering
\includegraphics[width=180mm]{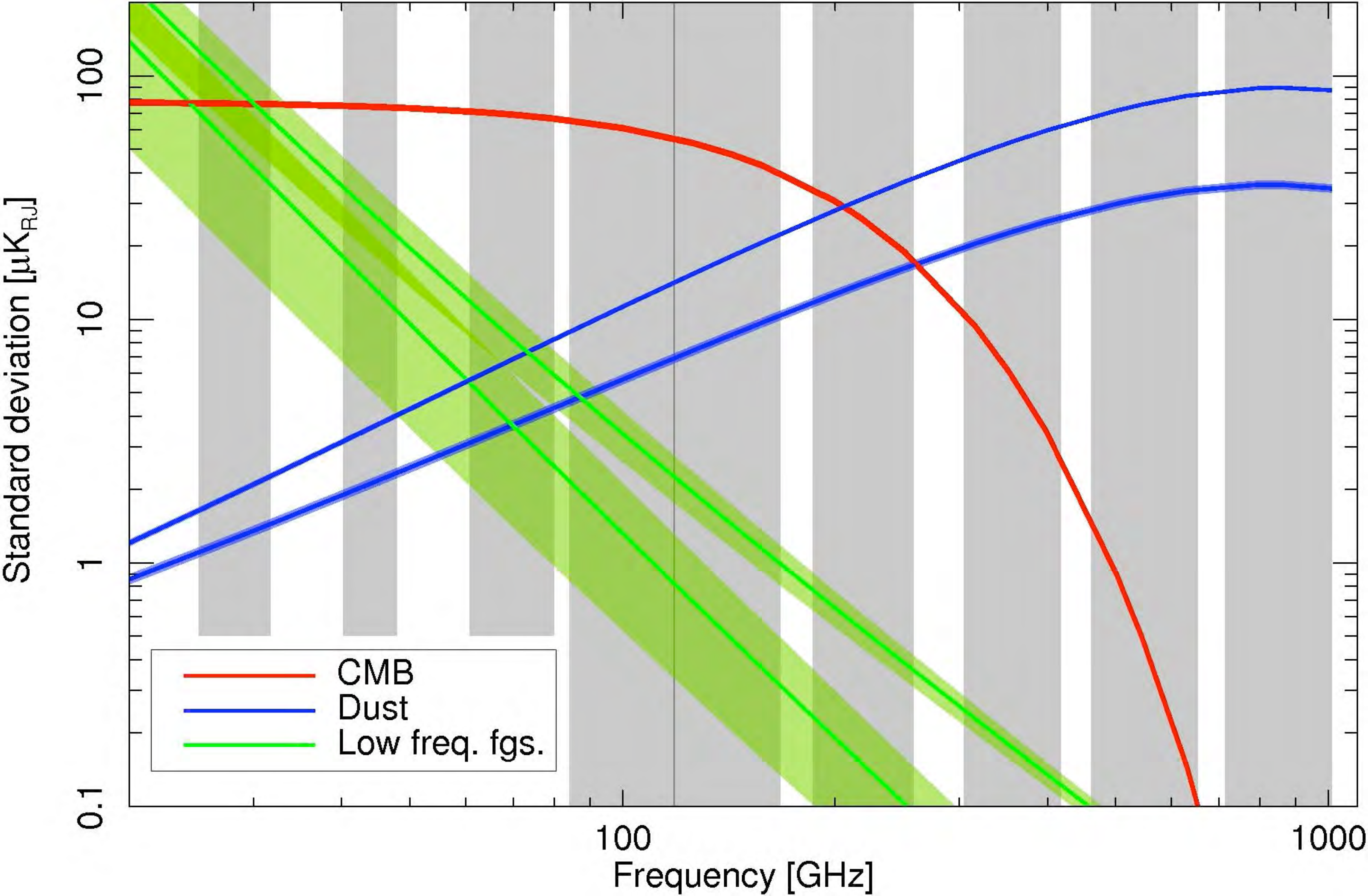} 
\caption{RMS fluctuations (in $\mu K_{RJ}$) of the diffuse components: the ``high-frequency Galactic component'', in blue; the ``low-frequency Galactic component'', in green, and the CMB in red, as obtained by the {\tt Commander} algorithm (Section~\ref{sec:CommFGs}). The maps have been smoothed to 35\arcm\ resolution.  The rms is calculated at high galactic latitudes, outside two masks covering 23\,\%  and 42\,\% of the sky around the galactic plane.  Point sources in the PCCS at 30 and 353\,GHz have also been masked. An uncertainty envelope is indicated, estimated from the difference of the half-ring-based foreground maps. For reference, the average rms level in the plane (i.e., using the complement of these masks) is $20\times$ higher. The grey shaded areas represent the frequency coverage of the Planck bands, based on equivalent-noise bandwidths.} 
\label{FigCommanderSpectra}
\end{figure*}

In the 2013 data release, we adopt the posterior mean as our signal estimate, and the posterior RMS as the corresponding uncertainty.  Mean and RMS maps are provided for each signal component and for each per-pixel spectral parameter.  Two caveats are in order regarding use of these products for further scientific analysis.  First, significant systematic uncertainties are associated with several of these estimates.  One example is the correlated HFI noise that is seen clearly in the thermal dust emissivity map; the products presented here do not take into account spatially correlated noise.  Second, the full posterior is significantly non-Gaussian due to the presence of non-Gaussian spectral parameters and the positivity amplitude prior, as well as strongly correlated between components.  The mean and RMS maps provided in this data release should therefore be understood as a convenient representation of the full posterior, rather than a precise description of each component; if very high statistical precision is required, one should use instead the original ensemble of individual Monte Carlo samples.

\begin{figure}
\centering
\includegraphics[width=88mm]{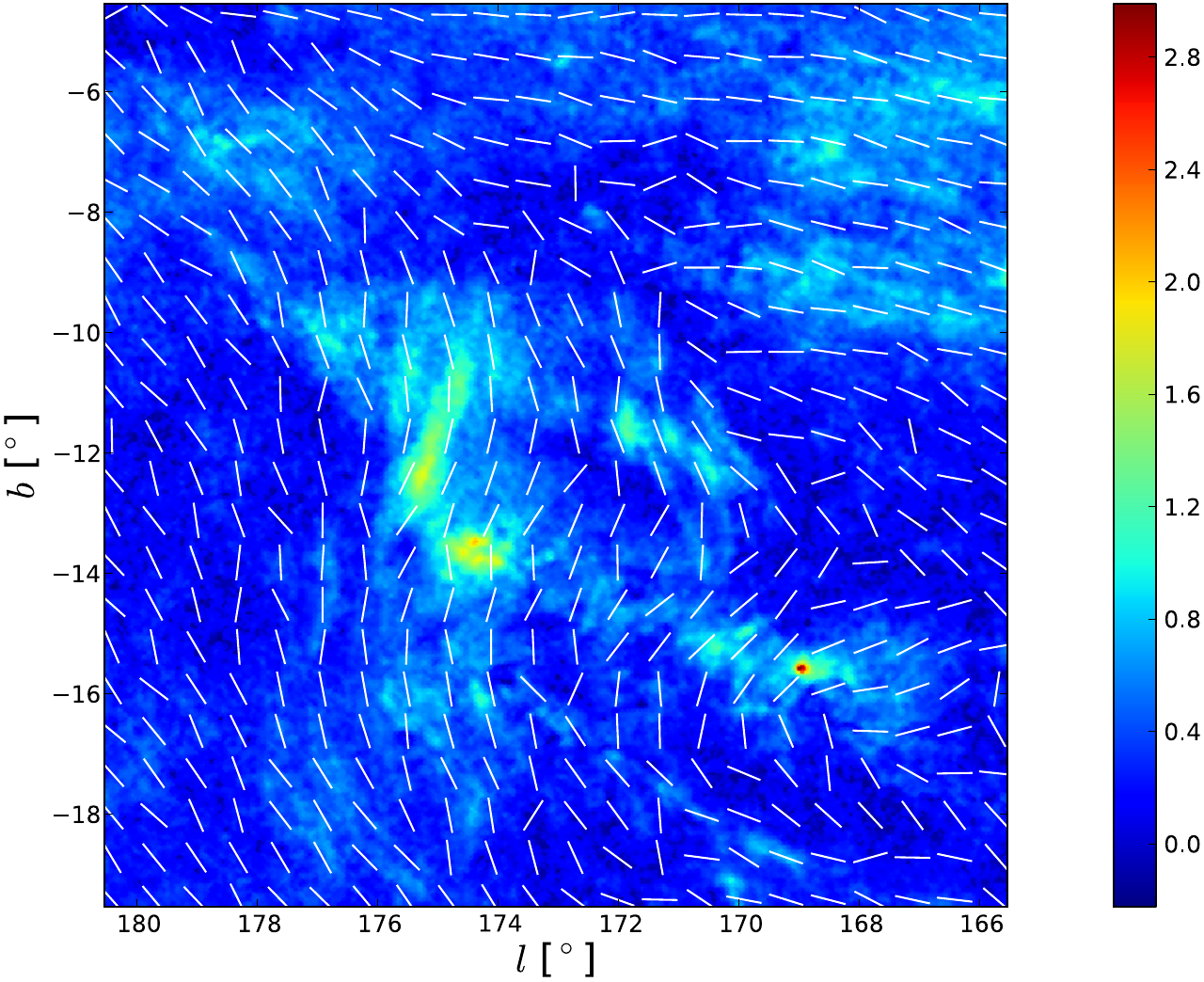}
\caption{Polarized intensity at 353\,GHz (in mK$_{CMB}$) and polarization orientation indicated as segments of uniform length, in the Taurus region.}
\label{FigTaurusP}
\end{figure}

\subsubsection{Thermal emission from Galactic dust}
\label{sec:Dust}

The CMB fades toward higher frequencies, whereas the thermal dust emission spectrum increases, and so dust becomes the dominant signal
at submillimetre wavelengths.  \Planck\ has multifrequency sensitivity in the ``dust channels'' covering the spectral range where this transition occurs up to 857\,GHz, for which the angular resolution is highest (Table~\ref{table:Instrument_performance}).  Dust emission is seen extending to high Galactic latitude in the wispy ``cirrus'' represented in bluish tones in Fig.~\ref{FigAllSky}.

Understanding the frequency dependence and spatial fluctuations of the intensity and polarization from thermal dust is important both in separating this foreground from the CMB and as an all-sky measure of column density.  This has motivated the development of the \Planck\ dust model (\citealt{planck2013-p06b}; Fig.~\ref{FigFGProducts}).  The use of a range of available ancillary data distinguishes this model from the dust component extracted using component separation techniques and frequencies below 353\,GHz, described in Sect.~\ref{sec:CommFGs}.

The \Planck\ dust model is based on the maps at 353, 545, and 857\,GHz, plus the {\it IRAS} 100\micron\ (3000\,GHz) data from the {\it IRIS} product \citep{IRIS05}.  As described in \cite{planck2013-p03}, the zero level of these maps can be set such that there is no dust emission where there is no atomic \ion{H}{i} gas column density (according to the LAB survey, \citealt{LAB05}).  

The dust temperature $T_{\rm d}$ is determined by multi-frequency fitting of a greybody SED for optically thin thermal dust, $I_\nu = \tau_\nu  B_{\nu}(T_{\rm d})$, where $\tau_\nu$ is the dust optical depth of the column of material and $B_{\nu}(T_{\rm d})$ is the Planck function for $T_{\rm d}$.  In the \Planck\ dust model there are three parameters, $T_{\rm d}$, $\tau_{353}$ at 353\,GHz, and $\beta$, the exponent of the assumed power-law frequency dependence of $\tau_\nu$. Conversion of $\tau_{353}$ to $E_{\rm B-V}$ is established by correlating the submillimetre optical depth with SDSS reddening measurements of quasars, a very similar approach to the one adopted by \citet{schlegel1998}.

All-sky maps of $T_{\rm d}$ and $\tau$ from \Planck\ were first presented by \citet{planck2011-7.0} using a fixed $\beta$.  In the new \Planck\ dust model \citep{planck2013-p06b}, the maps of $T_{\rm d}$ and $\tau_{353}$ are at 5\arcm\ resolution, while $\beta$ is estimated at 35\arcm.  This provides a much more detailed description of the thermal dust emission than the \cite{finkbeiner1999} model, which assumed constant $\beta$ and used a $T_{\rm d}$ map with an angular resolution of several degrees.

The high resolution of \Planck\ is a major improvement that results in a much more detailed mapping of column density structure, especially in denser regions of the ISM where the equilibrium temperature of big dust grains changes on small angular scales due to attenuation of the radiation field and also, it appears, to changes in the intrinsic dust opacity and its ability to emit.  Because of the increase in sensitivity, better control of systematic effects, and the combination of four intensity maps at 5\arcm\ resolution spanning the peak of the SED and into the Rayleigh-Jeans region, the new \Planck\ $E_{\rm B-V}$ product also provides a precise estimate of the dust column density even in the diffuse ISM where $T_{\rm d}$ does not vary as strongly on small scales.

\subsubsection{Polarized emission from Galactic dust}
\label{sec:DustPol}

As described in Sect.~\ref{subsec:polstatus}, \Planck\ polarization data are in a less mature state than temperature data, especially at low multipoles.   Nevertheless, strong polarized synchrotron and thermal dust emission from the Galaxy can already be imaged with high significance.  
A first set of Galactic polarization papers will be published shortly.  These papers will report results on the degree of dust polarization ($P/I$) over the whole sky, comparisons with maps of synchrotron polarization and Faraday rotation, the structure of the Galactic magnetic field and its coupling with interstellar matter, as well as turbulence in the diffuse interstellar medium (ISM).   With an angular resolution of 5\arcm, the maps also reveal the magnetic field structure in molecular clouds and star forming regions, and can be used to study which grains contribute to the observed polarization, where in the ISM they are aligned with the Galactic magnetic field, and with what efficiency.  

One highlight from these maps is the high degree of polarization of the dust emission from the diffuse interstellar medium, in many locations reaching $P/I > 15\,\%$ at 353\,GHz.  These studies will also address the statistics of $P/I$  and $\phi$ for selected fields towards nearby molecular clouds (e.g., Fig.~\ref{FigTaurusP}), the relationship with MHD simulations, the spectral dependence of the polarized emission, and a comparison of the degree of polarization at submillimetre  and visible (from stellar observations) wavelengths.

\subsubsection{CO extraction}
\label{sec:COmaps}

Rotational line emission from carbon monoxide (CO) in the interstellar medium is present in all HFI bands except 143\,GHz, most significantly  from the $J$=1$\rightarrow$0 (115\,GHz), $J$=2$\rightarrow$1 (230\,GHz), and $J$=3$\rightarrow$2 (345\,GHz) transitions. CO emission arises from the denser parts of the interstellar medium, and is concentrated at low and intermediate Galactic latitudes.  Three approaches to estimating CO emission have been evaluated and are described in \cite{planck2013-p03a}.  

{\sc Type~1} maps rely on the fact that each bolometer has a different responsivity to CO, largely due to its specific bandpass shape. The transmission at the CO transition frequency has been accurately measured on the ground, and can also be estimated by comparison to surveys made with dedicated observatories. Knowledge of the relative bolometer spectral response allows extraction of each CO line independently of the others and of any ancillary data. Being extracted from single-bolometer data, these maps have relatively low SNR; however, they are not affected by contamination from other channels, and thus can be used for unbiased removal of CO from frequency maps.

{\sc Type~2} maps are obtained using a multi-frequency component separation approach. Three maps are combined to extract independently maps of the $J$=1$\rightarrow$0 (100, 143, and 353\,GHz channels) and $J$=2$\rightarrow$1 (143, 217, and 353\,GHz channels) emission. Because channels are combined, the spectral behaviour of other foregrounds (free-free and dust) is needed as extra constraints to allow clean CO extraction. The {\sc Type~2} maps have higher SNR than the {\sc Type~1} maps, at the cost of residual contamination from other diffuse foregrounds.  These maps constitute a unique product for astrophysics that provides the excitation ratio for all parts of the sky where the CO intensity is strong enough.

The {\sc Type~3} map is determined by fixing the $J$=2$\rightarrow$1/$J$=1$\rightarrow$0 and $J$=3$\rightarrow$2/$J$=2$\rightarrow$1 line ratios.  The map is extracted using the full Commander-Ruler component separation pipeline (Sect.~\ref{sec:CommFGs}).  This yields a map of combined CO emission with very high SNR that can be used as a sensitive finding chart for low-intensity diffuse CO emission over the whole sky.  The line ratios can be determined from ground-based observations, or from a first iteration of the component separation algorithm with simplified assumptions; the latter is the route used by \Planck.

All three types of CO map have been extensively cross-checked internally.

\subsubsection{All-sky Sunyaev-Zeldovich emission}
\label{sec:SZDiff}

Using specialized component separation methods on \Planck\ maps from 100 to 857\,GHz, we have constructed an all-sky map that includes an estimate of the Compton $y$ parameter from the thermal Sunyaev-Zeldovich (tSZ) effect \citep{SZ}.  The angular power spectrum of this map, corrected for residual foregrounds, gives the first estimate of the tSZ power spectrum over a range of multipoles from $\ell=60$ to $\ell=1000$ \citep{planck2013-p05b}.   Diffuse thermal dust emission is the major foreground contaminant in the map at low multipoles ($\ell < 30$); at high multipoles ($\ell > 500$) the clustered CIB and unresolved radio and infrared point sources dominate.  At intermediate scales the tSZ dominates. The measured tSZ spectrum is composed of the total signal from resolved clusters in the \planck\ catalogue of SZ sources and from unresolved clusters of galaxies and hot diffuse gas (Sect.~\ref{sec:SZsci}).

The tSZ Compton parameter map is not released as a product to the community because of the complexity of the foreground contamination, which does not allow for direct use without a companion foreground model.

\subsection{Unresolved foregrounds at high galactic latitudes}
\label{SummaryFGs}

\begin{figure}
\centering
\includegraphics[width=0.5\textwidth]{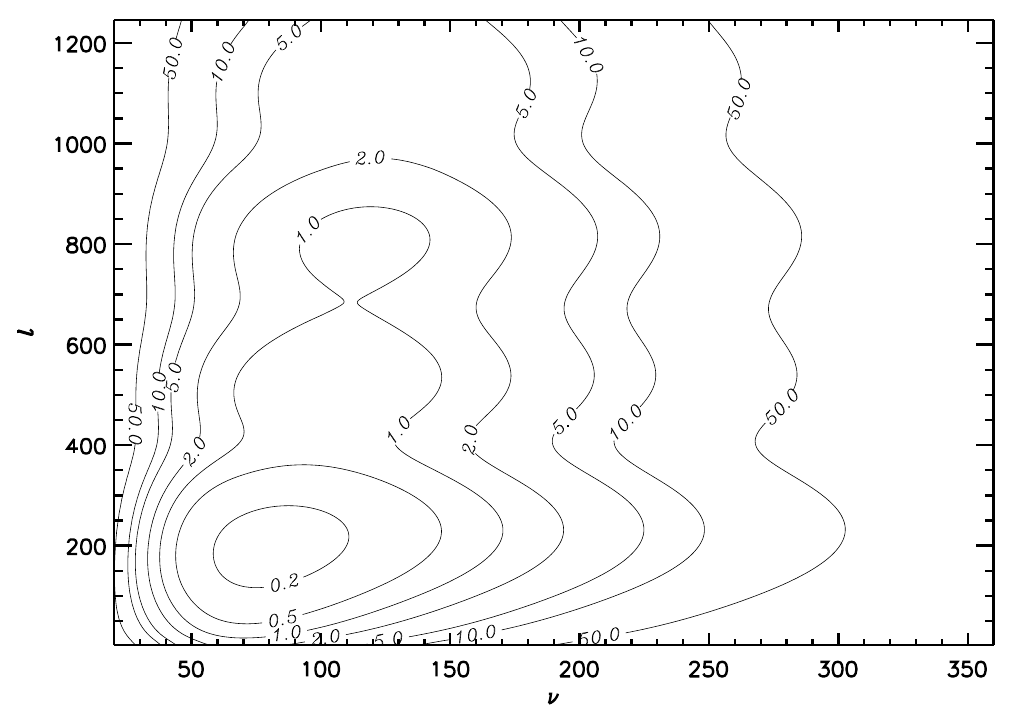} 
\caption{The contours show total foreground levels at high galactic latitudes ($f_{\rm sky} = 0.60$) as a fraction of the average CMB level in the frequency-multipole moment plane.  Minimum foreground contamination is found at $\ell\sim 200$ at frequencies between 70 and 100\,GHz. The ripples in $\ell$ follow the pattern of acoustic peaks in the CMB power spectrum.}
\label{FigFgContour}
\end{figure}

\begin{figure*}
\centering
\includegraphics[width=88mm]{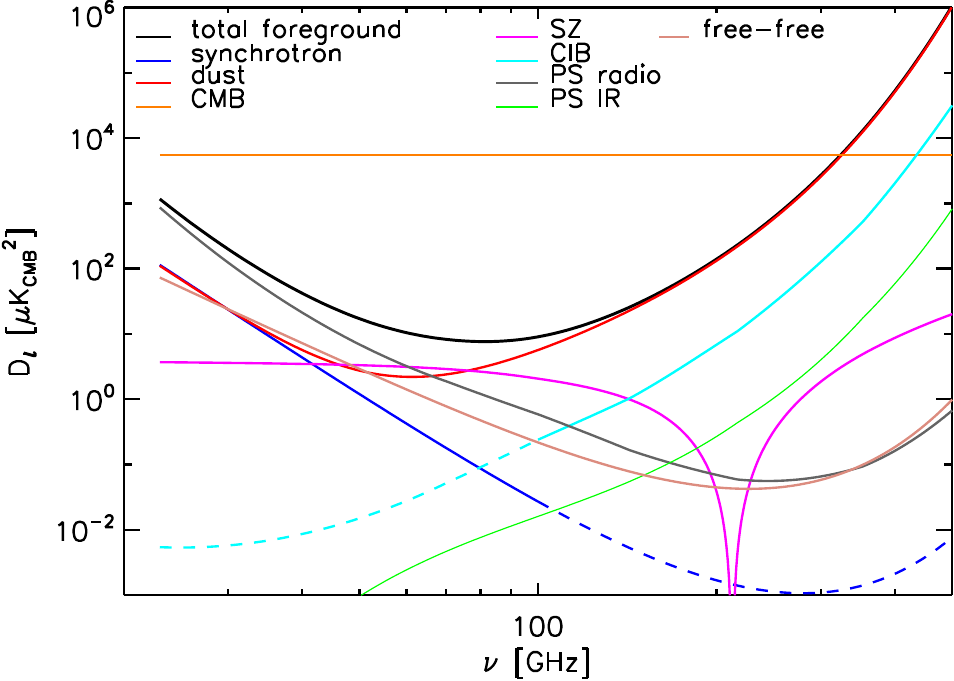}\hglue 3mm
\includegraphics[width=88mm]{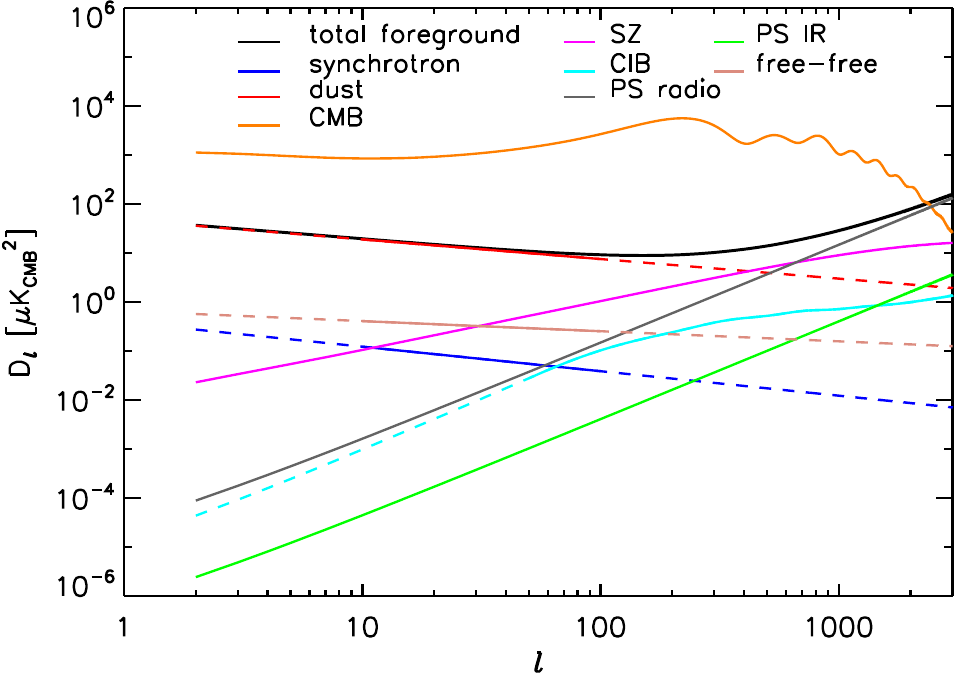}
\caption{{\it Left\/}: Frequency spectra of individual diffuse foregrounds at high galactic latitudes, estimated at $\ell = 200$, the angular scale at which CMB fluctuations are greatest and foreground fluctuations are relatively the least important (see Fig.~\ref{FigFgContour}).  The horizontal line gives the level of the CMB. {\it Right\/}: Angular power spectra of various foregrounds at 100\,GHz, along with the best-fit \Planck\ CMB spectrum. In both panels, solid lines show where the spectra are estimated from data, and dashed lines are extrapolations.}
\label{FigFgPS}
\end{figure*}

In this section we summarize our understanding of unresolved foregrounds at high galactic latitudes, largely determined by \Planck\ and described in previous sections.  Figure~\ref{FigFgContour} shows the total level of foregrounds as a fraction of the rms amplitude of CMB fluctuations over 56\,\% of the high-galactic-latitude sky (the mask used is G56, defined in \cite{planck2013-p08}). 
The foregrounds included are:
\begin{itemize}

\item Galactic thermal dust emission as modelled in \citet[see also Sect.~\ref{sec:Dust} and \citealt{planck2013-p06b}]{planck2013-XVII}. The angular power spectrum is based on the 353\,GHz map, from which the SMICA CMB has been subtracted, and has an $\ell^{-2.4}$ dependence. The SED of this component includes Anomalous Microwave Emission, which is determined by correlation with thermal dust.

\item Free-free emission as modelled in \cite{planck2013-XVII}. The angular power spectrum is normalized at 100\,GHz, and has an $\ell^{-2.2}$ and $\nu^{-4.28}$ dependence.

\item Synchrotron emission based on the Haslam 408\,MHz template, with $C_\ell \propto \ell^{-2.5} \nu^{-6}$. 

\item The Cosmic Infrared Background as determined in \citet[see also Sect.~\ref{sec:CIB}]{planck2013-pip56}, extrapolated analytically for $\ell <$ 50 and $\nu <$ 100 GHz.

\item Diffuse SZ emission, as modelled in \citet[see also Sect.~\ref{sec:SZsci}]{planck2013-p05b}.

\item Poisson noise from unresolved radio sources, based on the \Planck\ PCCS at 100\,GHz with a flux density threshold of 256\,mJy, extrapolated in frequency as $\nu^{-1.2}$.

\item Poisson noise from dusty galaxies based on the CIB model of \cite{planck2013-pip56}.

\end{itemize}

Figure~\ref{FigFgPS} shows two representative cuts through Fig.~\ref{FigFgContour}, at 100\,GHz and $\ell = 200$, close to where foregrounds are least important relative to the CMB (see Fig. \ref{FigFgContour}).  These figures show that at the lowest and highest CMB frequencies (70 and 217\,GHz), diffuse Galactic foregrounds dominate at low-$\ell$, with a minimum in the 70 to 100\,GHz range for the best half of the sky (see Figs.~\ref{FigCommanderSpectra} and \ref{FigFgContour}). 
Residual extragalactic foregrounds, composed of synchrotron emission from radio sources, SZ emission from clusters, and thermal dust emission from galaxies, have a more complicated behaviour in $\ell$ and frequency, as they contain Poisson terms and, for SZ and the CIB, correlated terms as well.  Over 60\,\% of the sky, extragalactic foregrounds  dominate at all multipoles larger than $\sim$200.  The frequency at which we find the minimum relative contribution of all foregrounds with respect to the CMB shifts from 70\,GHz at $\ell\sim 100$ to larger than 100\,GHz  at $\ell >$ 200 (with less than 1\,\% contamination up to $\ell\sim 800$).

%============================================================
\label{sec:CMBLensCosmo}
\section{\Planck\ 2013 cosmology results}
\label{sec:CMBcosmology}

\subsection{Parameter estimation, lensing, and inflation}
\label{sec:CMBPar}

Since their discovery, anisotropies in the CMB have contributed significantly to defining our cosmological model and measuring its key parameters.  The standard model of cosmology is based upon a spatially flat, expanding Universe whose dynamics are governed by General Relativity and dominated by cold dark matter and a cosmological constant ($\Lambda$).  The seeds of structure have Gaussian statistics and form an almost scale-invariant spectrum of adiabatic fluctuations.

\Planck's measurements of the cosmological parameters derived from the nominal mission are presented and discussed in
\citet{planck2013-p11}.  The most important conclusion from this paper is the excellent agreement between the \planck\ temperature spectrum at high $\ell$ and the predictions of the $\Lambda$CDM model.  All of our current observations can be fit remarkably well by a six parameter $\Lambda$CDM model (see Table~\ref{tab:params} for definitions), and we provide strong constraints on deviations from this model.  The best-fit cosmological parameters are not affected by foreground modeling uncertainties, and the best-fit model provides an excellent fit to the spectra from \planck, ACT, and SPT (see Fig.~\ref{FigAllCMBExps}). The ACT and SPT spectra are from \cite{2013arXiv1302.1841C}.

\begin{table*}[tmb]
\begingroup 
\newdimen\tblskip \tblskip=5pt
\caption{Cosmological parameters used in our 6-parameter model. For each, we give the symbol, prior range, and summary definition.
The top block contains parameters with uniform priors that are varied in the
MCMC chains. The ranges of these priors are listed in square brackets.
The lower blocks define various derived parameters. A more complete table of parameters can be found in \cite{planck2013-p11}. Best-fit values are given in Table \ref{tab:LCDMparams}. }
\label{tab:params}
\vskip -5mm
\footnotesize 
\setbox\tablebox=\vbox{ %
\newdimen\digitwidth 
\setbox0=\hbox{\rm 0}
\digitwidth=\wd0
\catcode`*=\active
\def*{\kern\digitwidth}
\newdimen\signwidth
\setbox0=\hbox{+}
\signwidth=\wd0
\catcode`!=\active
\def!{\kern\signwidth}
\halign{\hbox to 2.7cm{#\leaderfil}\tabskip=0.4cm&
   \hfil#\hfil\tabskip=0.6cm&
   #\hfil\tabskip=0pt\cr
\noalign{\doubleline}
\omit\hfil Parameter\hfil&\omit\hfil Prior range\hfil&\omit\hfil Definition\hfil\cr
\noalign{\vskip 3pt\hrule\vskip 3pt}
$\Omega_{\mathrm{b}} h^2$&   $[0.005, 0.1]$& Baryon density today\cr
$\Omega_{\mathrm{c}}  h^2$& $[0.001, 0.99]$& Cold dark matter density today\cr
$100\theta_{\mathrm{MC}}$&    $[0.5, 10.0]$& $100\,{\times}$ approximation to $\rstar/D_{\rm A}$ (CosmoMC)\cr
$\tau$&                       $[0.01, 0.8]$& Thomson scattering optical depth due to reionization\cr
$\ns$&                         $[0.9, 1.1]$& Scalar spectrum power-law index ($k_0 = 0.05$\,Mpc\mo)\cr
$\ln(10^{10}\As)$            & $[2.7, 4.0]$& Log power of the primordial curvature perturbations ($k_0 = 0.05$\, Mpc\mo)\cr
\noalign{\vskip 5pt\hrule\vskip 3pt}
$\Omega_\Lambda$&                          & Dark energy density divided by the critical density today\cr
$\sigma_8$&                                & RMS matter fluctuations today in linear theory\cr
$z_{\mathrm{re}}$&                         & Redshift at which Universe is half reionized\cr
$H_0$&                             [20,100]& Current expansion rate in $\rm{km}\, \rm{s}^{-1}\, Mpc^{-1}$\cr
Age/Gyr&                                 & Age of the Universe today (in Gyr)\cr
$100\theta_{\rm eq}$&                      & $100\,\times$ angular size of the comoving horizon at matter-radiation equality\cr
$r_{\mathrm{drag}}=\rs(z_{\mathrm{drag}})$&& Comoving size of the sound horizon at $z = z_{\mathrm{drag}}$\cr
\noalign{\vskip 3pt\hrule\vskip 3pt}}}
\endPlancktablewide
\endgroup
\end{table*}

In some cases we find significant changes compared to previous CMB experiments, as discussed in detail in \citet{planck2013-p11}.
In particular, when we compare models based on CMB data only we find that the \Planck\ best-fit model retrieves lower $\Omega_\Lambda$ (by $\sim$6\,\%), higher $\Omega_{\rm baryons}$ (by $\sim$9\,\%), and higher $\Omega_{\rm CDM}$ (by $\sim$18\,\%) than the corresponding WMAP9\footnote{We compare the model [\Planck\ +WP+highL] of Table 5 in \citet{planck2013-p11} with [WMAP+eCMB] of Table 4 of \citet{hinshaw2012}. } model. However, when adding BAO to both sets of data, the gap reduces by a factor of $\sim$3 in all three components.

\begin{figure*}
\centering
\includegraphics[width=0.9\textwidth]{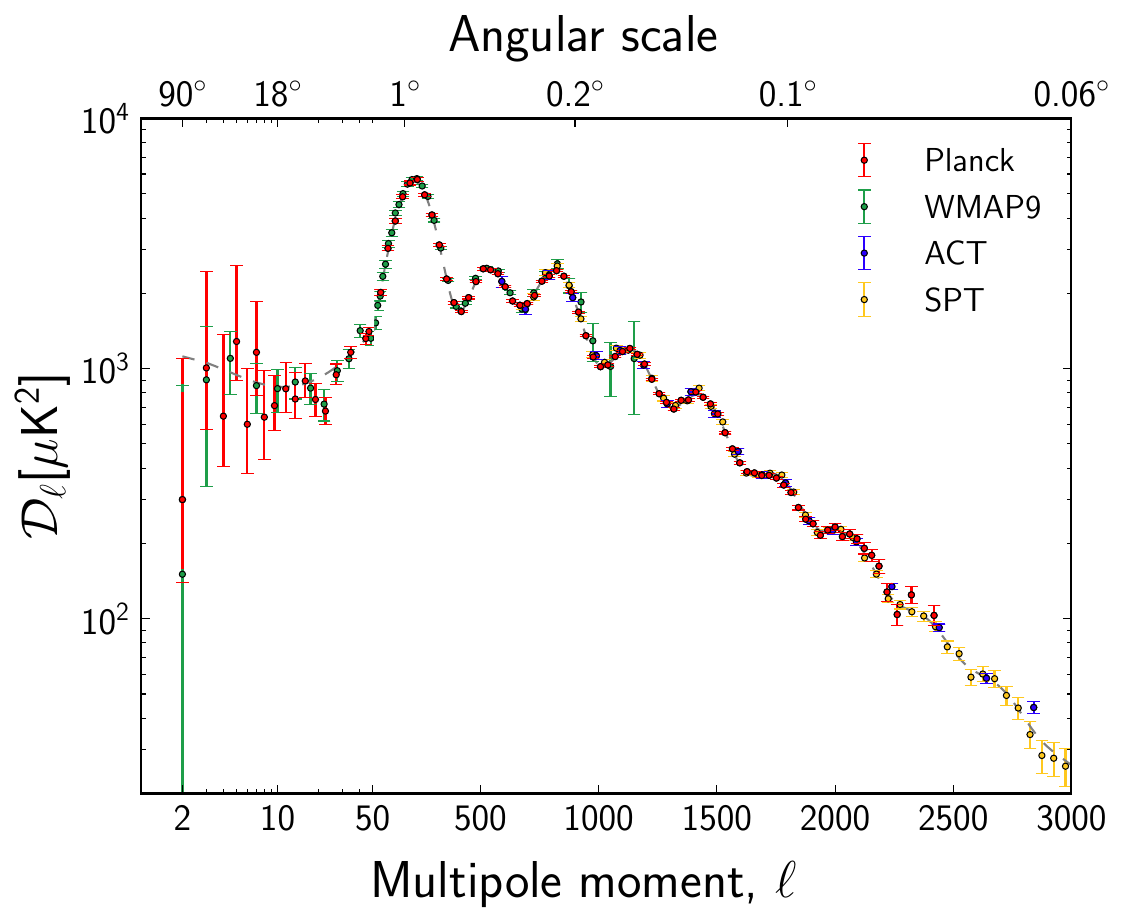}
\caption{Measured angular power spectra of \Planck, WMAP9, ACT, and SPT. The model plotted is \Planck's best-fit model including \Planck\ temperature, WMAP polarization, ACT, and SPT (the model is labelled [Planck+WP+HighL] in \citet{planck2013-p11}).  Error bars include cosmic variance. The horizontal axis is logarithmic up to $\ell = 50$, and linear beyond.  
}
\label{FigAllCMBExps}
\end{figure*}

Among the constraints that we determine, several are notable.  The angular size of the acoustic scale is determined to  $0.06\,\%$, as $\theta_\star=(1.19355\pm 0.00078)^\circ$, which leads to a $0.3\,\%$ constraint in the $\Omega_{\rm m}-h-\Omega_{\rm b}h^2$ subspace for $\Lambda$CDM models \citep[all confidence intervals are $68\,\%$;][]{planck2013-p11}.  For $\Lambda$CDM, the matter and baryon densities are well determined, with the latter being consistent with recent results from big-bang nucleosynthesis (BBN).  We find excellent consistency with BBN even in extensions to the six-parameter model.  The predictions of the baryon density from these two methods involve all of the known forces of nature, and this highly non-trivial consistency provides strong evidence for the universality of those laws.

Lensing of the CMB enters the \Planck\ parameter estimation results discussed in~\citet{planck2013-p11}  in two ways. First, the power spectrum of the temperature anisotropies is modified at the few percent level by lensing, with the primary effect being a smoothing of the acoustic peaks on angular scales relevant for \Planck. We  detect this smoothing effect at  $10\,\sigma$, and include it in our parameter constraints.  Second, the measurements of the power spectrum of the reconstructed gravitational lensing potential, described in~\citet{planck2013-p12} and Sect.~\ref{sec:CMBLensing}, can be combined with the main \Planck\ likelihood developed in~\citet{planck2013-p08} (see also Sect.~\ref{sec:Like}). The lensing power spectrum measurements condense the cosmological signal contained in the non-Gaussian 4-point function of the CMB anisotropies in a near-optimal way. Combining the lensing likelihood with the main \Planck\ likelihood is therefore equivalent to a joint analysis of the anisotropy power spectrum and that part of the 4-point function due to lensing.

The expected lensing power spectrum is tightly constrained in the six-parameter \lcdm\ model by the \Planck\ temperature power spectrum and the \WMAP\ low-$\ell$ polarization data. The best-fitting model predicts a lensing power spectrum in good agreement with the \Planck\ lensing reconstruction measurement, further validating the predictions of the \lcdm\ model, calibrated on the CMB fluctuations at $z\approx 1100$. These predictions include clustering and the evolution of the geometry at low redshift. We express the amplitude of the lensing power spectrum in terms of a phenomenological power spectrum amplitude parameter, $\Aphiphi$, which scales the theoretical 4-point function (due to lensing) at each point in parameter space. From \Planck's best-fit model, the expected value of this scaling parameter is 1.0; for the nominal mission we find $\Aphiphi = 0.99\pm 0.05$ (68\,\% CL;~\citealt{planck2013-p11}).  

Without the low-$\ell$ polarization data, and in the absence of lensing, the amplitude of the primordial power spectrum, $\As$, and the optical depth to reionization, $\tau$, would be degenerate with only the combination $\As e^{-2\tau}$ being well determined by the (unlensed) temperature power spectrum. However, lensing partially breaks this degeneracy, since the lensing power spectrum is independent of the optical depth. Combining the temperature power spectrum with the lensing likelihood, we determine $\tau = 0.089 \pm 0.032$ (68\,\% CL) \emph{from the temperature anisotropies alone}. This constraint is consistent, though weaker, than that from \WMAP\ polarization~\citep{2012arXiv1212.5226H}.
Importantly, the lensing route does not depend on the challenging issue of removing large-scale polarized emission from our Galaxy that is critical for the \WMAP\ measurement. At 95\,\% confidence, we can place a lower limit on the optical depth $\tau < 0.04$, which exceeds the value for instantaneous reionization at $z=6$, further supporting the picture that reionization is an extended process.

Beyond the six-parameter \lcdm\ model, the \Planck\ lensing measurements strengthen the evidence reported by ACT~\citep{2013arXiv1301.0824S} and SPT~\citep{2012ApJ...756..142V,2012arXiv1210.7231S} for dark energy from the CMB alone in models with spatial curvature. Closed models with low energy density in dark energy can be found that produce unlensed CMB power spectra nearly identical to the best-fitting \lcdm\ model. This ``geometric'' degeneracy is partially broken by lensing, since the closed models predict too much lensing power. Even without using the \planck\ lensing reconstruction, the $10\sigma$ detection of the smoothing of the temperature power spectrum allows \planck, used in combination with ACT and SPT at high-$\ell$ (to better constrain extragalactic foregrounds) and \WMAP\ large scale polarization, to break the geometrical degeneracy, and provides evidence for dark energy {\it purely from the CMB} \citep{planck2013-p12}. Adding the lensing likelihood, we constrain any departures from spatial flatness at the percent level: $\Omega_K = -0.0096^{+0.010}_{-0.0082}$ (68\,\% CL) for the same data combination, improving earlier CMB-only constraints~\citep{2012arXiv1210.7231S} by around a factor of two, and setting our determination of dark energy from temperature anisotropies data alone to  $\Omega_\Lambda = 0.67^{+0.027}_{-0.023}$ (68\,\% CL). Tighter constraints from the combination of \Planck\ and other astrophysical data are given in~\citet{planck2013-p11}.

Within the minimal, six-parameter model the expansion rate is well determined, independent of the distance ladder.  One of the most striking results of the nominal mission is that the best-fit Hubble constant $H_0=(67\pm 1.2)\;{\rm km}\,{\rm s}^{-1}\,{\rm Mpc}^{-1}$, is lower than that measured using traditional techniques, though in agreement with that determined by other CMB experiments
(e.g., most notably from the recent WMAP9 analysis where \citealt{Hinshaw:12} find $H_0 =  (69.7\pm 2.4)\; {\rm km}\,{\rm s}^{-1}\,{\rm Mpc}^{-1}$ consistent with the \planck\ value to within $\sim 1\,\sigma$).  \citet{Freedman:12}, as part of the {\it Carnegie Hubble Program}, use mid-infrared observations with the Spitzer Space Telescope to recalibrate secondary distance methods used in the HST Key Project.   These authors find $H_0=(74.3\pm 1.5 \pm 2.1)\;{\rm km}\,{\rm s}^{-1}\,{\rm Mpc}^{-1}$ where the first error is statistical and the second systematic.  A parallel effort by \citet{Riess:2011yx} used the Hubble Space Telescope observations of Cepheid variables in the host galaxies of eight SNe Ia to calibrate the supernova magnitude-redshift relation.  Their `best estimate' of the Hubble constant, from fitting the calibrated SNe magnitude-redshift relation is, $H_0=(73.8\pm 2.4)\;{\rm km}\,{\rm s}^{-1}\,{\rm Mpc}^{-1}$ where the error is $1\,\sigma$ and includes known sources of systematic errors.  At face value, these measurements are discrepant with the current \planck\  estimate at about the $2.5\,\sigma$ level.  This discrepancy is discussed further in \citet{planck2013-p11}.

Extending the Hubble diagram to higher redshifts, we note that the best-fit $\Lambda$CDM model provides strong predictions for the distance scale.  This prediction can be compared to the measurements provided by studies of Type Ia SNe and baryon acoustic oscillations (BAO).  Driven in large part by our preference for a higher matter density, we find mild tension with the (relative) distance scale inferred from compilations of SNe \citep{Conley:11,Suzuki:12}.  In contrast our results are in excellent agreement with the BAO distance scale compiled in \citet{Anderson:12} covering the redshift range 0.1 to 0.7.

\begin{table*}
\caption{\label{tab:LCDMparams} Cosmological parameter values for the \Planck-only best-fit  6-parameter \LCDM\ model (\Planck\ temperature data plus lensing) and for the \Planck\ best-fit cosmology including external data sets (\Planck\ temperature data, lensing, \WMAP\ polarization [\WP] at low multipoles, high-$\ell$ experiments, and BAO, labelled [Planck+WP+highL+BAO] in \citet{planck2013-p11}). The six parameters fit are above the line; those below are derived from the same model.  Definitions and units for all parameters can be found in Table \ref{tab:params} and \citet{planck2013-p11}.}
\begin{center}
\begingroup
\openup 5pt
\newdimen\tblskip \tblskip=5pt
\nointerlineskip
\vskip -6mm
\footnotesize
\setbox\tablebox=\vbox{
    \newdimen\digitwidth
    \setbox0=\hbox{\rm 0}
    \digitwidth=\wd0
    \catcode`"=\active
    \def"{\kern\digitwidth}
    \newdimen\signwidth
    \setbox0=\hbox{+}
    \signwidth=\wd0
    \catcode`!=\active
    \def!{\kern\signwidth}
\halign{
\hbox to 0.9in{$#$\leaderfil}\tabskip=3.0em&
  \hfil$#$\hfil\tabskip=2.0em&
  \hfil$#$\hfil\tabskip=3.0em&
  \hfil$#$\hfil\tabskip=2.0em&
  \hfil$#$\hfil\tabskip=0pt\cr
\noalign{\doubleline}
\omit&  \multispan2\hfil \Planck\ (CMB+lensing)\hfil&\multispan2\hfil \Planck+WP+highL+BAO\hfil\cr
\noalign{\vskip -3pt}
\omit&\multispan2\hrulefill&\multispan2\hrulefill\cr
 \omit\hfil Parameter\hfil&\omit\hfil Best fit\hfil&\omit\hfil 68\,\% limits\hfil&\omit\hfil Best fit\hfil&\omit\hfil 68\,\% limits\hfil\cr
\noalign{\vskip 3pt\hrule\vskip 5pt}
\Omega_{\mathrm{b}} h^2&  0.022242& 0.02217\pm 0.00033&  0.022161&0.02214\pm 0.00024\cr
\Omega_{\mathrm{c}} h^2&  0.11805&  0.1186\pm 0.0031&    0.11889& 0.1187\pm 0.0017\cr
100\theta_{\mathrm{MC}}&  1.04150&  1.04141\pm 0.00067&  1.04148& 1.04147\pm 0.00056\cr
\tau&                     0.0949&   0.089\pm 0.032&      0.0952&  0.092\pm 0.013\cr
n_\mathrm{s}&             0.9675&   0.9635\pm 0.0094&    0.9611&  0.9608\pm 0.0054\cr
\ln(10^{10} A_\mathrm{s})&3.098&    3.085\pm 0.057&      3.0973&  3.091\pm 0.025\cr
\noalign{\vskip 5pt\hrule\vskip 3pt}
\Omega_\Lambda&           0.6964&   0.693\pm 0.019&      0.6914&  0.692\pm 0.010\cr
\sigma_8&                 0.8285&   0.823\pm 0.018&      0.8288&  0.826\pm 0.012\cr
z_{\mathrm{re}}&          11.45&    10.8^{+3.1}_{-2.5}&  11.52&   11.3\pm 1.1\cr
H_0&                      68.14&    67.9\pm 1.5&         67.77&   67.80\pm 0.77\cr
\mathrm{Age}/\mathrm{Gyr}&13.784&   13.796\pm 0.058&     13.7965& 13.798\pm 0.037\cr
100\theta_\ast&           1.04164&  1.04156\pm 0.00066&  1.04163& 1.04162\pm 0.00056\cr
r_{\mathrm{drag}}&        147.74&   147.70\pm 0.63&      147.611& 147.68\pm 0.45\cr
\noalign{\vskip 5pt\hrule\vskip 3pt}}}
\endPlancktable
\endgroup
\end{center}
\end{table*}

The \planck\ data, in combination with polarization measured by \WMAP, high-$\ell$ anisotropies from ACT and SPT and other, lower redshift data sets, provide strong constraints on deviations from the minimal model.  The low redshift measurements provided by BAO allow us to break some degeneracies still present in the \planck\ data and significantly tighten constraints on cosmological parameters in these model extensions.  The ACT and SPT data help to fix our foreground model at high $\ell$.  The combination of these experiments provides our best constraints on the standard 6-parameter model; values of some key parameters in this model are summarized  in Table \ref{tab:LCDMparams}.

From an analysis of an extensive grid of models, we find no strong evidence to favour any extension to the base $\Lambda$CDM cosmology, either from the CMB temperature power spectrum alone or in combination with the \planck\ lensing power spectrum and other astrophysical datasets.  For the wide range of extensions that we have considered, the posteriors for extra parameters generally overlap the fiducial model within $1\,\sigma$.  The measured values of the $\Lambda$CDM parameters are relatively robust to the inclusion of different parameters, though a few do broaden significantly if additional degeneracies are introduced.  When the \planck\ likelihood does provide marginal evidence for extensions to the base $\Lambda$CDM model, this comes predominantly from a deficit of power (compared to the base model) in the data at $\ell<30$.

The primordial power spectrum is well described by a power-law over three decades in wave number, with no evidence for ``running'' of the spectral index.  The spectrum does, however, deviate significantly ($6\,\sigma$) from scale invariance, as predicted by most models of inflation (see below).  The unique contribution of \planck, compared to previous experiments, is that the departure from scale invariance is robust to changes in the underlying theoretical model.

We find  an effective number of neutrino-like relativistic degrees of freedom of $N_{eff} = 3.36\pm0.34$, compatible with the standard value of 3.046, implying no need 
 for extra relativistic species beyond the three species of (almost) massless neutrinos and photons.  The main effect of massive neutrinos is a suppression of clustering on scales larger than the horizon size at the non-relativisitic transition. This affects both $C_L^{\phi\phi}$, with a damping for $L>10$, and $C_\ell^{TT}$, reducing the lensing-induced smoothing of the acoustic peaks.  Using \Planck\ data in combination with polarization measured by \WMAP\ and high-$\ell$ anisotropies from ACT and SPT allows for a constraint of $\sum m_\nu < 0.66\,\mathrm{eV}$ (95\,\% CL) based on the [\planck+\WP+\highL] model.  Curiously, this constraint is weakened by the addition of the lensing likelihood $\sum m_\nu < 0.85\,\mathrm{eV}$ (95\,\% CL), reflecting mild tensions between the measured lensing and temperature power spectra, with the former preferring larger neutrino masses than the latter. Possible origins of this tension are explored further in~\cite{planck2013-p11} and are thought to involve both the $C_{L}^{\phi\phi}$ measurements and features in the measured $C_{\ell}^{TT}$ on large scales ($\ell < 40$) and small scales $\ell > 2000$ that are not fit well by the \lcdm+foreground model.  The SNR on the lensing measurement will improve with the full mission data, including polarization, and it will be interesting to see how this story develops.

The combination of large lever arm, sensitivity to isocurvature fluctuations, and non-Gaussianity makes \planck\ particularly powerful at probing inflation.  The constraints on inflationary models are presented in \citet{planck2013-p17} and are consistent with a single, weakly coupled, neutral scalar field driving the accelerated expansion and generating curvature perturbations.  The data can be explained using a field with a canonical kinetic term and slowly rolling down a featureless potential, and we find no compelling evidence calling for any extension to this simple explanation.

Of the models considered, those with locally concave potentials are favored and occupy most of the region in the $n_{\rm s}$-$r$\, plane allowed at 95\,\% confidence level (see Fig.~\ref{Fignsr}). Power law inflation, hybrid models driven by a quadratic term, and monomial large field potentials with a power larger than two lie outside the 95\,\% confidence contours.  The quadratic large field model, in the past often cited as the simplest inflationary model, is now at the boundary of the 95\,\% confidence contours of Planck +
WP + CMB high $\ell$ data.

\begin{figure*}
\centering
\includegraphics[width=180mm]{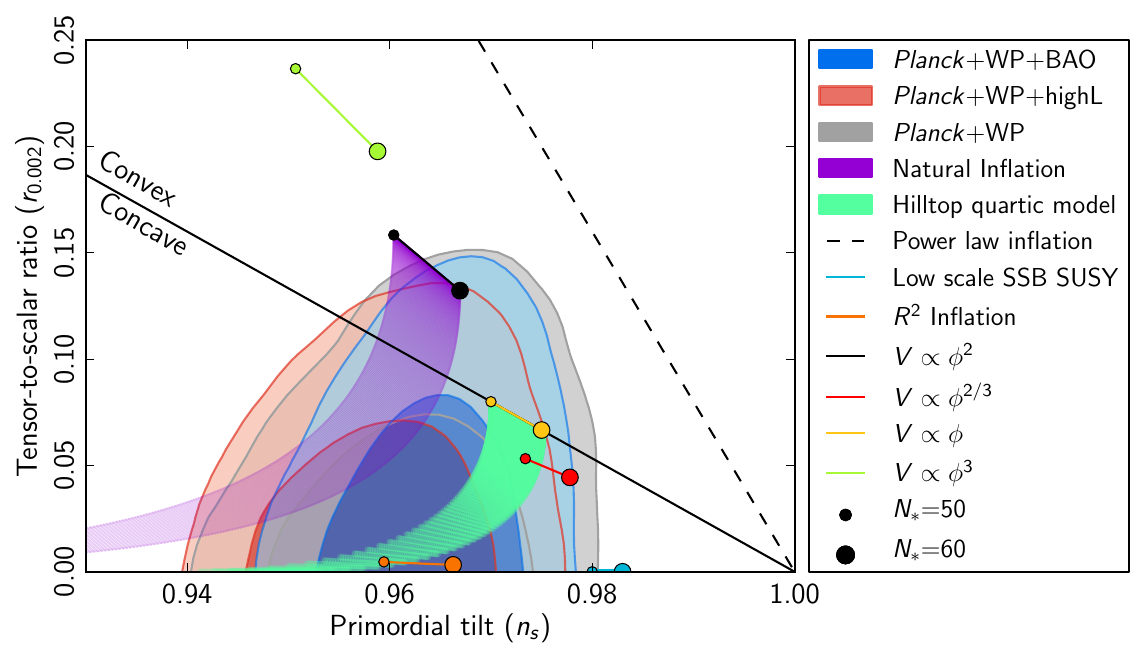}
\caption{Marginalized 68\,\% and 95\,\% confidence levels for $n_s$ (the scalar spectral index of primordial fluctuations) and $r_{0.002}$  (the tensor to scalar power ratio at the pivot scale $k = 0.002$\,Mpc\mo) from Planck+WP, alone and combined with high-$\ell$ and BAO data, compared to the theoretical predictions of selected inflationary models.}
\label{Fignsr}
\end{figure*}

The axion and curvaton scenarios, in which the CDM isocurvature mode is uncorrelated or fully correlated with the adiabatic mode, respectively, are not favored by \Planck, which constrains the contribution of the isocurvature mode to the primordial spectra at $k = 0.05$Mpc\mo\ to be less than 3.9\,\% and 0.25\,\% (95\,\% CL), respectively.

The \planck\ results come close to the tightest upper limit on the tensor-to-scalar amplitude possible from temperature data alone.
The precise determination of the higher acoustic peaks breaks degeneracies that have weakened earlier measurements.  The bound ($r<0.11$ at 95\,\% CL) implies an upper limit for the energy scale of standard inflation of $1.9\times 10^{16}\,$GeV (95\,\%CL).

The power spectrum of the best fit base $\Lambda$CDM cosmology has a higher amplitude than the observed power spectrum at multipoles $\ell<30$.  The low-$\ell$ difference is in turn related to the preference for a higher lensing amplitude when fitting to the temperature anisotropy power spectrum through a chain of parameter degeneracies \citep[see][for discussion]{planck2013-p11}.
There are other indications for `anomalies' at low $\ell$ (Sect.~\ref{sec:CMBNG}, \citet{planck2013-p09}), which may be indicative of new physics operating on the largest scales; however, the interpretation of such anomalies is difficult in the absence of a compelling theoretical framework.  In addition, our determination of the power spectrum amplitude is in weak tension
with that derived from the abundance of rich clusters found with the Sunyaev-Zeldovich effect in the \planck\ data \citep{planck2013-p15} and from measurements of cosmic shear from the CFHTLenS survey \citep{Heymans:12,Erben:12}.

\subsection{Isotropy and Gaussianity}
\label{sec:CMBNG}

Two of the fundamental assumptions of the standard cosmological model, that the initial fluctuations are statistically isotropic
and Gaussian, are rigorously examined in \citet{planck2013-p09} and \citet{planck2013-p09a}, using the four CMB maps described in Sect.~\ref{subsec:CMBmapNG}.  Realistic simulations incorporating essential aspects of the \Planck\ measurement process have been used to support the analysis.  Deviations from isotropy have been found in the data that are robust against changes in the component separation algorithm or mask used, or the frequency examined.  Many of these anomalies were previously observed in the \textit{WMAP} data on large angular scales (e.g., an alignment between the quadrupole and octopole moments, an asymmetry of power between two preferred hemispheres, and a region of significant decrement, the so-called Cold Spot), and are now confirmed at similar levels of significance ($\sim3\sigma$) but a higher level of confidence.   In spite of the presence of strong non-Gaussian and anisotropic emission coming from Galactic and extragalactic sources, the consistency of the tests performed on the four CMB maps produced by the component separation algorithms strongly favors a cosmological origin for the anomalies. Moreover, the agreement between \WMAP\ and the two independent instruments of \Planck\ argues against possible explanations based on systematic artifacts.

On the other hand, we find little evidence for non-Gaussianity, with the exception of a few statistical signatures that seem to be associated with specific anomalies.  In particular, we find that the quadrupole-octopole alignment is also connected to a low observed variance of the CMB signal with respect to the standard $\Lambda$CDM model. In addition, the hemispherical asymmetry is now found to persist to smaller angular scales, and can be described in the low-$\ell$ regime at a statistically significant level by a phenomenological dipole modulation model.  It is plausible that some of these features may be reflected in the angular power spectrum of the data, which shows a deficit of power on these scales. Indeed, when the two opposing hemispheres defined by the preferred direction are considered separately, the power spectrum shows a clear power asymmetry, as well as oscillations between odd and even modes that may be related to parity violation and phase correlations also detected in the data.  While these analyses represent a step forward in building an understanding of the anomalies, a satisfactory explanation based on physically motivated models is still lacking.

The search for specific types of non-Gaussianity (NG) in the statistics of the CMB anisotropies provides important clues to the physical origin of cosmological perturbations.  Indeed, perturbations generated during inflation are expected to display specific forms of NG. Different inflationary models, firmly rooted in modern theoretical particle physics, predict different {\it amplitudes} and {\it shapes} of NG.  Thus, constraints on primordial NG are complementary to constraints on the scalar spectral index of curvature perturbations and the tensor-to-scalar ratio, lifting the degeneracy among inflationary models that predict the same power-spectra.  The level of NG predicted by the simplest models of inflation, consisting of a single slowly-rolling scalar field, is low and undetectable even by \Planck. However, extensions of the simplest paradigm generically lead to levels of NG in CMB anisotropies that should be detectable. A detection of primordial NG would rule out \emph{all} canonical single-field slow-roll models of inflation, pointing to physics beyond the simplest inflation model. Conversely, a significant upper bound on the level of primordial NG, as we have obtained, severely limits extensions of the simplest paradigm. 

Inflationary NG can be characterized by the dimensionless non-linearity parameter $f_{\rm NL}$ \citep{planck2013-p09a}, which measures the amplitude of primordial NG of quadratic type in the comoving curvature perturbation mode.  We have estimated $f_{\rm NL}$ for various NG shapes, including the three fundamental ones, {\it local}, {\it equilateral}, and {\it orthogonal}, predicted by different classes of inflationary models.  Results for these three fundamental shapes, obtained using a suite of optimal bispectrum estimators \citep{planck2013-p09a}, are reported in Table~\ref{tab:fNL_KSW}, which gives independent estimates for each contribution.  The reported values have been obtained after marginalizing over the Poisson bispectrum contribution of diffuse point-sources and subtracting the bias due to the secondary bispectrum arising from the coupling of the Integrated Sachs-Wolfe effect (ISW; Sect.~\ref{sec:ISW}) and the weak gravitational lensing of CMB photons \citep{planck2013-p14}.   We also obtain constraints on key, primordial, non-Gaussian paradigms, including non-separable single-field models, excited initial states (non-Bunch-Davies vacua), and directionally-dependent vector field models, and we provide an initial survey of scale-dependent features and resonance models.  The absence of significant non-Gaussianity implies that the speed of sound of the inflaton field in these models must be within two orders of magnitude of the speed of light.

Moreover, we derive bispectrum constraints on a selection of specific inflationary mechanisms, including both general single-field inflationary models and multifield ones.  Our results lead to a lower bound on the speed of sound, $c_{s} > 0.02$ ($95\,\%$ CL), in the effective field theory parametrization of the inflationary model space.  Moving beyond the bispectrum, \Planck\ data also provide an upper limit on the amplitude of the trispectrum in the local NG model, $\tau_{\rm NL} < 2800$ ($95\,\%$ CL). 

The \Planck\ data have been used to provide stringent new constraints on cosmic strings and other defects \citep{planck2013-p20}. Using CMB power-spectrum forecasts for cosmic strings, we obtain new limits $G\mu/c^2 < 1.5 \times 10^{-7}$ for Nambu strings and $G\mu/c^2 < 3.2 \times 10^{-7}$ for field theory strings. Tighter constraints for joint analysis with high-$\ell$ data are also described, along with results for textures and semi-local strings. Complementary non-Gaussian searches using different methodologies also find no evidence for cosmic strings, with somewhat weaker constraints.

Alternative geometries and non-trivial topologies have also been analyzed \citep{planck2013-p19}. The Bianchi~VII$_{\rm h}$ models, including global rotation and shear, have been constrained, with the vorticity parameter $\omega_0<10^{-9}H_0$ (95\,\% CL). Topological models are constrained by the lack of matched circles or other evidence of large-scale correlation signatures, limiting the scale of the fundamental domain to the size of the diameter of the scattering surface in a variety of specific models.

\begin{table}[tmb] 
\begingroup
\newdimen\tblskip \tblskip=5pt
\caption{
Separable template-fitting estimates of primordial $f_{\rm NL}$ for local, equilateral, and orthogonal shapes, as obtained from the 
\texttt{SMICA} foreground-cleaned map, after marginalizing over the Poisson point-source bispectrum contribution and subtracting the ISW-lensing bias.  Uncertainties are $1\sigma$. Constraints for each shape are lower by factors ranging from 2 (equilateral shape) to 3 (local) compared with the \WMAP\ 9-year results.  \Planck\ shrinks the combined constraint volume in the space of the three standard bispectrum templates by a factor $\sim$21.}                           
\label{tab:fNL_KSW}                           
\nointerlineskip
\footnotesize
\setbox\tablebox=\vbox{
   \newdimen\digitwidth 
   \setbox0=\hbox{\rm 0} 
   \digitwidth=\wd0 
   \catcode`*=\active 
   \def*{\kern\digitwidth}
   \newdimen\signwidth 
   \setbox0=\hbox{+} 
   \signwidth=\wd0 
   \catcode`!=\active 
   \def!{\kern\signwidth}
\halign{\hfil#\hfil\tabskip=2em&
        \hfil#\hfil&
        \hfil#\hfil\tabskip 0pt\cr
\noalign{\doubleline}
\multispan3\hfil $f_{\rm NL}$\hfil\cr
\noalign{\vskip 0pt}
\multispan3\hrulefill\cr
Local&Equilateral&Orthogonal\cr
\noalign{\vskip 4pt\hrule\vskip 6pt}
$2.7\pm5.8$& $-42\pm75$& $-25\pm39$\cr
\noalign{\vskip 5pt\hrule}}}
\endPlancktable
\endgroup
\end{table}

\subsection{CMB polarization}
\label{sec:PolCMB}

The current data release and scientific results are based on temperature data only. \Planck\ measures polarization from 30 to 353\,GHz, and both DPCs routinely produce polarization products.  The analysis of polarization data is more complicated than that of temperature data: there are few celestial polarization sources that can be used for calibration;  polarized astrophysical foregrounds dominate the polarized CMB over the whole sky; and the detection of polarized signals is subject to specific systematic effects, such as leakage of total intensity into polarization through calibration errors between detectors in a polarized pair.    

These issues are not yet resolved at a level satisfactory for cosmological analysis at large angular scales ($\ell < $100).  At smaller angular scales and high Galactic latitudes, however, systematic effects are sub-dominant, and CMB polarization is already being measured by \Planck\ with unprecedented sensitivity.  Moreover, strong polarized synchrotron and thermal dust emission from the Galaxy are currently being imaged with high significance (Sect.~\ref{sec:DustPol}).

\Planck's ability to measure polarization is well illustrated by the use of stacking around CMB peaks (Fig.~\ref{fig:stack_pol_spots}; compare with \citealt{hinshaw2012}).  Adiabatic scalar fluctuations result in a specific polarization pattern around cold and hot spots.  We degrade ILC estimates (100--353\,GHz; \citealt{2004ApJ...612..633E}) of the CMB $I$, $Q$, and $U$ maps to HEALPix $N_{\rm side}=512$ and smooth to 30\arcm.  On the 71\,\% of the sky outside the \Planck\ component separation mask \citep{planck2013-p06}, we find  11\,396 cold spots and 10\,468 hot spots, consistent with the \LCDM\ \Planck\ best fit model prediction of $4\pi f_{\rm sky}\bar{n}_{\rm peak}=11\,073$ each.  Around each of these temperature extrema, we extract $5^\circ\times5^\circ$ square maps, which are stacked.  The stacked $Q$ and $U$ maps are rotated in the temperature extrema radial frame $Q_r(\theta)$ and $U_r(\theta)$ \citep{1997PhRvD..55.7368K}. In this reference frame the standard model predicts $Q_r(\theta)$ alternating between positive (radial polarization) and negative (tangential polarization) values and $U_r(\theta)=0$ \citep{baccigalupi1999}.

\begin{figure*}[t!]
\centering
\includegraphics[height=6.5cm,keepaspectratio]{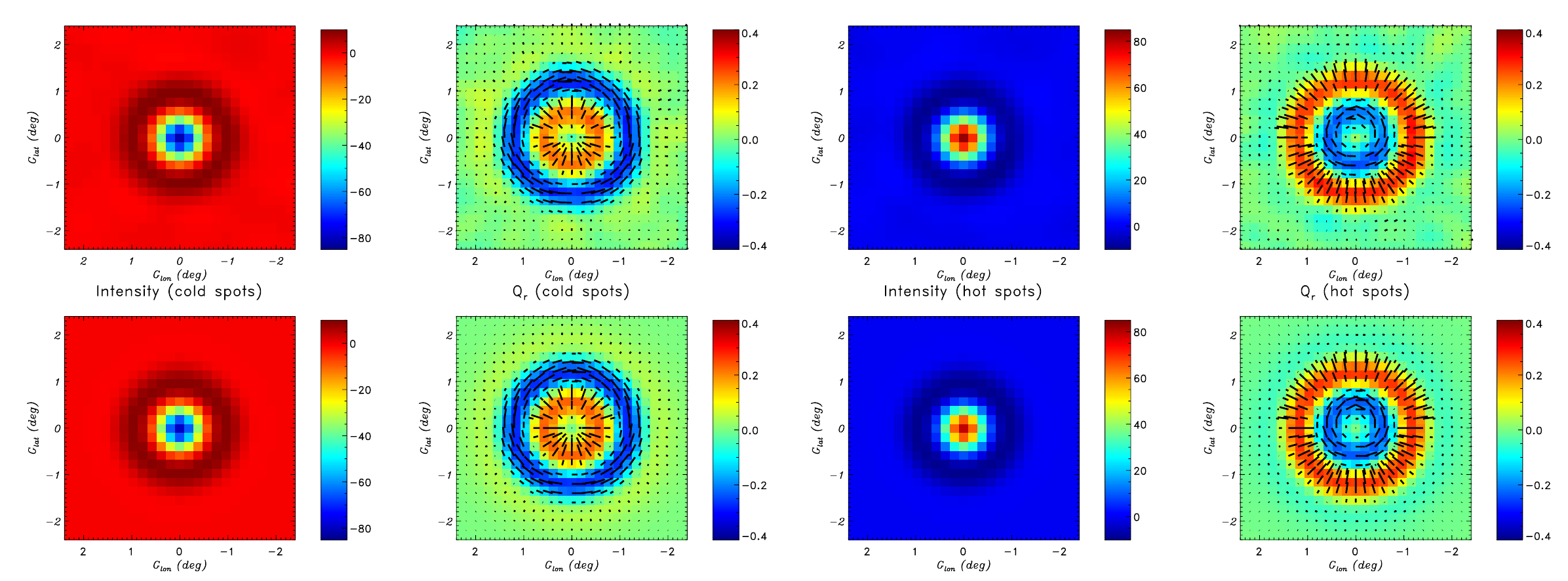}
\caption{Stacked maps of CMB intensity $I$ and polarization $Q_r$ at the position of the temperature extrema, at a common resolution of 30\arcm.   Maps stacked on CMB cold spots are on the left; maps stacked on hot spots are on the right.  Measured data on the top row are compared to the \Planck\  best-fit \LCDM\ model prediction on the bottom row.}
\label{fig:stack_pol_spots}
\end{figure*}

Figure~\ref{fig:stack_pol_spots} compares stacked $I$ and $Q_r$ maps for cold and hot spots computed from the \Planck\ data with those computed from the  \Planck\ best-fit \LCDM\ model.  The data are in excellent agreement with the best-fit model. The combined best fit amplitude is $0.999\pm0.010$ (68\,\% CL), a detection with a statistical significance greater than 95$\sigma$. 

The most interesting cosmological signal visible in polarization is the very large-scale ($\ell < 10$) $E$-mode peak due to reionization, at a typical brightness level well below one microkelvin. At the present stage of analysis, and with the data currently available, there are unexplained residuals in the Survey-to-Survey difference maps that are comparable to or larger than this.  For these reasons, we are delaying the use of CMB polarization measurements from \Planck\ for cosmological analysis until we have firmer understanding and control of such systematic effects.

\subsection{The ISW effect} 
\label{sec:ISW}

In the spatially flat Universe established by \Planck, the detection of the integrated Sachs-Wolfe (ISW) effect provides complementary evidence of the accelerated expansion of the Universe governed by some form of Dark Energy. The high sensitivity, high resolution, and full-sky coverage of \Planck\ has permitted us, for the first time, to obtain evidence of the ISW directly from CMB measurements, via the non-Gaussian signal induced by the cross-correlation of the secondary anisotropies due to the ISW itself and the lensing clearly detected by \Planck\ \citep{planck2013-p12}. Following this approach, we report an ISW detection of $2.5\sigma$ from the CMB alone.

In addition, we have also confirmed \citep{planck2013-p14} the ISW signal by cross-correlating the clean CMB maps produced by \Planck\ with several galaxy catalogues, which act as tracers of the gravitational potential.  Combining information from all the Surveys, this standard technique provides an overall detection of $3\,\sigma$. This figure is somewhat weaker than previous claims made from \WMAP\ data~\citep[e.g.,][]{Ho2008,Giannantonio2012}.  Differences do not seem to be related to the CMB data itself, but rather to the way in which the uncertainties are computed and, especially, to the characterization of the galaxy catalogues.  A small fraction of these differences (around 0$.3\,\sigma$) may be due to the different cosmological models determined by each experiment, in particular, the lower values of $H_0$ and $\Omega_\Lambda$ reported by \Planck.  Clear agreement with previous detection claims ($\lesssim 3\,\sigma$) using the NVSS data is reported.  The ISW amplitude estimation made with \Planck\ is in good agreement with the theoretical expectation (which depends on such characterizations), whereas deviations of more than $1\,\sigma$ were found in previous work.  These results give support and robustness to our findings.

The ISW signal induced by isolated features in the large-scale structure of the universe has also been studied. In particular, we stacked CMB fluctuations at the positions of voids and super-clusters, obtaining a clear detection ($>3\,\sigma$ and almost $3\,\sigma$ for voids and clusters, respectively) of a secondary anisotropy. The results are compatible with previous claims made with \WMAP\ data~\citep{Granett2008a}, and the most likely origin of the secondary anisotropy is the time evolution of the gravitational potential associated with those structures. However, the signal initially detected is at odds in scale and amplitude with expectations of a pure ISW effect. Using more recent void catalogues leads to the detection of a signal at up to $2.5\,\sigma$ with scales and amplitudes more consistent with expectations of the ISW effect. Taking advantage of the large frequency coverage of \Planck, we have confirmed that the stacked signal is stable from 44 to 353\,GHz, supporting  the cosmological origin of this detection.

\subsection{The cosmic infrared background}
\label{sec:CIB}

CIB anisotropies are expected to trace large-scale structures and probe the clustering properties of galaxies, which in turn are linked to those of their host dark matter halos. Because the clustering of dark matter is well understood, observations of anisotropies in the CIB constrain the relationship between dusty, star-forming galaxies and the dark matter distribution. Correlated anisotropies also depend on the mean emissivity per comoving unit volume of dusty, star-forming galaxies, and can be used to measure the star formation history.

The extraction of CIB anisotropies in \Planck\ \citep{planck2011-6.6, planck2013-pip56} is limited by our ability to separate the CIB from the CMB and Galactic dust.  At multipole $\ell$=100, the  power spectrum of the CIB anisotropies has an amplitude less than 0.2\,\%  of the CMB power spectrum at 217\,GHz, and less than 25\,\% of the dust power spectrum in very diffuse regions of the sky ($N_{\rm HI}<2.5\times10^{20}$\,cm$^{-2}$) at 857\,GHz.  Using \ion{H}{i} data from three radio telescopes (Parkes, GBT, and Effelsberg) and cleaning the CMB using the 100\,GHz map as a template, it has been possible to obtain new measurements of the CIB anisotropies with \Planck. The CIB has been extracted from the maps on roughly 2300\,deg$^2$ \citep{planck2013-pip56}. Auto- and cross-power spectra have been computed, from 143 to 3000\,GHz, using both \Planck\ and {\it IRAS}. Two approaches have been developed to model the power spectra. The first uses only the linear part of the clustering and gives strong constraints on the evolution of the star formation rate throughout cosmic time. The second is based on a parameterized relation between the dust-processed infrared luminosity and (sub-)halo mass, and probes the interplay between baryonic and dark matter up to very high redshifts, complementing current and foreseeable optical or near-infrared measurements.

The CIB, not the CMB, is the dominant  extragalactic signal at 857 and 545\,GHz, and even at 353\,GHz for $\ell>1100$.  Comprising redshifted thermal radiation from ultraviolet-heated dust enshrouding young stars, the CIB contains much of the energy from processes involved in structure formation.  According to current models, the dusty star-forming galaxies that give rise to the CIB have a redshift distribution peaked between $z\sim1$ and $z\sim 3$, and tend to live in $10^{11.5}$--$10^{13}\,\Msolar$ dark matter halos. 
At all redshifts, the dominant contribution to the SFR density is from halos of mass $\sim10^{12}$\,$\Msolar$.

\subsection{Lensing and the cosmic infrared background}
\label{sec:LensCIB}

\Planck's multi-frequency observations provide information on both the integrated history of star formation via the CIB and the distribution of dark matter via CMB lensing.  

Gravitational lensing by large-scale structure produces small shear and magnification effects in the observed fluctuations, which can be exploited to reconstruct an integrated measure of the gravitational potential along the line of sight. This ``CMB lensing potential'' is sourced primarily by dark matter halos located at $1\lesssim z \lesssim 3$, roughly halfway between ourselves and the last scattering surface. 

The conjunction of these two unique probes allows us to measure directly the connection between dark and luminous matter in the high redshift ($1 \le z \le 3$) Universe \citep{planck2013-p13}.  We use a three-point statistic optimized to detect the correlation between these two tracers, and report the first detection of the correlation between the CIB and CMB lensing using \Planck\ data only. The well-matched redshift distribution of these two signals leads to a detection significance with a peak value of $42\,\sigma$ at 545\,GHz.  Equivalently, we measure a correlation as high as 80\,\% across these two tracers.  Our full set of multi-frequency measurements (both CIB auto- and CIB-lensing cross-spectra) is consistent with a simple halo-based model, with the mean halo mass that is most efficient at hosting star formation being $\log_{10}\left(M/\Msolar\right) = 12.6$.  Leveraging the frequency dependence of our signal, we isolate the high redshift contribution to the CIB, and constrain the star formation rate (SFR) density at $z\geq 1$. We measure directly the SFR density with around 4$\,\sigma$ significance for three redshift bins between $z=1$ and 7, thus opening a new window into the study of the formation of stars at early times.

\begin{figure*}[!t]
 \includegraphics[width=0.9\textwidth]{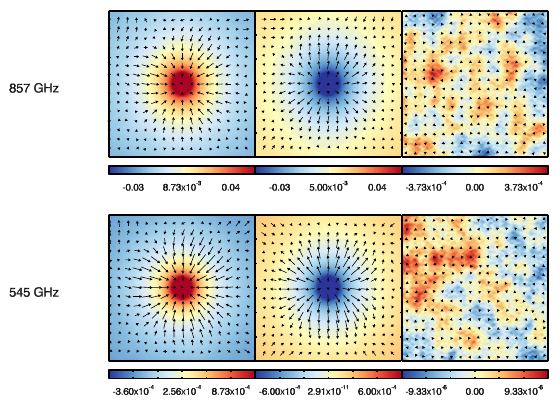}
  \caption[]{Temperature maps $1\deg \times 1\deg$ at 545 and 857\,GHz stacked on the 20\,000 brightest peaks (left column) and valleys (centre column), and on 20\,000 random map locations (right column). The temperature scale is in kelvin. The arrows indicate the lensing deflection angle deduced from the gradient of the band-pass-filtered lensing potential map \citep{planck2013-p12} stacked on the same peaks. The longest arrow corresponds to a deflection of 6\parcs3, which is only a fraction of the total deflection angle because of our filtering. This stacking allows us to visualize in real space the lensing of the CMB by the galaxies that generate the CIB. The small offset between the peak of the lensing potential and the CIB is due to noise in the stacked lensing potential map. We use the same random locations for both 545 and 857\,GHz,  hence the similar pattern seen in the top and bottom right panels.}
  \label{fig:stacking}
\end{figure*}

Figure~\ref{fig:stacking} shows the real-space correlation between the observed temperature and the lens deflection angles. This figure allows us to visualize the correlation between the CIB and the CMB lensing deflection angles for the first time.  These images were generated using a stacking technique, as described in \citet{planck2013-p13}.  We select $\sim$20\,000 local maxima and an equal number of local minima, stack them in one-degree squares, then take the gradient of the stacked lensing potential to calculate the deflection angles, shown in the figure as arrows. The result of stacking over the maxima, minima, and random points is displayed from left to right. The strong correlation seen already in the cross-power spectrum is clearly visible in both the 545 and 857\,GHz extrema, while stacking on random locations leads to a lensing signal consistent with noise.  As expected, we see that the temperature maxima of the CIB, which contain a larger-than-average number of galaxies, deflect light inward, i.e., they correspond to gravitational potential wells, while temperature minima trace regions with fewer galaxies and deflect light outward, i.e., they correspond to gravitational potential hills.

\subsection{Cosmology from \Planck\ Sunyaev--Zeldovich emission}
\label{sec:SZsci}

Clusters of galaxies are good tracers of the evolution and content of the Universe.  \citep{planck2013-p15} gives constraints obtained with a well defined sample of 189 clusters (Sect.~\ref{sec:PSZ}) for which we have computed the selection function.  This sample has $\hbox{SNR} \ge 7$   to ensure 100\,\% reliability and to maximize the number of redshifts (188).  Using a relation between mass and SZ signal based on comparison to X-ray measurements, we derive constraints on the matter power spectrum amplitude $\sigma_8$ and matter density parameter $\Omega_\mathrm{m}$ in a flat $\Lambda$CDM model.  Assuming a bias between the X-ray determined mass and the true mass in the range 0--30\,\%, motivated by comparison of the observed mass scaling relations to those from several sets of numerical simulations, we find that  $\sigma_8(\Omega_{\mathrm{m}}/0.27)^{0.3}=0.764 \pm 0.025$, with one dimensional ranges $\sigma_8=0.75\pm 0.03$ and $\Omega_{\mathrm{m}}=0.29\pm 0.02$. This result appears to be robust against the SNR cut, choice of sub-sample, mass function, or completeness assumptions. 

In addition to the above analysis based on cluster counts, we can derive cosmological constraints from the power spectrum of tSZ emission (see Sect.~\ref{sec:SZDiff}, and \citealt{planck2013-p05b}).  We have compared the \Planck\ angular power spectrum of the diffuse thermal SZ emission (tSZ) to theoretical models in order to set cosmological constraints. The two analyses exhibit a similar degeneracy relation between $\sigma_8$ and $\Omega_{\mathrm{m}}$. In particular, we measure  $\sigma_8(\Omega_{\mathrm{m}}/0.28)^{3.2/8.1}=0.784\pm 0.016$, with one-dimensional ranges $\sigma_8=0.74\pm 0.06$ and $\Omega_{\mathrm{m}}=0.33\pm 0.06$, in agreement with the constraints derived from SZ cluster counts.

The tSZ effect secondary anisotropies are expected to be non-Gaussian, thus extra and independent cosmological information can be
extracted from the higher-order moments of their distribution and from their bispectrum.  By computing the one-dimensional probability distribution
function of the tSZ map, we find $\sigma_{8} = 0.779 \pm 0.015$, compatible with the bispectrum-based estimate of $\sigma_{8} = 0.74 \pm 0.04$.

While these analyses show good consistency on the constraints from the SZ signal detected by \Planck\ (and with other cluster measurements), they favour somewhat low values of $\sigma_8$ and $\Omega_\mathrm{m}$ as compared to the CMB analysis (Sect.~\ref{sec:CMBPar}). This tension can be alleviated either by relaxing our assumption on the bias between X-ray mass and true mass, or by assuming massive neutrinos.      Although additional neutrino species are not required by analysis of fluctuations in the CMB (Sect.~\ref{sec:CMBPar}, the upper limits on $\Sigma m_\nu$ given there do not rule out additional massive neutrinos at a level that would reduce the tension in $\sigma_8$ and $\Omega_{\rm m}$.

\section{Summary and Conclusions}
\label{sec:summary}

We have summarized the data products and scientific results of \Planck's first 15.5 months of science operations.  Full descriptions of all aspects of the \Planck\ data analysis and science results in this 2013 release are given in the accompanying papers (Planck Collaboration 2013 II -- XXXI). 

\Planck\ has been a tremendous technical success.  The satellite and its complex cryogenic payload have operated without interruption over a lifetime  longer than initially planned by a factor of three, with payload performance as good as or better than expected from pre-launch ground testing. 

The 2013 data release fulfills the promise made at mission inception in 1995 by delivering: (a)~a set of nine well-characterized frequency maps with sub-percent calibration accuracy across the CMB channels; (b)~a map of the temperature anisotropies of the CMB limited only by unresolved foregrounds down to an angular resolution of 5\arcm; (c)~a catalogue of compact Galactic and extragalactic sources that represents an important improvement over the Early Release Compact Source Catalogue released in January 2011; (d)~a catalogue of galaxy clusters detected via the Sunyaev-Zeldovich effect, which increases by a factor of $\sim$10 the number of clusters detected by this technique; and (e)~a first-generation set of maps of diffuse foregrounds that includes the main sources of Galactic emission --- thermal dust and the cosmic infrared background at high frequencies and synchrotron, free-free, and anomalous emission at low frequencies. 

In addition to these long-planned products, the 2013 release includes: an all-sky map of dust opacity that represents an important improvement over the best previous {\it IRAS}-based product; an all-sky map of the CMB lensing deflection field, the first ever in its category; all-sky maps of the integrated emission of carbon monoxide, an important tracer of the interstellar medium; and the first measurement of the angular power spectrum of the diffuse Sunyaev-Zeldovich emission over a large part of the sky.

The main cosmological results of \Planck\ can be summarised as follows:
\begin{itemize}

\item Using a likelihood approach that combines \Planck\ CMB and lensing data, data from ACT and SPT at high $\ell$s to constrain foregrounds, and the \WMAP\ polarized likelihood function at low $\ell$s to constrain $\tau$, we have estimated the values of a six-parameter  \LCDM\ model with the highest accuracy ever. These estimates are highly robust, as demonstrated by the use of multiple methods based both on likelihood and on component-separated maps. 

\item The parameters of the \Planck\ 6-parameter \LCDM\ model are significantly different from those previously estimated. In particular, we find a weaker cosmological constant (by $\sim$2\,\%), more baryons (by $\sim$3\,\%), and more cold dark matter (by $\sim$5\,\%). The spectral index of primordial fluctuations is firmly established to be below unity, even when extending the \LCDM\ model with additional parameters.

\item We find no significant improvements to the best-fit model when extending the set of parameters beyond six, implying no need for new physics to explain the \Planck\  measurements.

\item The \Planck\ best-fit model is in excellent agreement with the most current BAO data. However, it requires a Hubble constant that is significantly lower ($\sim$67 km s$^{-1}$ Mpc$^{-1}$) than that determined from traditional measurement techniques, raising the possibility of systematic effects in the latter. 

\item An exploration of parameter space beyond the basic set leads to: (a)  establishing a value for the effective number of relativistic species (neutrinos) consistent with the standard value of 3.046;  (b) constraining the flatness of space-time to a level of 0.1\% (including BAO data); and (c) setting significantly improved constraints on the total mass of neutrinos (at $\sum m_\nu < 0.85\,\mathrm{eV}$), the abundance of primordial helium, and the running of the spectral index of the power spectrum \citep{planck2013-p11}.

\item We find no evidence at the current level of analysis for tensor modes, for a dynamical form of dark energy, or for time-variations of the fine structure constant  \citep{planck2013-p11}. 

\item We find some tension between the amplitude of matter fluctuations ($\sigma_8$) derived from CMB data and that derived from Sunyaev-Zeldovich data; we attribute this tension to uncertainties in cluster physics and measurements that affect the latter.

\item We find important support for single-field slow-roll inflation via our constraints on running of the spectral index, curvature, and $f_{NL}$. 

\item The \Planck\  data squeeze the region of allowed standard inflationary models, preferring a concave potential.  Power law inflation, the simplest hybrid inflationary models, and simple monomial models with $n>2$, do not provide a good fit to the data.

\item We find no evidence for statistical deviations from isotropy at $\ell >600$, to very high precision \citet{planck2013-p09}.

\item We do find evidence for deviations from isotropy at low multipoles, confirming the existence of the so-called \WMAP\ anomalies, the hemispheric asymmetry and the Cold Spot. 

\item We find a coherent deficit of power at multipoles between 20 and 30 with respect to our best-fit \LCDM\ model.  

 \end{itemize}

These results highlight the maturity and precision being achieved in our understanding of the Universe, but at the same time reveal small differences between the data and the best-fit model whose significance is not yet fully understood.

Other results for which the current \Planck\ data are making unique contributions are:
\begin{itemize}

\item A $25\sigma$ detection of the distortion of the CMB due to lensing by intervening structure yields a highly significant map over most of the sky of the integrated distribution of mass back to the CMB surface of last scattering.  The detection of lensing helps \Planck\ to break parameter degeneracies, in particular to constrain the reionization optical depth without the help of polarization data.

\item The first detection at high significance ($42\sigma$) of the cross-correlation between CMB lensing and the cosmic infrared background, which allows us to constrain the star formation rate at high redshifts.

\item The measurement of the angular power spectrum of the cosmic infrared background over a large area and at frequencies as low as 217\,GHz, which allows us to constrain the properties of dark matter halos at high redshifts.

\item The first power spectrum of the diffuse Sunyaev-Zeldovich emission over the range $60 \le \ell \le 1000$, used to constrain the amplitude of matter fluctuations ($\sigma_8$).

\item A $2.5\sigma$ detection of the Integrated Sachs-Wolfe effect via its cross-correlation with \Planck-detected lensing, providing independent evidence for $\Omega_\Lambda\sim$0.7.
  
\end{itemize}

The \Planck\ 2013 release does not include polarization products.   Our current cosmological analysis relies not at all on \Planck\ polarization data, and only mildly on \WMAP\ polarization data.  However, we have shown that quite basic processing of the CMB polarization already yields angular power spectra in excellent agreement with the \Planck\ best-fit cosmology derived from temperature data only.  Analysis of the stacking of hot and cold CMB peaks shows spectacular agreement with expectations, and demonstrates the potential of the \Planck's CMB polarization measurements. A number of papers on polarized dust emission are due to be published within a few months. All these points show that the processing of \Planck\ polarization data is well advanced, and progressing towards the goal of releasing polarized data and associated results in 2014.

%________________________________________________________________

\begin{acknowledgements}

\Planck\ is too large a project to allow full acknowledgement of
all contributions by individuals, institutions, industries, and
funding agencies. The main entities involved in the mission operations are as
follows. 
The European Space Agency operates the satellite via its Mission
Operations Centre located at ESOC (Darmstadt, Germany) and coordinates
scientific operations via the \Planck\ Science Office located at ESAC (Madrid, Spain).
Two Consortia, comprising around 100
scientific institutes within Europe, the USA, and Canada, and funded by
agencies from the participating countries, developed the
scientific instruments LFI and HFI, and 
continue to operate them via Instrument Operations Teams located in Trieste
(Italy) and Orsay (France).
The Consortia are also responsible for scientific
processing of the
acquired data. The Consortia are led by the Principal Investigators:
J.-L. Puget in France for HFI (funded principally by CNES and CNRS/INSU-IN2P3)
and N.  Mandolesi in Italy for LFI (funded principally via ASI). 
NASA's US \Planck\ Project, based at JPL and involving
scientists at many US institutions, contributes 
significantly to the efforts of these two Consortia. 
A third Consortium, led by H. U. Norgaard-Nielsen and supported by the 
Danish Natural Research Council, contributed to the reflector programme.
The author list
for this paper has been selected by the \Planck\ Science Team from the Planck Collaboration, and
is composed of individuals from all of the above entities who have
made multi-year contributions to the development of the mission. It
does not pretend to be inclusive of all contributions to \Planck. 
A description of the Planck Collaboration and a list of its members, 
indicating which technical or scientific activities they have been involved in, 
can be found at \emph{(http://www.rssd.esa.int/index.php?project=PLANCK\& page=Planck\_ Collaboration)}. 
The Planck Collaboration acknowledges the support of: ESA; CNES, and CNRS/INSU-IN2P3-INP (France); 
ASI, CNR, and INAF (Italy); NASA and DoE (USA); STFC and UKSA (UK); 
CSIC, MICINN, and JA (Spain); Tekes, AoF, and CSC (Finland); DLR and MPG (Germany); 
CSA (Canada); DTU Space (Denmark); SER/SSO (Switzerland); RCN (Norway); SFI (Ireland); 
FCT/MCTES (Portugal); and PRACE (EU).

\end{acknowledgements}

\bibliographystyle{aa}

% we load the bilbiographic file containing the entries
%\bibliography{Planck_bib,Planck_bib_Missionentries}
%  IMPORTANT !!! Planck_bib is just a dynamic link to the actual file which is located under "/Volumes/Macintosh\ HD/Users/jtauber/Work/Planck/Publications/Planck_Publication_Management/Repositories/BibTex/Planck_bib.bib" or
% "~/Work/Planck/Publications/Planck_Publicaiton_Management/Repositories/Repositories/BibTex/Planck_bib.bib"
\bibliography{Planck_bib,Planck_bib_Missionentries}

\raggedright
\end{document}